\documentclass[%
 reprint,
superscriptaddress,
preprintnumbers,
nofootinbib,
 amsmath,amssymb,
 aps,
]{revtex4-2}

\usepackage[colorlinks]{hyperref}
\usepackage{graphicx}
\usepackage{dcolumn}
\usepackage{bm}
\usepackage[dvipsnames]{xcolor}
\usepackage{placeins}
\usepackage{slashed}
\usepackage{multirow}
\usepackage{array}
\usepackage{tabularx}
\usepackage{mathtools}
\usepackage{wasysym}
\usepackage{enumitem}
\usepackage{mathrsfs} 

\newcommand{\eq}{\mathrm{eq}}
\newcommand{\eff}{\mathrm{eff}}
\newcommand{\vrel}{v_{\mathrm{rel}}}

\definecolor{red}{rgb}{0.9, 0,0}
\definecolor{cerulean}{rgb}{0., 0.62,0.9}
\definecolor{navy}{rgb}{0.05, 0.05,0.8}
\definecolor{orange}{rgb}{0.8, 0.4, 0.}

\renewcommand{\eqref}[1]{Eq.~(\ref{#1})}
\newcommand{\mgfive}{ MadGraph5\_aMC@NLO }
\newcommand{\fig}{Fig.}
\newcommand{\sect}{Sec.}
\newcommand{\tabl}{Table}

\begin{document}

\title{Closing the window on WIMP Dark Matter}

\author{Salvatore Bottaro}
\affiliation{Scuola Normale Superiore, Piazza dei Cavalieri 7, 56126 Pisa, Italy}
\affiliation{INFN, Sezione di Pisa, Largo Bruno Pontecorvo 3, I-56127 Pisa, Italy}
\author{Dario Buttazzo}
\affiliation{INFN, Sezione di Pisa, Largo Bruno Pontecorvo 3, I-56127 Pisa, Italy}
\author{Marco Costa}
\affiliation{Scuola Normale Superiore, Piazza dei Cavalieri 7, 56126 Pisa, Italy}
\affiliation{INFN, Sezione di Pisa, Largo Bruno Pontecorvo 3, I-56127 Pisa, Italy}
\author{Roberto Franceschini}
\affiliation{Universit\`a degli Studi and INFN Roma Tre, Via della Vasca Navale 84, I-00146, Rome}
\author{Paolo Panci}
\affiliation{Dipartimento di Fisica E. Fermi, Universit\`a di Pisa, Largo B. Pontecorvo 3, I-56127 Pisa, Italy}
\affiliation{INFN, Sezione di Pisa, Largo Bruno Pontecorvo 3, I-56127 Pisa, Italy}
\author{Diego Redigolo}
\affiliation{CERN, Theoretical Physics Department, Geneva, Switzerland.}
\affiliation{INFN, Sezione di Firenze Via G. Sansone 1, 50019 Sesto Fiorentino, Italy}
\author{Ludovico Vittorio}
\affiliation{Scuola Normale Superiore, Piazza dei Cavalieri 7, 56126 Pisa, Italy}
\affiliation{INFN, Sezione di Pisa, Largo Bruno Pontecorvo 3, I-56127 Pisa, Italy}

\preprint{CERN-TH-2021-XXX}

\date{\today}

\begin{abstract}
{We study scenarios where Dark Matter is a weakly interacting particle (WIMP) embedded in an ElectroWeak multiplet. In particular, we consider real SU(2) representations with zero hypercharge, that automatically avoid direct detection constraints from tree-level $Z$-exchange. We compute for the first time \emph{all the calculable thermal masses} for  scalar and fermionic WIMPs, including  Sommerfeld enhancement and bound states formation at leading order in gauge boson exchange and emission.
WIMP masses of few hundred TeV are shown to be compatible both with $s$-wave unitarity of the annihilation cross-section, and perturbativity.
We also provide theory uncertainties on the masses for all multiplets, which are shown to be significant for large SU(2) multiplets.
We then outline a strategy to probe these scenarios at future experiments. Electroweak 3-plets and 5-plets have masses up to about 16~TeV and can efficiently be probed at a high energy muon collider. We study various experimental signatures, such as single and double gauge boson emission with missing energy, and disappearing tracks, and determine the collider energy and luminosity required to probe the thermal Dark Matter masses.
Larger multiplets are out of reach of any realistic future collider, but can be tested in future $\gamma$-ray telescopes and possibly in large-exposure liquid Xenon experiments.} 
\end{abstract}

\maketitle


\section{Introduction}

The possibility that Dark Matter (DM) is a new weakly interacting massive particle (WIMP), thermally produced in the early Universe and freezing out through $2\!\to\! 2$ annihilations into Standard Model (SM) states, remains one of the main motivations for new physics in the 10~GeV~--~100~TeV range. Under these simple assumptions, the lower bound on the WIMP mass comes from astrophysical constraints on DM annihilations into SM products~\cite{Leane:2018kjk}, while the upper bound is a consequence of 
$s$-wave unitarity of the DM annihilation cross-section~\cite{Griest:1989wd}.  

A particularly interesting possibility within this framework, because of its minimality and predictive power, is that the DM is the lightest neutral component of one electroweak (EW) multiplet. In particular, fermionic and scalar $n$-plets of SU(2) with odd $n$ and zero hypercharge automatically avoid strong constraints from direct detection searches, and will be taken here as a minimal realization of the EW WIMP scenario.
The lightest particle in any such representation can be made stable by enforcing a symmetry acting on the DM only (for multiplets with $n\geq 5$ such a symmetry arises accidentally in the renormalizable Lagrangian). However, we shall see that in general this can require additional assumptions about the completion of the theory at some high UV scale.

\begin{figure}
\includegraphics[width=0.45\textwidth]{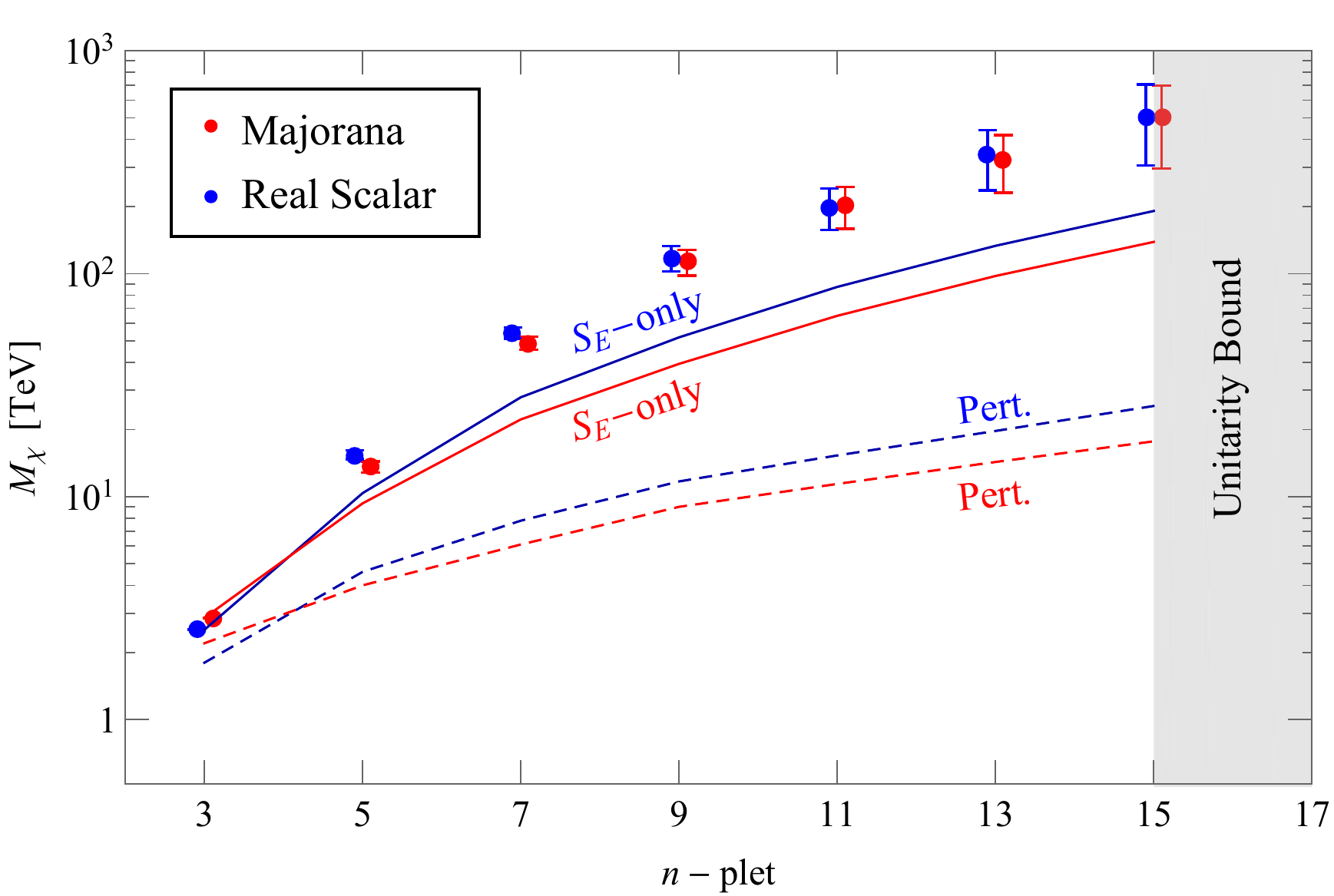}
\caption{Summary of the thermal masses for Majorana fermion (red) and real scalar WIMPs (blue) including both Sommerfeld enhancement (SE) and bound state formation (BSF). The solid lines are the thermal masses with SE. The dashed lines are the thermal masses for the hard annhilation cross-section. The gray shaded region is excluded by $s$-wave perturbative unitarity including BSF. }
\label{fig:summary}
\end{figure}

The main purpose of this paper is to precisely determine the WIMP freeze-out predictions in a systematic way.
For any given $n$-plet, computing the EW annihilation cross-section in the early Universe allows to infer the WIMP cosmological abundance. By requiring it to match the measured value of the DM abundance today, $\Omega_{\rm DM} h^2 = 0.11933\pm 0.00091$ \cite{Planck:2018vyg}, the mass of the $n$-plet can be univocally determined. These mass predictions are an essential input to assess if and how the future experimental program will be able to fully test the EW WIMP scenario.
In contrast to previous papers on the subject~\cite{Cirelli:2005uq,Cirelli:2007xd,Cirelli:2009uv,Hambye:2009pw}, our approach here is to minimize the theory assumptions and fully classify the \emph{calculable} freeze-out predictions. Because of its infrared-dominated nature, the calculability of freeze-out depends purely on the partial wave unitarity of the total annihilation cross-section~\cite{Griest:1989wd}, which we re-analyze here 
for EW $n$-plets. All in all, demanding perturbative unitarity requires $n\leq13$ for both bosonic and fermionic DM. Approaching this boundary the theory uncertainty on the mass prediction grows as shown in \fig~\ref{fig:summary}. Stronger constraints on $n$ can be imposed by demanding the EW interactions to remain perturbative up to scales 
well above the thermal DM mass.

The effects of Sommerfeld enhancement (SE) and of bound state formation (BSF) are known to significantly affect the freeze-out predictions and
need to be included. The first effect has long been recognized to lead to an enhancement of the annihilation cross-section at small relative velocities~\cite{Hisano:2003ec,Hisano:2006nn,ArkaniHamed:2008qn,Cassel:2009wt}.  The effects of BSF for WIMP freeze-out have been first computed in Ref.~\cite{Mitridate:2017izz} for the $n=5$ fermionic multiplet (see Ref.s~\cite{vonHarling:2014kha,Cirelli:2016rnw} for earlier computations in other contexts). Here we extend their treatment to fermionic and scalar representations of arbitrary high $n$, up to the break-down of perturbative unitarity.  At growing $n$, we find that bound states (BS) are more tightly bound, with their ionization rate being exponentially suppressed. At the same time, the multiplicity of accessible BS channels grows significantly. These two effects result in an increase of the annihilation cross-section compared to the estimates of Ref.~\cite{Smirnov:2019ngs}. 

The freeze-out mass predictions are summarized in \tabl~\ref{table:summary} and \fig~\ref{fig:summary} for the real $n$-plets considered here.
With masses ranging from several TeV to tens or hundreds of TeV, most of the EW WIMP candidates are still out of reach of present experiments, but could be tested in the future, thanks to the forthcoming progress in collider physics and DM detection experiments.
With the mass predictions at hand, we thus commence a systematic survey of the WIMP phenomenology: $i)$ at very high energy lepton colliders with 10 to 30 TeV center of mass energy~\cite{Delahaye:2019omf,ALEGRO:2019alc}; $ii)$ at direct detection experiments with 100 tons/year of exposure like DARWIN~\cite{Schumann:2015cpa,Aalbers:2016jon}; $iii)$ at high-energy $\gamma$-ray telescopes like CTA~\cite{Lefranc:2016fgn,Acharyya:2020sbj,Silverwood:2014yza,Lefranc:2015pza}.    
We first examine the reach of a hypothetical future muon collider, studying in detail for which values of center-of-mass energy and integrated luminosity the EW $3$-plets and $5$-plets can be fully probed through direct production. We instead find direct production of the EW multiplets with $n>5$ to be beyond the reach of any realistic future machine (this is in contrast with the results of the recent study~\cite{Han:2020uak} due to the increase of the thermal mass of the 7-plet with the inclusion of BSF effects).
These larger $n$-plets are possibly within the reach of large exposure direct detection experiments, and will probably be tested more easily with future high energy $\gamma$-ray telescopes. A careful study of the expected signals in indirect detection is left for a future work~\cite{futureind}.       

This paper is organized as follows. In \sect~\ref{sec:WIMP} we summarize the EW WIMP paradigm, in \sect~\ref{sec:cosmo} we illustrate the main features of our freeze-out computation, and in \sect~\ref{sec:unitaritybound} we discuss the unitarity bound assessing the theory uncertainties. These three sections provide a full explanation on the results of \tabl~\ref{table:summary} and \fig~\ref{fig:summary}. In \sect~\ref{sec:collider} we discuss the implications of our study for a future muon collider, while in \sect~\ref{sec:directandindirect} we briefly re-examine the reach of direct and indirect detection experiments in light of our findings. In \appendixname~\ref{app:boundstates} we give further details on the nature of next-to-leading order corrections and we detail the BS dynamics for the 7-plet. \appendixname~\ref{app:collider} contains further information on the collider studies.  

\begin{table*}[t]
\renewcommand{\arraystretch}{1.3}
\begin{center}
 \begin{tabular}{ c | c |  c | c | c | c }
            DM spin             &EW n-plet & $M_{\chi}$ (TeV) & $(\sigma v)^{J=0}_{\text{tot}}/(\sigma v)^{J=0}_{\text{max}}$& $\Lambda_{\text{Landau}}/M_{\text{DM}}$ &  $\Lambda_{\text{UV}}/M_{\text{DM}}$ \\ \hline
  \multirow{6}{*}{Real scalar} 
                                               & $3$ & $2.53\pm 0.01$ & --  & $2.4\times 10^{37}$  &$4\times 10^{24}$*\\
                                               &  $5$ & $15.4\pm0.7$ & 0.002 &  $7\times 10^{36}$  & $3\times 10^{24}$\\ 
                                               &  $7$ & $54.2\pm 3.1$ &0.022& $7.8\times 10^{16}$ &$2\times 10^{24}$\\
                                               &  $9$ & $117.8\pm 15.4$ &0.088 & $3\times 10^{4}$& $2\times 10^{24}$\\
                                               &  $11$ &  $199\pm 42$                    &    0.25    &   62  &  $1\times 10^{24}$                                         \\
                                               &  $13$ & $ 338 \pm 102$ &0.6 & 7.2  & $2\times 10^{24}$  \\
                                              \hline
     \multirow{6}{*}{Majorana fermion}
     & $3$ & $2.86\pm 0.01$ & -- & $2.4\times 10^{37}$ & $2\times 10^{12}$*\\
                                                          & $5$ & $13.6\pm0.8$ &0.003 & $5.5\times10^{17}$ & $3\times 10^{12}$  \\
                                                          & $7$ & $48.8\pm 3.3$ &0.019 & $1.2\times 10^4$ &$1\times 10^{8}$\\
                                                           &$9$ & $113\pm 15$ &0.07&41 &$1\times 10^{8}$\\
                                                          &$11$ & $202\pm 43$ & 0.2 & 6  &$1\times 10^{8}$\\ 
                                                          &$13$ & $324.6\pm 94$ & 0.5 & 2.6 &$1\times 10^{8}$\\                                                          
  \end{tabular}
 \caption{Freeze-out mass predictions for WIMP DM in real EW multiplets with $Y=0$. The annihilation cross-section includes both the contribution of SE and BSF. We provide a measure of how close the DM annihilation cross-section is  to the unitarity bound for $s$-wave annihilation $(\sigma v)^{J=0}_{\text{max}}=4\pi/M_{\text{DM}}^2v$. Approaching the unitarity bound, the error on the WIMP mass grows proportionally to the enhancement of the next-to-leading order (NLO) contributions estimated in \eqref{eq:1loop}. We derive the scale where EW gauge coupling will develop a Landau pole by integrating-in the WIMP multiplet at its freeze-out mass. The stability of both scalar and fermionic DM can always be enforced by requiring a $\mathbb{Z}_2$ symmetry in the DM sector to forbid DM decays. This symmetry forbids the scalar and fermionic 3-plets decay at renormalizable level as indicated by the~*. The value of the UV cut-off $\Lambda_{\text{UV}}$ gives an idea of the required \emph{quality} for this symmetry to make DM stable and avoid stringent bounds on decaying DM ($\tau_{\text{DM}}>10^{28}\text{sec}$)~\cite{Cohen:2016uyg}: a new physics scale lower than $\Lambda_{\text{UV}}$ would require a $\mathbb{Z}_2$ to explain DM stability, while a cut-off higher than $\Lambda_{\text{UV}}$ would make DM stability purely accidental. \label{table:summary}}
\end{center}
\end{table*}

\section{Which WIMP?}\label{sec:WIMP}

We summarize here the logic of our WIMP classification very much inspired by previous papers on the subject~\cite{Cirelli:2005uq,Cirelli:2007xd,Cirelli:2009uv,Hambye:2009pw,DelNobile:2015bqo}. Requiring the neutral DM component to be embedded in a representation of the EW group imposes that $Q=T_3+Y$, where $T_3=\text{diag}\left(\frac{n+1}{2}-i\right)$ with $i=1,\dots, n$, and $Y$ is the hypercharge. At this level, we can distinguish two classes of WIMPs: $i)$ real EW representations with $Y=0$ and odd $n$; $ii)$ complex EW representations with arbitrary $n$ and $Y=\pm\left( \frac{n+1}{2}-i\right)$ for $i=1,\dots, n$.  Here we focus on the first class of WIMPs, which is particularly interesting because the DM does not couple to the $Z$-boson at  tree level, avoiding strong constraints from direct detection experiments. Other possibilities will be discussed elsewhere. 

At the renormalizable level, the extensions of the SM that we consider are
\begin{align}
\mathscr{L}_{\text{s}} &=\frac{1}{2}\left(D_\mu \chi\right)^2-\frac{1}{2}M_{\chi}^2\chi^2-\frac{\lambda_H}{2}\chi^2  \vert H\vert^2-\frac{\lambda_\chi}{4}\chi^4\label{eq:scalarWIMP}\, , \\
\mathscr{L}_{\text{f}} &=\frac{1}{2}\chi \left (i\bar{\sigma}^{\mu} D_\mu-M_{\chi}\right)\chi\, , \label{eq:fermionWIMP}
\end{align}
for scalars and fermions, respectively,
where $D_\mu=\partial_\mu-i g_2 W_\mu^a T^a_\chi$ is the covariant derivative, and $T^a_\chi$ are generators in the $n$-th representation of SU(2). The Lagrangian for the real scalar in \eqref{eq:scalarWIMP} also admits quartic self-coupling and Higgs-portal interactions at the renormalizable level. The latter is bounded from above by direct detection constraints (see Fig.~\ref{fig:dd} right) and gives a negligible contribution to the annihilation cross-section.\footnote{No other quartic coupling is allowed since $\chi T^a_\chi \chi$ identically vanishes. Indeed, $(T^a_\chi)_{ij}$ is antisymmetric in $i,j$, being the adjoint combination of two real representations, while $\chi_i\chi_j$ is symmetric.}

The neutral component and the component with charge $Q$ of the EW multiplet are splitted by radiative contributions from gauge boson loops. In the limit $m_W\ll M_{\text{DM}}$ these contributions are non-zero and independent on $M_{\chi}$. This fact can be understood  by computing the Coulomb energy of a charged state at distance $r\gtrsim 1/m_W$ or the IR mismatch (regulated by $m_W$) between the self-energies of the charged and neutral states. The latter can be easily computed at 1-loop~\cite{Cheng:1998hc,Feng:1999fu,Gherghetta:1999sw},
\begin{equation}
M_Q-M_0\simeq \frac{Q^2\alpha_{\text{em}}m_W}{2(1+\cos\theta_W)}=Q^2\times \left(167\pm4\right)\text{ MeV}\ ,\label{eq:splitting}
\end{equation}
with the uncertainty dominated by 2-loop contributions proportional to $\alpha_2^2m_t/16\pi$. These have been explicitly computed in Ref.s~\cite{Ibe:2012sx,McKay:2017xlc} giving a precise prediction for the lifetime of the singly-charged component, which decays to the neutral one mainly by emitting a charged pion with
\begin{equation}
c\tau_{\chi^+}\simeq \frac{120\text{ mm}}{T(T+1)}\ ,
\end{equation}
where $2T+1=n$. The suppression of the lifetime with the size of the EW multiplet can be understood in the $M_\chi \gg m_W$ limit where the mass splitting between the charged and neutral components is independent of $n$ while the coupling to $W$ is controlled by $\sqrt{T(T+1)/2}$. As we will discuss in \sect~\ref{sec:collider_dt}, the production of a singly charged DM component at colliders gives the unique opportunity of probing EW multiplets with $n=3$ and $n=5$ through disappearing tracks~\cite{Cirelli:2005uq,Low:2014cba,Cirelli:2014dsa,Capdevilla:2021fmj,Han:2020uak}.    

Interestingly, the IR generated splitting from gauge boson loops is not modified substantially by UV contributions. The latter are generated only by dimension 7 (dimension 6) operators if the DM is a Majorana fermion (real scalar) and can be written as
\begin{equation}
\Delta\mathscr{L}_I\supset\frac{c_I}{\Lambda_{\text{UV}}^{n_I}}\chi^a \chi^b(H^{\dagger} T^a H) (H^{\dagger} T^b H)\ ,\\
\end{equation}
with $n_I=3,2$ for $I=f,s$. This corresponds to a splitting $\Delta M_I\simeq c_I v^4/\Lambda_{\text{UV}}^{n_I}M_{\chi}^{3-n_I}$ which is always negligible with respect to the residual error on the 2-loop splitting for $\Lambda_{\text{UV}}\gtrsim 100\text{ TeV}$ and $c_I\sim\mathcal{O}(1)$. 

We now move to discuss DM stability. In the case of the EW 3-plet, the renormalizable operators $\chi H^{\dagger}H$ and $\chi H L$, for scalars and fermions, respectively, can induce fast DM decay. We assume these operators to be forbidden by a symmetry (e.g.\ a discrete $\mathbb{Z}_2$-symmetry) acting only on the DM sector.
For all the other $n$-plets with $n\geq 5$, instead, $\mathbb{Z}_2$-odd operators are accidentally absent at renormalizable level.

Higher dimensional operators that break the $\mathbb{Z}_2$-symmetry are in general expected to be generated at the ultraviolet cut-off scale $\Lambda_{\text{UV}}$.
We sketch here the operators of lowest dimension that can induce the decay of scalar and fermionic WIMPs for generic $n$:
\begin{widetext}
\begin{align}
&\mathscr{L}_{\rm s}\supset
\frac{C_{1}^{(s)}}{\Lambda_{\text{UV}}^{n-4}} \chi (H^\dag H)^{\!\frac{n-1}{2}}
+ \frac{C_{2}^{(s)}}{\Lambda_{\rm UV}^{n-4}}\chi W_{\mu\nu}W^{\mu\nu}(H^\dag H)^{\frac{n-5}{2}} + \cdots
+ \frac{C_{w}^{(s)}}{\Lambda_{\rm UV}^{n-4}}\chi (W_{\mu\nu}W^{\mu\nu})^{\frac{n-1}{4}}
+ \frac{C_{3\chi}^{(s)}}{\Lambda_{\text{UV}}}\chi^3 H^{\dagger}H, \label{eq:scalardecay}\\
&\mathscr{L}_{\rm f}\supset
\frac{C_{1}^{(f)}}{\Lambda_{\text{UV}}^{n-3}} (\chi H L) (H^{\dagger}\!H)^{\!\frac{n-3}{2}}
+ \frac{C_{2}^{(f)}}{\Lambda_{\text{UV}}^{n-3}} (\chi \sigma^{\mu\nu} H L) W_{\mu\nu} (H^{\dagger}\!H)^{\!\frac{n-5}{2}}
+ \cdots
+ \frac{C_{w}^{(f)}}{\Lambda_{\text{UV}}^{n-3}} (\chi H L) (W_{\mu\nu}W^{\mu\nu})^{\!\frac{n-3}{4}}
+ \frac{C_{3\chi}^{(f)}}{\Lambda_{\text{UV}}^3} \chi^3 H L ,\label{eq:fermiondecay}
\end{align}
\end{widetext}
where SU(2) contractions are implicit, and the dots indicate operators of the same dimension with different combinations of $W$ and $H$ fields.\footnote{If $(n-1)/4$ is not integer, the operator with the highest number of $W$ fields in \eqref{eq:scalardecay} is $\chi (H^\dag H)(W_{\mu\nu}W^{\mu\nu})^{\frac{n-3}{4}}$. Similarly, for the fermions in \eqref{eq:fermiondecay} it is $(\chi\sigma_{\mu\nu}HL) W^{\mu\nu}(W_{\rho\sigma}W^{\rho\sigma})^{\frac{n-5}{4}}$.} Higher-dimension operators with additional SM fields or derivatives are of course also possible.
The first operators in the two equations above are just the renormalizable operators of the 3-plet case ``dressed'' with extra Higgs insertions.
The dominant contribution to the decay width at tree-level always comes from the operator with the highest number of $W$ insertions (namely $(n-3)/2$ for fermions and $2\lfloor (n-1)/4\rfloor$ for scalars).
Notice that for fermionic DM, dipole-like operators with an odd number of $W$ fields can always be constructed.
In the last operator in both \eqref{eq:scalardecay} and \eqref{eq:fermiondecay}, $\chi^3$ is the unique isospin triplet constructed out of three SU(2) irreducible representations of odd isospin~\cite{DiLuzio:2015oha,DelNobile:2015bqo}. 
These operators contribute to the WIMP decay at one-loop as
\begin{align}
& \Gamma_{\rm s,f}\sim \frac{M_\chi }{2048 \pi^5} \!\left(\frac{\alpha_2 (n^2-1)}{4\pi}\right)^{\!\!\!\frac{n-3}{2}} \!\left[ C_{3\chi}^{(s,f)} \!\left(\frac{M_\chi}{\Lambda_{\rm UV}} \right)^{\!\! q}\right]^2\!,\!\!\!\!
&
\end{align}
where the exponent $q=1$ (3) holds for scalars (fermions).
For both scalar and fermionic WIMPs these are the dominant contributions for multiplets with $n > 5$.
More precise results for specific $n$-plets have been computed in Ref.s~\cite{DiLuzio:2015oha,DelNobile:2015bqo} but do not modify our conclusions.
For \emph{all} the scalar $n$-plets, DM decay is induced by a dimension 5 operator, and the required scale for stability is well above $M_{\text{Pl}}$. As a consequence, the stability of scalar WIMPs can be determined only by understanding the subtle issues related to the fate of discrete symmetries in quantum gravity~\cite{Banks:2010zn}. For fermionic representations, DM decay is instead induced by dimension 6 operators for $n\leq5$, and dimension 7 operators for $n>5$, and the DM stability can be determined within quantum field theory.   

A lower bound on $\Lambda_{\text{UV}}$ is obtained by requiring the DM lifetime to be long enough to circumvent cosmological bounds~\cite{Audren:2014bca,Aubourg:2014yra} ($\tau_{\text{DM}}\gtrsim10^{19} \text{ sec}$) or astrophysical bounds on the decay products of decaying DM~\cite{Cohen:2016uyg,Ando:2015qda,Cirelli:2012ut} ($\tau_{\text{DM}}\gtrsim10^{28}\text{ sec}$). We can then quantitatively measure the required \emph{quality} of the $\mathbb{Z}_2$-symmetry by considering the ratio between the minimal $\Lambda_{\text{UV}}$ allowed by the constraints and the WIMP freeze-out mass. A naive dimensional analysis (NDA) estimate of $\Lambda_{\rm UV}$, assuming all the Wilson coefficients to be $\mathcal{O}(1)$, is given in \tabl~\ref{table:summary} for all the relevant $n$-plets.

Requiring perturbativity of the EW gauge coupling above the WIMP thermal mass can provide an upper bound on the dimension of the SU(2) representation.
Indeed, large SU(2) $n$-plets will make the EW gauge coupling run faster in the UV, eventually leading to a Landau pole. In \tabl~\ref{table:summary} we provide the value of the scale $\Lambda_{\text{Landau}}$ such that $g_2(\Lambda_{\text{Landau}})=4\pi$. We integrate the RGE equations for the SM gauge couplings at 2-loops and integrate-in the $n$-plet at the WIMP thermal mass.\footnote{Our results are compatible with the ones found in Ref.~\cite{DiLuzio:2015oha} (where $\chi$ is integrated-in at $M_Z$) given that $\Lambda_{\text{Landau}}/M_{\text{DM}}$ is approximately independent on $M_{\text{DM}}$. } Comparing $\Lambda_{\text{Landau}}$ and $\Lambda_{\text{UV}}$, we see that the stability of the fermionic $n$-plets with $n\leq5$ only depends on physics in a regime where the EW coupling is still perturbative. Instead, the stability of $n$-plets with $n>5$ requires specifying a UV completion for the EW gauge group that does not give rise to the dangerous operators of \eqref{eq:scalardecay} and \eqref{eq:fermiondecay}. In this sense, the Majorana 5-plet studied in Ref.~\cite{Cirelli:2005uq} is special, because it can be made accidentally stable by raising the scale $\Lambda_{\rm UV}$, without any further assumption on the nature of the UV completion at $\Lambda_{\rm Landau}$.

Requiring $\Lambda_\text{UV}/M_\chi\gtrsim 10$ to ensure perturbativity of the theory up to well above the WIMP mass
would select $n\leq9$ for fermions, and $n\leq 11$ for scalars. However, requiring a large hierarchy between $\Lambda_{\text{Landau}}$ and $M_\chi$ is not necessary to ensure the calculability of thermal freeze-out, which depends only on EW processes at energies much below the DM mass. A more robust upper bound on the dimension of the SU(2) $n$-plets will be derived in \sect~\ref{sec:unitaritybound}, analyzing the $s$-wave unitarity of the annihilation cross-section. This bound will require $n\leq 13$ for both fermionic and scalar WIMPs.     

Finally, let us comment on the EW WIMPs in complex representation of SU(2). For odd $n$, complex multiplets with $Y=0$ are allowed. Their freeze-out dynamics shares many similarities to the one of the real multiplets discussed here and has been partially discussed in Ref.~\cite{DelNobile:2015bqo}. For complex representations with $Y\neq0$, direct detection constraints can be circumvented only by introducing a splitting between the two Weyl spinors forming the Dirac pair. The required splitting can be generated via dimension 5 operators above the WIMP thermal mass leaving the freeze-out predictions unaffected. A full classification of these WIMP scenarios and their phenomenological probes is left for a future work~\cite{futurecoll}. 

\section{WIMP cosmology}\label{sec:cosmo}

The determination of the DM thermal mass hinges on a careful computation of the DM annihilation cross-section in the non-relativistic regime. In particular, the potential generated by EW gauge boson exchange between DM pairs is attractive for isospins $I\lesssim \sqrt{2}n$ resulting into Bound State Formation (BSF) through the emission of an EW gauge boson in the final state. The energy of the emitted gauge boson is of the order of the Bound State (BS) binding energy $E_{B_I} \simeq \frac{\alpha_\eff^2 M_\chi}{4n_B^2}-\alpha_{\text{eff}} m_W$, where $n_B$ is the BS energy level, $\alpha_{\rm eff}$ is the effective weak coupling defined in \eqref{eq:sommi}, and we neglected corrections of order $m_W^2/M_\chi^2$. In the non-relativistic limit, and at leading order in gauge boson emission, the  BSF process 
\begin{equation}
\chi_i + \chi_j \rightarrow \text{BS}_{i'j'} +V^a 
\end{equation}
is encoded in the effective dipole Hamiltonian described in Ref.~\cite{Mitridate:2017izz,Harz:2018csl} which dictates the BS dynamics and it is written for completeness in \appendixname~\ref{app:boundstates}. 

The BS dynamics relevant for DM freeze-out is well described by the unbroken phase of SU(2) so that the configuration of the DM pair can be decomposed into eigenstates of the isospin $I$ of the pair 
\begin{equation}
|\chi\chi\rangle_{II_z}=\mathcal{C}(II_z|ij)|\chi_i\chi_j\rangle,\,\,\, I_z\in\left[-\frac{I-1}{2},\frac{I-1}{2}\right],
\end{equation}
where $\mathcal{C}(II_z|ij)$ are the Clebsch-Gordan coefficients and $I$ is the dimension of the isospin representation.  
Denoting with $L$ and $S$ the total angular momentum and the spin, the isospin-Lorentz structure of the dipole Hamiltonian enforces the following selection rules: $i)$ $\Delta S=0$ because the dipole Hamiltonian is spin-independent; $ii)$ $|\Delta L| = 1$ because the dipole operator transform as a vector under rotations; $iii)$ $|\Delta I| =2$ because a single, G-parity odd weak boson is emitted. 

Since we are dealing with real representations, spin-statistics imposes further restrictions on the allowed quantum numbers, depending on the fermionic or scalar nature of the wave function. In particular we have
\begin{equation}
\label{eq:spin-stat}
(-1)^{L+S+\frac{I-1}{2}}=1\ ,
\end{equation}
which implies that for scalars $n_Bs$ ($n_Bp$) bound states, $i.e.$ with $L=0$ ($L=1$), can exist only with even (odd) $\frac{I-1}{2}$, while for fermions odd (even) $\frac{I-1}{2}$ states with $L=0$ are forced to have $S=1$ ($S=0$). 

We are now ready to describe the system of coupled Boltzmann equations for the evolution of the number densities of DM and BS. Following \cite{Mitridate:2017izz}, we will discuss how this coupled system can be reduced to a single equation for the DM number density with an effective annihilation cross-section. The Boltzmann equations for DM and BS read 
\begin{widetext}
\begin{subequations}
\label{eq:boltzmann}
\begin{align}
& z\frac{\mathrm{d}Y_\mathrm{DM}}{\mathrm{d}z}=-\frac{2s}{H}\langle\sigma_{\text{ann}}\vrel\rangle\left[Y_{\mathrm{DM}}^2-(Y_{\mathrm{DM}}^{\eq})^2\right]-\frac{2s}{Hz}\sum_{B_I}\langle\sigma_{B_I}\vrel\rangle\left[Y_{\mathrm{DM}}^2-(Y_{\mathrm{DM}}^{\eq})^2\frac{Y_{B_I}}{Y_{B_I}^{\eq}}\right] \label{eqb:dm}\ ,\\
&z\frac{\mathrm{d}Y_{B_I}}{\mathrm{d}z}=Y_{B_I}^{\eq}\left\{\frac{\langle\Gamma_{B_I\text{,break}}\rangle}{H}\left[\frac{Y_{\mathrm{DM}}^2}{(Y_{\mathrm{DM}}^{\eq})^2}-\frac{Y_{B_I}}{Y_{B_I}^{\eq}}\right]+\frac{\langle\Gamma_{B_I\text{,ann}}\rangle}{H}\left[1-\frac{Y_{B_I}}{Y_{B_I}^{\eq}}\right]+\sum_{B_J}\frac{\langle\Gamma_{B_I\rightarrow B_J}\rangle}{H}\left[\frac{Y_{B_J}}{Y_{B_J}^{\eq}}-\frac{Y_{B_I}}{Y_{B_I}^{\eq}}\right]\right\}\label{eqb:bs}\ ,
\end{align}
\end{subequations}
\end{widetext}
where $B_{I,J,\dots}$ labels the different bound states, $z=\frac{M_\chi}{T}$, $s$ is the entropy density and $Y=\frac{n}{s}$ is the number density per co-moving volume. 

The dynamics of a  given BS $B_{I}$ in the plasma is described by \eqref{eqb:bs} and depends  on: $i)$ its ionization rate $\langle\Gamma_{B_I\text{,break}}\rangle$; $ii)$ its annihilation rate into SM states $\langle\Gamma_{B_I\text{,ann}}\rangle$; $iii)$ its decay width into other bound states $\langle\Gamma_{B_I\rightarrow B_J}\rangle$.  The ionization rate $\langle\Gamma_{B_I\text{,break}}\rangle \equiv n_\gamma \langle\sigma_{I\text{,break}} \vrel\rangle$ encodes the probability of a photons from the plasma to break the BS $B_I$. Assuming thermal equilibrium, detailed balance relates the cross-section for the BS breaking $\langle\sigma_{I\text{,break}} \vrel\rangle$ to the BSF cross-section $\langle\sigma_{B_I}\vrel\rangle$
\begin{equation}
\langle\Gamma_{B_{I}\text{,break}}\rangle =\frac{g_\chi^2}{g_{B_{I}}}\frac{(M_\chi T)^\frac{3}{2}}{16\pi^\frac{3}{2}}e^{-\frac{E_{B_I}}{T}}\langle\sigma_{B_I} \vrel\rangle\ ,\label{eq:ionrate}
\end{equation}
where $g_{B_{I}}$ and $g_\chi$ count the number of degrees of freedom of the bound state $B_I$ and of the DM multiplet, respectively.  If either the BS decay or the annihilation rate satisfies $\Gamma \gg H$, we can neglect the LHS in \eqref{eqb:bs}, obtaining algebraic relations between the DM and the BS yields. 

Plugging these relations into \eqref{eqb:dm}, we arrive at the final form of the DM Boltzmann equation
\begin{equation}
\label{eq:single_boltzmann}
\frac{\mathrm{d}Y_{\mathrm{DM}}}{\mathrm{d}z}=-\frac{\langle\sigma_{\eff}\vrel\rangle s}{Hz}(Y_{\mathrm{DM}}^2-Y_{\mathrm{DM}}^{\eq,2})\ ,
\end{equation}
where
\begin{equation}
\langle\sigma_{\eff}\vrel\rangle\equiv S_{\text{ann}}(z)+\sum_{B_J}S_{B_J}(z),
\end{equation}
and we defined the effective cross-section as the sum of the direct annihilation processes, $S_{\text{ann}}$, and the ones which go through BSF, $S_{B_J}$. In particular, $S_{\text{ann}}$ can be written as 
\begin{equation}
S_{\text{ann}}=\sum_I \langle S_E^I\sigma_{\text{ann}}^I\vrel\rangle\ ,
\end{equation}
where $\sigma_{\text{ann}}^I$ is the hard cross-section for a given isospin channel $I$, $S_E^I$ is the Sommerfeld enhancement (SE) of the Born cross-section, and $\vrel$ is the relative velocity of the two DM particles. In the limit of small relative velocity between the DM particles (but larger than $m_W/M_\chi$), the SE factor can be approximated as
\begin{equation}
S_E^I \approx \frac{2\pi \alpha_{\eff}}{\vrel}\ ,\quad \text{where}\quad \alpha_\eff \equiv  \frac{I^2+1-2n^2}{8} \alpha_2 \ .\label{eq:sommi}
\end{equation}
The finite mass effects modify the behavior of the SE at  $v_{\text{rel}}\lesssim m_W/M_\chi$ and are included in our full computation (see Ref.~\cite{Cassel:2009wt} for explicit formulas). However, \eqref{eq:sommi} will be enough to estimate the behavior of the SE at the temperatures most relevant for freeze-out.

Analogously we can factorize the BSF processes as 
\begin{equation}
S_{B_J}=\sum_{I,l}\langle S_E^I S_{B_J}^{I,l}\rangle R_{B_J}\ ,\label{eq:rbs}
\end{equation}
where $S_{B_J}^{I,l}$ is the ``hard'' BSF cross-section of the state $B_J$ starting from a free state with angular momentum $l$ and isospin $I$ multiplied by the SE factor of that particular isospin channel as defined in \eqref{eq:sommi}. Explicit expressions for this can be found in Ref.~\cite{Mitridate:2017izz,Harz:2018csl}. $R_{B_J}$ gives instead the effective annihilation branching ratio into SM states which depends on the detailed BS dynamics ($i.e.$ annihilation, ionization and decay). In particular, $R_{B_J}$ approaches 1 once the temperature of the plasma drops below the binding energies of the bound states involved in the decay chains.
In the case of a single BS, $R_{B_J}$ takes a rather intuitive form
\begin{equation}
\label{eq:singleBS}
R_{B_J}=\frac{\langle\Gamma_{\text{ann}}\rangle}{\langle\Gamma_{\text{ann}}\rangle+\langle\Gamma_{\text{break}}\rangle}\ ,
\end{equation}
which applies to $1s_I$ and $2s_I$ BS with $I\leq 5$. The latter, once formed, annihilate directly into pairs of SM vectors and fermions, with rates $\Gamma_{\text{ann}}\simeq \alpha_{\text{eff}}^5/n_B^2 M_\chi$. These BS together make up for more of the $50\%$ of the BSF cross-section. More complicated examples of BS dynamics will be illustrated in \appendixname~\ref{eq:7plet} where we detail the case of the EW $7$-plet.

\begin{figure}[]
\centering
\includegraphics[width=0.49\textwidth]{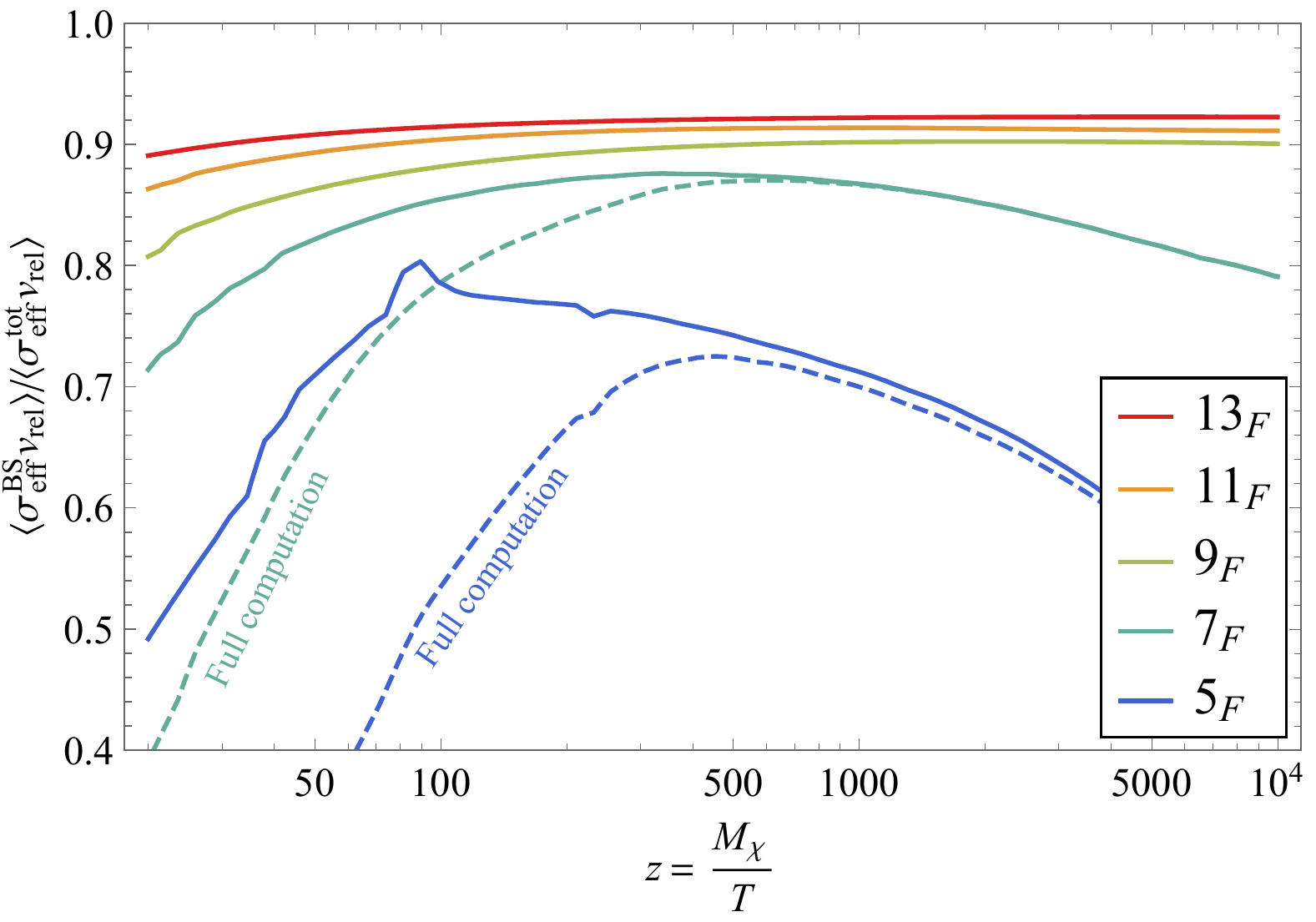}
\caption{Effective cross-section for BSF normalized over the total annihilation cross-section as a function of $z=M_\chi/T$ assuming vanishing ionization rates, $i.e.$ $R_{\text{BS}}=1$ (see \eqref{eq:rbs} and below). The dashed lines for the fermionic 5-plet (dark blue) and 7-plet (cyan) show the deviation of the real bound state dynamics from the approximation of vanishing ionization rates. For $n>5$ the error due to the $R_{\text{BS}}=1$ is subdominant compared to the virtual and real effects at NLO in gauge boson emission.}
\label{fig:bs_n}
\end{figure}

While the effect of BSF has already been computed for the fermionic $5$-plet in Ref.~\cite{Mitridate:2017izz}, here we include it for the first time for all the real WIMP candidates with $n\geq 7$. For larger EW multiplets, we find the relative effect of BS dynamics on the total cross-section increases, as can be seen from \fig~\ref{fig:bs_n}.

This is the consequence of two effects: $i)$ the binding energy grows at large $n$, suppressing the ionization rate with respect to the annihilation one; $ii)$ at larger $n$ the number of attractive channels increases and thus the BS multiplicity per energy level grows linearly with $n$. For example, for $n=5$ the attractive channels have $I=1,3, 5$, for $n=7$ BS with $I=7, 9$ can also form. The relevance of these higher isospin channels was not recognized in \cite{Smirnov:2019ngs}, where only the $I=1,3$ channels were included, significantly underestimating the thermal mass already for $n=7$. In \appendixname~\ref{eq:7plet} we show explicitly the relative contributions coming from the different isospin channels for the 7-plet. The 7-plet thermal mass was computed including all the BS up to $3s$ and $2p$ but we checked that the contribution from $4s$ and $3p$ BS is negligible.  

As we increase the dimension of the multiplet, the bound states become more tightly bounded and the effect of the ionization rate becomes smaller. This can be explicitly seen from \eqref{eq:ionrate} where the binding energy controls the Boltzmann suppression of the ionization rate. For this reason, we only account for the detailed BS dynamics for $n\leq7$ while for $n>7$ we set the annihilation branching ratios to 1. We assume, as explicitly checked for the 7-plet, that the formation cross sections for $4s$ and $3p$ BS are negligible. In fact, the cross sections of BS differring only for their principal quantum number have the same parametric dependence on $n$, so that the hierarchy between different energy levels is independent on $n$. Close to the unitarity bound limit, excited states with larger angular momentum can become important. However, their long lifetimes and small binding energies limit their contributions to the thermal mass. Moreover, since the typical velocity inside the bound state is $\alpha_\eff/n_B$, relativistic corrections can also be important. We leave the discussion of these contributions to a future work.

 In \appendixname~\ref{eq:7plet} we estimate the error on the WIMP mass due to this approximation by comparing its effect on the thermal masses of $5$-plet and the $7$-plet against the full computation. We find a shift in mass  $\Delta M_{DM}\simeq 5$ TeV for both $n=5$ and $n=7$  resulting in a smaller relative error for $n=7$, as expected.  We keep 5 TeV as an estimate of the error induced by this approximation for the larger multiplets. As we will discuss in the next Section, the uncertainty for $n\geq 7$ will be anyhow dominated by the next-to-leading order (NLO) contributions to the SE which are not included here. 

Finally, we comment on the theory uncertainty on the mass prediction for the 5-plet. This is dominated by the approximate treatment of EW symmetry breaking effects in computation of the BSF cross-sections. The SU(2)-symmetric approximation fails once the DM de Broglie wavelength becomes of the order of $m_W$ (i.e. for $z\simeq 10^4$ for $n\geq 5$). After the EW phase transition, Coulomb and Yukawa potentials appear at the same time so that employing either the Coulomb or the Yukawa centrifugal correction to the SE (see Ref.~\cite{Cassel:2009wt}) overestimate and underestimate, respectively, the freeze out cross-section. This gives us a rough way of determining the theory uncertainty: $i)$ to set the lower bound on the freeze-out mass we include BSF in $\sigma_{\eff}$ until $z=10^4$ with the centrifugal correction coming from the Yukawa; $ii)$ to set the upper bound we push the effect of BSF, neglecting the vector masses in the centrifugal correction, to arbitrary large values of $z$. We observe that the abundance saturates already for $z\approx 10^5$. This procedure gives the uncertainty for the 5-plet in \tabl~\ref{table:summary} which is different than the one quoted in Ref.~\cite{Mitridate:2017izz}, where the BS contribution was switched off at $z=10^4$, underestimating the effect of BSF. 

\section{The WIMP Unitarity Bound}\label{sec:unitaritybound}

We now analyze the constraint of perturbative unitarity on the annihilation cross-section, including bound state formation. The perturbative unitarity of the S-matrix sets an upper bound on the size of each partial wave contribution to the total annihilation cross-section\,\footnote{This constraint was derived for $e^+e^-$ annihilations in \cite{Cabibbo:1961sz,Cabibbo:1974ke} and then used for the first time in the DM context in \cite{Griest:1989wd}. It can be checked that this constraints is not modified in the presence of long range interactions~\cite{Landau:1991wop}.}   
\begin{equation}
(\sigma_\eff \vrel)^{J}\leq \frac{4\pi(2J+1)}{M_\chi^2 \vrel}\ ,
\end{equation}
where $\vec{J} = \vec{L} + \vec{S}$ is the total angular momentum. The stronger inequality comes from the $s$-wave channel (i.e. $J=0$) which can be written as
\begin{equation}
\label{eq:unit_bound}
(\sigma_\text{ann} \vrel)+\sum_{B_J} f_{B_J}^0 (\sigma_{B_J} \vrel)\leq \frac{4\pi}{M_\chi^2 \vrel}\ ,
\end{equation}
where $f_{B_i}^0$ selects the BS contributions that can be formed by $J=0$ initial wave, which are limited by the selection rules discussed in the previous Section. 

For a scalar WIMP selecting the $s$-wave implies $L=0$, and only BS in $p$-orbitals can contribute to the $s$-wave cross-section with $f_{\text{BS}}^0=1$. The spin statistics of the wave function in \eqref{eq:spin-stat} forces these BS to have odd $(I-1)/2$. In practice, the $s$-wave unitarity bound for scalars is determined solely by the SE.   
For fermionic WIMP selecting the $s$-wave implies the same selection rules of the scalar when $S=0$. Additional contributions arise from $S=1$ $s$-orbital states, whose isospin must be odd due to Fermi statistics. In this case, the projection onto the $J=0$ wave gives $f_{\text{BS}}^0=\frac{1}{9}$.

Solving the constraint in \eqref{eq:unit_bound} we find that $s$-wave unitarity is violated for $n\geq 15$ for both fermion and scalar WIMPs. In both cases the $s$-wave cross-section is largely dominate by the SE.  We checked that a similar constraint can be obtained by looking at the $p-$wave unitarity, where the cross-section is instead dominated by the formation of $1s$ BS. 

The selection rules that regulates the BS dynamics derive from the dipole Hamiltonian which is written for completeness in \eqref{eq:eff_ham}. These selection rules are only broken by NLO contributions in gauge boson emission which can be estimated as  
\begin{equation}
\label{eq:1loopBSF}
\frac{\Delta \sigma_{\text{BSF}}^{\text{NLO}}}{\sigma_{\text{BSF}}^{\text{LO}}}\sim \frac{\alpha_\eff^3}{64\pi }\ ,
\end{equation}
where the extra $\alpha_{\text{eff}}^2$ correctly accounts for the phase space suppression in the limit of small velocities as detailed in \appendixname~\ref{app:LOandNLO}. As a result, the LO selection rules apply all the way till the breaking of perturbative unitarity. 

\begin{figure*}[]
\centering
\includegraphics[width=\textwidth]{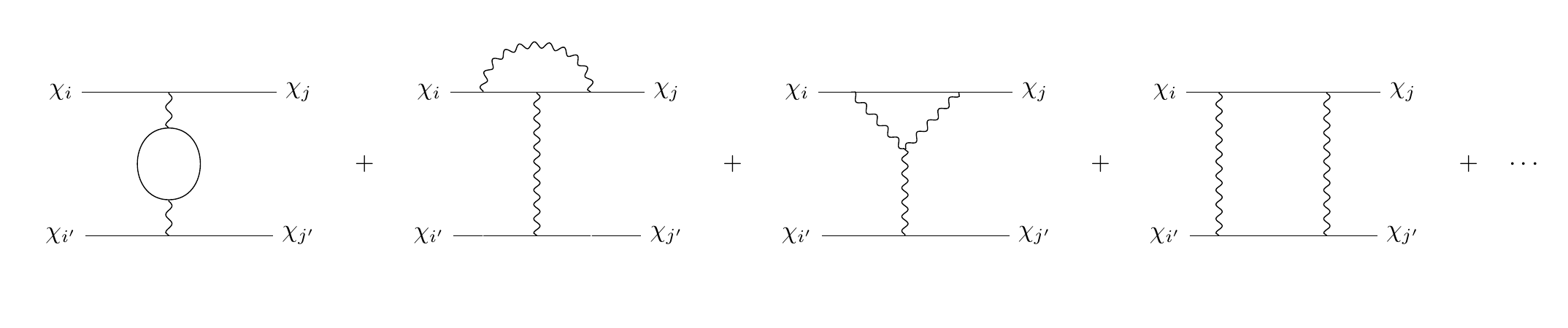}
\caption{Examples of Feynman diagrams contributing at NLO to the non-relativistic potentials as estimated in \eqref{eq:1loop}.}
\label{fig:1loop}
\end{figure*}

Interestingly, the upper bound on $n$ from perturbative unitarity derived from \eqref{eq:unit_bound} is significantly stronger than the one derived from the perturbative unitarity of the Born cross-section   which is violated for $n\geq 38$ (i.e. $\alpha_{\text{eff}}\geq4\pi$). This suggests that because of SE, the ratio between the NLO and the LO cross-section should appreciably deviate from the NDA scaling of the Born cross-section: $\sigma_{\text{Born}}^{\text{NLO}}/\sigma_{\text{Born}}^{\text{LO}}\sim \alpha_{\eff}/4\pi$. Estimating the NLO correction to the potentials controlling the SE we indeed get 
\begin{equation}
\label{eq:1loop}
\frac{\Delta V_{\text{NLO}}}{V_{\text{LO}}}\sim \frac{\alpha_\eff}{4\pi }\log\left(\frac{m_W\sqrt{z}}{M_\chi}\right)\ ,
\end{equation}
where the NLO potential is resumming ladder diagrams like the ones in \fig~\ref{fig:1loop}, and where we substituted the de Broglie length $1/M\vrel \approx \sqrt{z}/M_\chi$ as the typical lenght scale for the annihilation process. Our estimate above matches the explicit NLO computation of the SE for the 3-plet in Ref.~\cite{Beneke:2020vff}. Requiring this correction to be $\lesssim1$ across the freeze-out temperatures leads to a similar upper bound on $n$ than the one inferred from perturbative unitarity. 

We use the estimate above to assess the theory uncertainty on the WIMP thermal masses in \tabl~\ref{table:summary}. Indeed, \eqref{eq:1loop} results in a correction to the Sommerfeld factor $S_E$, which affects both $S_\text{ann}$ and $S_{B_J}$ as introduced in \eqref{eq:single_boltzmann}.  We find that neglecting the NLO contribution dominates the DM mass theory uncertainty for $n\geq 7$. The uncertainty grows as we increase the dimensionality of the multiplet becoming as large as $\mathcal{O}(30\%)$ for $n=13$.
   
Finally, we compare our results to the ones obtained in Ref.~\cite{Smirnov:2019ngs}. Numerically, the upper bound on the WIMP mass corresponding to the saturation of the unitarity bound is roughly $500\pm200$ TeV, which is the expected thermal mass for $n=15$ as can be seen from \fig~\ref{fig:summary}. The unitarity boundary was set instead to 150 TeV for $n=13$ in Ref.~\cite{Smirnov:2019ngs} without a quoted theory uncertainty. Beside the numerical differences, our computation differ from the one in Ref.~\cite{Smirnov:2019ngs} in two crucial instances: $i)$ at large $n$ we find that large isospin channels enhance significantly the BSF cross-section making the WIMP DM mass \emph{heavier} than in Ref.~\cite{Smirnov:2019ngs} at fixed $n$; $ii)$ we find that including BSF does not accelerate by much the saturation of the unitarity bound because of the selection rules of the dipole Hamiltonian at LO. As we discussed above, the LO selection rules are not lifted by NLO corrections until the boundary of perturbative unitarity is reached. These two effects together push the heaviest calculable WIMP mass very close to the PeV scale appreciably enlarging the EW WIMP scenarios beyond the reach of any realistic future collider.      

\section{WIMP at high energy lepton colliders}\label{sec:collider}

We now look at the possible detection strategies for direct production of WIMPs at collider experiments. From the results in \tabl~\ref{table:summary} one can immediately see that DM masses $\gtrsim 50$ TeV are required to achieve thermal freeze-out for EW multiplets with $n > 5$.
Pair-production of these states would require center-of-mass energies exceeding 100 TeV, which are unlikely to be attained at any realistic future facility.
On the other hand, multiplets with $n\leq 5$ have thermal masses in the few TeV range, potentially within the reach of present and future colliders. 

Direct reach on these dark matter candidates at hadron colliders is limited by  the absence of QCD interactions for the DM candidates, which can be produced only via electro-weak interactions. As such the limits at the LHC (see e.g.~\cite{Ostdiek:2015aa}) are rather far from the interesting thermal mass targets and only a future $pp$ collider may have the reach for some low-$n$ candidates if collisions around 100~TeV can be attained~\cite{Cirelli:2014dsa,Low:2014sh,1910.11775v2}. Lepton colliders tend to have reach mainly through indirect effects, e.g. the modification of the angular distributions in simple $f\bar{f}$ production at center of mass energies below the threshold to produce the DM pair. The reach in this case is up to  masses a factor a few above the center of mass energy~\cite{Harigaya:2015qq,DiLuzio:2018jwd}.

A very-high-energy lepton collider, such as a muon collider, would be the perfect machine to hunt for these WIMPs, due to its large center-of-mass energy, relatively clean collision environment, and the capability of pair-producing weakly interacting particles {\it up to kinematical threshold}.
Here we consider in particular a future muon collider with center-of-mass energy of 10~TeV or more and the baseline integrated luminosity of \cite{Delahaye:2019omf}
\begin{equation}
\mathcal{L}\simeq 10\;\textrm{ab}^{-1} \cdot \left(  \frac{\sqrt{s}}{10\;\textrm{ TeV}}  \right)^{2} \label{eq:lumi}.
\end{equation}
While such a machine is currently not feasible, various efforts to overcome the technological challenges are ongoing.
Early developments on machine performances \cite{1808.01858v2,Palmer_2014} found the luminosity \eqref{eq:lumi} to be achievable for $\sqrt{s}\lesssim 6$~TeV, and further development to push it to larger energies is currently in progress \cite{MuonColliderCollaboration}.

We consider various search channels for EW 3-plets and 5-plets, and determine the minimal center-of-mass energy and luminosity required to directly probe the freeze-out predictions.
First, we detail in \sect~\ref{sec:missingp} the prospects for the observation of DM as undetected carrier of momentum recoiling against one or more SM objects.
We systematically study all the ``mono-V'' channels, where DM is recoiling against a SM gauge boson $V=\gamma,Z,W$.
We also investigate double vector boson production, that we dub ``di-V'' channels, where requiring a second SM gauge boson in the final state could help ameliorating the sensitivity.
Second, in \sect~\ref{sec:collider_dt} we study the reach of disappearing track searches -- which are robust predictions of WIMPs in real EW representations as discussed in \sect~\ref{sec:WIMP} -- recasting the results of \cite{Capdevilla:2021fmj}.
Notice that our study is in principle applicable both to high-energy $\mu^+\mu^-$ and $e^+e^-$ colliders, even though soft QED radiation, beam-strahlung, and the presence of beam-induced backgrounds could affect the results in different ways.

The projections for direct production derived here have to be contrasted with similar studies in the context of future high energy proton machines~\cite{Cirelli:2014dsa,Low:2014cba} (which are limited by the partial reconstruction of the collision kinematics) or electron-positron machines~\cite{Fox:2011fx,Bartels:2012ex} (which are limited by the moderate center-of-mass energy and hence more effective to hunt for lighter DM candidates) .

Complementary studies have also considered indirect probes of WIMPs at future high energy lepton colliders, focusing on the modifications of Drell-Yan processes~\cite{DiLuzio:2018jwd}. Given the freeze-out masses of \tabl~\ref{table:summary}, EW $n$-plets with $n>5$ are beyond the reach of any realistic future collider both \emph{directly} and \emph{indirectly}, even though a definitive statement about indirect observables would require further studies.

\subsection{WIMPs as missing momentum}\label{sec:missingp}

\begin{figure*}[t]
\centering
Mono-$W$ reach --- Majorana 3-plet \hfil\qquad\qquad\qquad\qquad\qquad Mono-$W$ reach --- Majorana 5-plet \\[-8pt]
\includegraphics[width=0.45\textwidth]{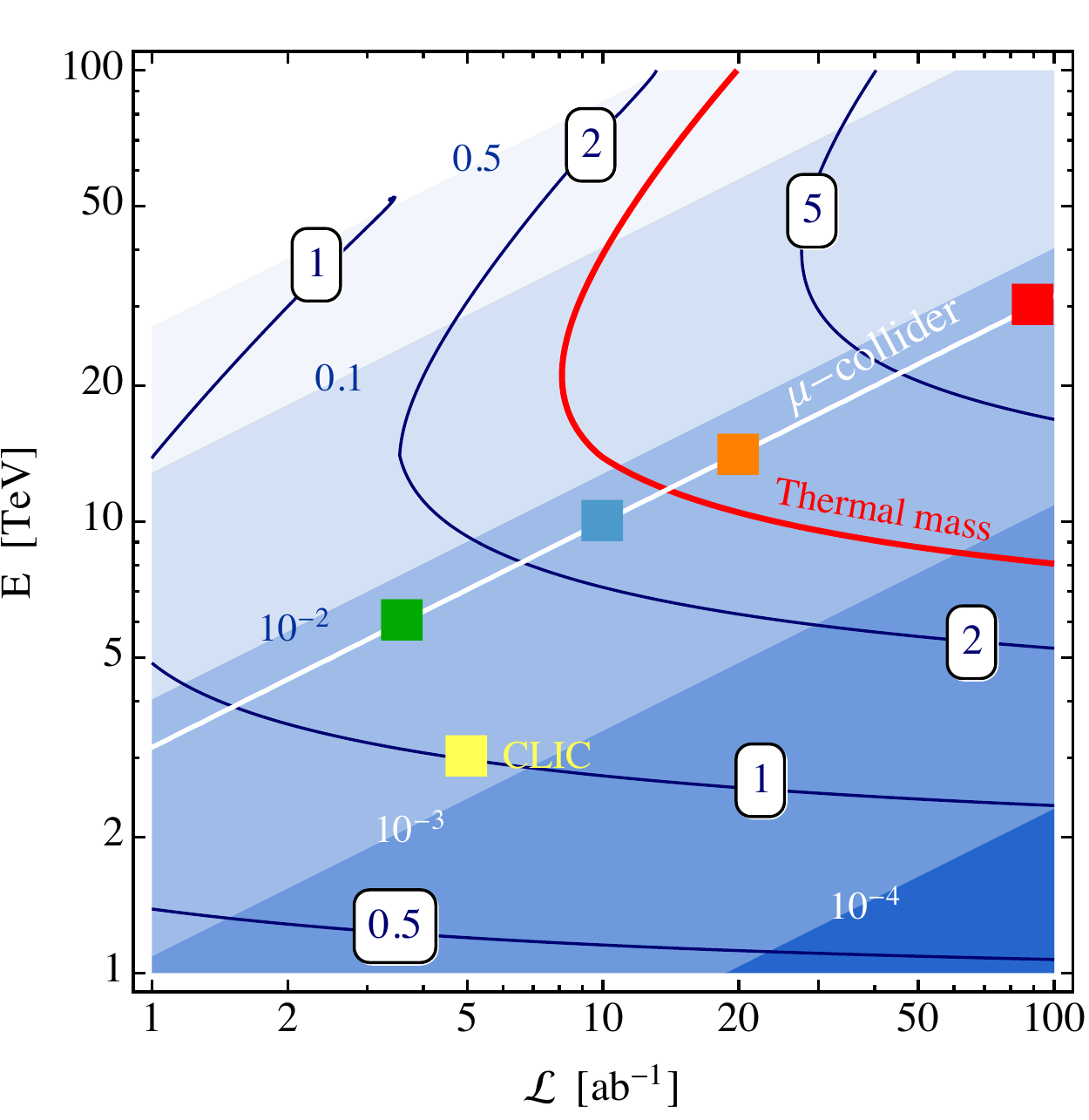}\hfill%
\includegraphics[width=0.45\textwidth]{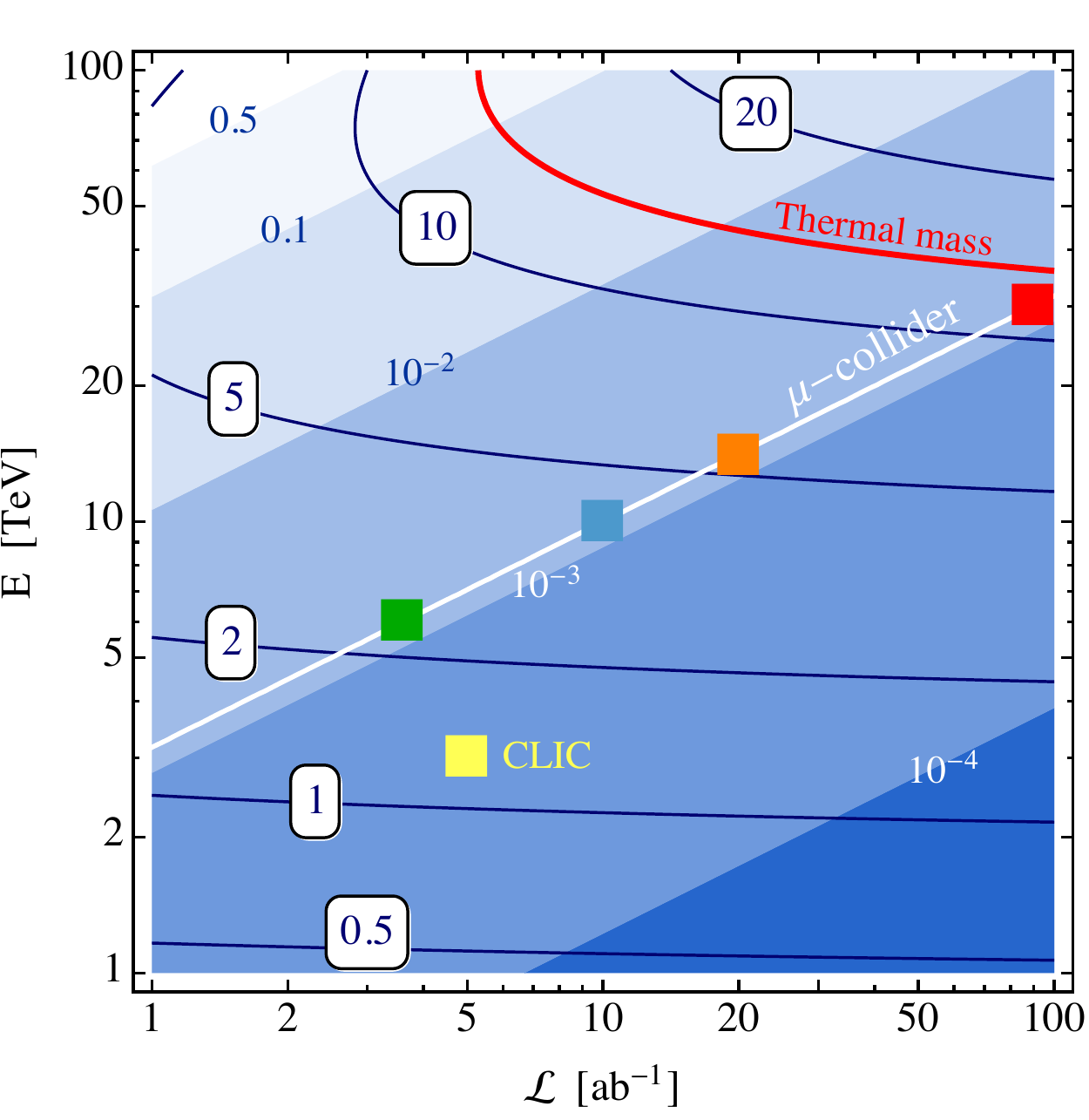}
\caption{Reach from mono-$W$ searches at a muon collider, as a function of collider center-of-mass energy $\sqrt{s}$ and integrated luminosity $\mathcal{L}$. The blue contours show the 95\% C.L. reach on the WIMP mass; the prediction from thermal freeze-out is shown as a red line. The precision of the measurement is shown by the blue shadings. Systematic uncertainties are assumed to be negligible. The white line corresponds to the luminosity scaling \eqref{eq:lumi}, with various collider benchmarks shown as colored squares: $\sqrt{s}=6\text{ TeV}$ green, $\sqrt{s}=10\text{ TeV}$ blue, $\sqrt{s}=14\text{ TeV}$ orange and $\sqrt{s}=30\text{ TeV}$ red. The yellow square corresponds to the 3 TeV CLIC~\cite{deBlas:2018mhx}. 
{\bf Left:} Majorana 3-plet. {\bf Right:} Majorana 5-plet.\label{fig:lumi_vs_energy}}
\end{figure*}

We perform a full study of the different channels to observe DM as undetected carrier of momentum.  The generic strategy is to measure a hard SM particle or a set of particles $X$ recoiling against a pair of invisible objects,
\begin{equation}
    \ell^+\ell^- \to \chi^i \chi^{j} + X \, .\label{eq:monoVgen}
\end{equation}
Notice that we treat all the components $\chi^i$ of the EW multiplet as invisible, assuming the soft decay products of the charged states to be undetected.
Additional soft SM radiation is also implicit in \eqref{eq:monoVgen}.
The prospects for the ``mono-photon'' topology at a future muon collider have been already studied in \cite{Han:2020uak}.  Here, we want to extend this analysis by enlarging the set of SM objects recoiling against the invisible DM multiplets.

\medskip

\noindent {\bf Mono-V.}\; We start by considering ``mono-V'' scattering processes 
where $V=\gamma,Z,W$ is a generic EW gauge boson that accompanies the production of $\chi$ states from the $n$-plet,
\begin{align}
& \text{mono-$\gamma$:} & &\ell^+\ell^- \to \chi^i \chi^{-i} + \gamma \ ,\label{monogamma} \\
& \text{mono-$Z$:} & &\ell^+\ell^- \to \chi^i \chi^{-i} + Z   \, , \label{monoz}\\
&  \text{mono-$W$:} & &\ell^+\ell^- \to \chi^i \chi^{-i\mp 1} + W^\pm   \, .\label{monoW}
\end{align}

The main contribution to all these processes comes from initial- and final-state radiation of a vector boson, which have sizeable rates because of the large weak charge of the DM multiplet and the weak charge of the beams.\footnote{The mono-Higgs signal has a much lower cross-section due to the suppression of initial- and final-state radiation. Furthermore, final-state radiation is model-dependent for scalar DM.} We sum over all components of the multiplet $\chi^i$, but the dominant signal corresponds to the production of the state with largest electric charge ($i = \pm n$), subsequently decaying into DM plus soft SM particles.

For each of these signals, the corresponding SM background is dominated by a single process,
\begin{align}
& \text{mono-$\gamma$ bkg:} & &\ell^+ \ell^- \to \gamma \nu \bar{\nu}\ ,\label{monogammabkd} \\
& \text{mono-$Z$ bkg:} & &\ell^+ \ell^- \to Z \nu \bar{\nu}   \, , \label{monozbkd}\\
&  \text{mono-$W$ bkg:} & &\ell^+ \ell^- \to W^\mp \nu \; + \; \ell^\pm (\mathrm{lost})  \, ,\label{monoWbkd}
\end{align}
where the missing transverse momentum is carried by neutrinos; the mono-$W$ background also requires a lost charge along the beam.

We simulate signal and background events with \mgfive~\cite{Alwall:2011uj,Alwall:2014hca}, for different DM mass hypotheses and different collider energies. 
The $W$ and $Z$ bosons are assumed to be reconstructed from all their visible decay products and are treated as single objects.
We impose basic acceptance cuts on the rapidity and transverse momentum of the vectors, requiring $|\eta_V| < 2.5$ and $p_{{\rm T},V} > 10$~GeV.
Other detector effects are neglected.      

We then perform a cut-and-count analysis, estimating the significance of the signal as
\begin{equation}
    \text{significance} = \frac{S}{\sqrt{S+B+\epsilon_{{\rm sys}}^2\left(S^2+B^2\right)}}\ , \label{eq:Significance}
\end{equation}
where $S,B$ are the numbers of physical signal and background events, and $\epsilon_{\rm sys}$ parametrizes the systematic uncertainties. The signal is isolated from the background employing the kinematics of the visible object, parametrized in terms of its transverse momentum $p_{{\rm T},V}$, its pseudo-rapidity $\eta_V$, and the missing invariant mass (MIM) which
is a function of the energy of the visible particle itself
\begin{equation}
    \mathrm{MIM}=\left( s + m_V^2 - 2\sqrt{s}E_V \right)^{1/2} \,.
\end{equation}
We select events with $\text{MIM}\geq 2M_\chi$, $p_{{\rm T},V} \geq p_{{\rm T},V}^{\rm cut}$, $|\eta_V|~\leq~\eta_V^{\rm cut}$, where the $p_{\rm T}$ and $\eta$ selection cuts are chosen to maximize the significance for each value of $M_\chi$. The precise values of the selection cuts, together with the expected number of events and the reach of the various search channels, are given in \tabl~\ref{tab:Majorana} in the Appendix.

The background rates for mono-$\gamma$ and mono-$Z$ are very similar, with fiducial cross-sections of around $3$ pb that depend weakly on the collider energy.
As already pointed out in~\cite{Han:2020uak} for the mono-$\gamma$ case, the optimal reach on $M_\chi$ is obtained for low signal-to-noise ratios -- in other words, systematic uncertainties could be important. For this reason, we present results for different values of $\epsilon_{\rm sys} = 0, 1\permil, 1\%$. We point out that in presence of larger systematic uncertainties, the optimal selection cuts are stronger (as can be seen in \tabl~\ref{tab:Majorana}) and lead to higher values of $S/B$.

\begin{figure*}
\begin{centering}
\includegraphics[width=0.46\textwidth]{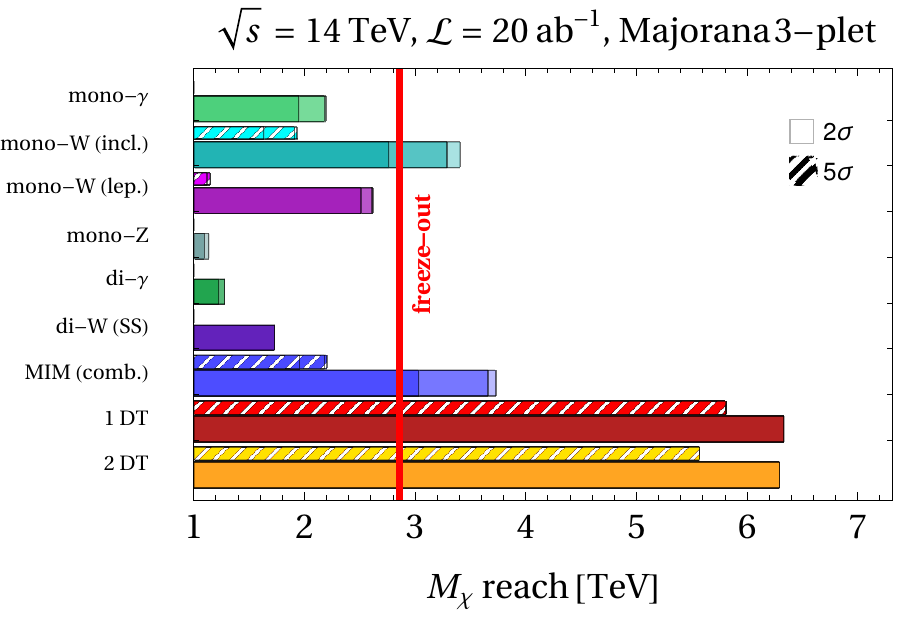}\qquad\qquad
\includegraphics[width=0.46\textwidth]{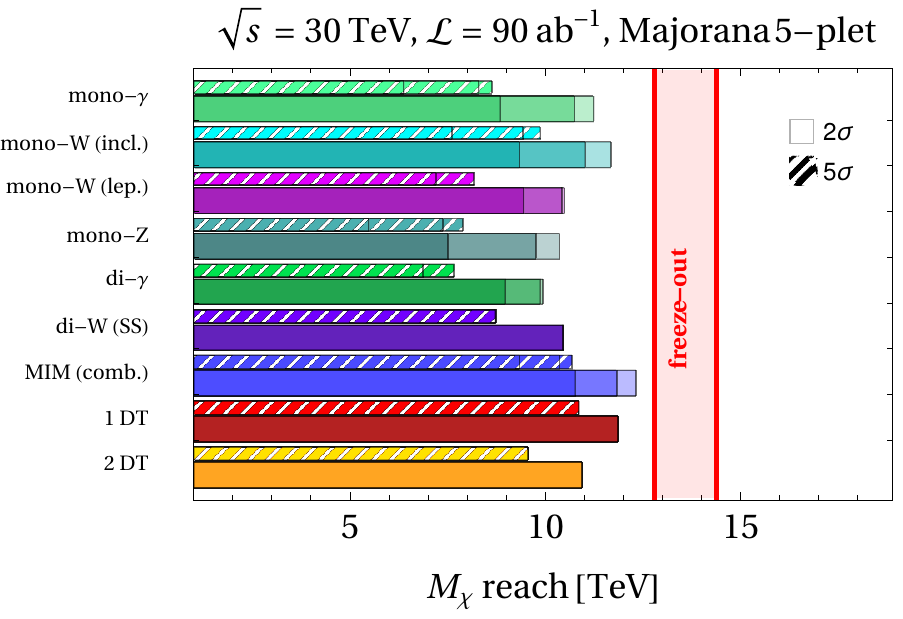}
\end{centering}
\caption{Different bars show the $2\sigma$ (solid wide) and $5\sigma$ (hatched thin) reach on the WIMP mass at a muon collider for different search channels. The first seven bars show the channels discussed in \sect~\ref{sec:missingp} where DM would appear as missing invariant mass (MIM) recoiling against one or more SM objects: mono-gamma, inclusive mono-W, leptonic mono-W, mono-Z, di-gamma, same sign di-W, and the combination of all these MIM channels (blue). The last two bars show the reach of disappearing tracks as discussed in \sect~\ref{sec:collider_dt}, requiring at least 1 disappearing track (red), or at least 2 tracks (orange).  All the results are shown assuming systematic uncertainties to be 0 (light), $1\permil$ (medium), or 1\% (dark). The vertical red bands show the freeze-out prediction. {\bf Left:} Majorana 3-plet for $\sqrt{s} = 14\,{\rm TeV}$ and $\mathcal{L} = 20\,{\rm ab}^{-1}$.  {\bf Right:} Majorana 5-plet for $\sqrt{s} = 30\,{\rm TeV}$ and $\mathcal{L} = 90\,{\rm ab}^{-1}$.\label{fig:barchart}}
\end{figure*}

The mono-W differs from the other two channels. The SM background
is dominated by vector boson fusion (VBF) processes, that lead to forward leptons (lost along the beam pipe) and $W$ bosons. The signal is instead made of events where the $W$ is radiated from the initial or final states, leading to a more central distribution.
The cut on $p_{{\rm T},W}$ 
can efficiently suppress the VBF background, with a lesser impact on the signal compared to the mono-$\gamma$ or mono-$Z$ cases.
As a consequence, we find that the mono-$W$ search has the best sensitivity among the various mono-X channels.
The 95\% C.L.\ exclusion reach on $M_\chi$ for a Majorana 3-plet and 5-plet is shown in \fig~\ref{fig:lumi_vs_energy} as a function of collider center-of-mass energy $\sqrt{s}$ and luminosity $\mathcal{L}$.
We also show the expected values of $S/B$ for the excluded signal in absence of systematic errors, which are rather low also for the mono-$W$ search.

Due to the presence of initial-state radiation, the $W$ boson of the signal has a preference for being emitted in the forward (backward) direction, measured with respect to the flight direction of the $\ell^-$ beam, if its charge is negative (positive). Since the charge of the $W$ boson is potentially observable for leptonic decays, we can envisage a strategy to isolate the signal from the background using the full distribution in $\eta_W$ (instead of its absolute value). We thus also perform an analysis of leptonic mono-$W$ events, where we impose the additional cut $\eta_{W^{\pm}} \lessgtr 0$.
We find the reach of this search to be weaker than the one of the inclusive mono-$W$ because of the small leptonic branching ratio.
However, the leptonic mono-$W$ search possesses signal-free regions of the $\eta_W$ distribution which would allow for an {\it in situ} calibration of the background from the data itself, leading to possible reduction of the systematic uncertainties.

\medskip

{\bf Di-V.}\;
We now consider scattering processes with multiple emission of vector bosons. While generally being suppressed by higher powers of the gauge coupling constant, these processes can be enhanced for large center-of-mass energies, and for multiplets with large weak charge. They can therefore provide very useful handles to probe WIMPs in the regimes where the mono-V searches have very low signal-to-noise ratios.
Of course, a too large rate for multiple boson radiation would indicate the breakdown of the perturbative expansion, requiring the resummation of large logarithms. We have checked that for the EW 3-plet and 5-plet, and for the energies under consideration here, the fixed-order computations are still accurate.

First, we consider the di-photon process
\begin{equation}
    \ell^+\ell^- \to \chi^i \chi^{-i} + \gamma\gamma   \, . \label{doublegamma}
\end{equation}
We apply the same acceptance cuts of the mono-$\gamma$ analysis, and in addition we require a separation $\Delta R_{\gamma\gamma}>0.4$ between the two photons.
We employ the same event selection strategy of the mono-$\gamma$ case, using as variables $\eta_X$, $p_{\mathrm{T},X}$, where $X$ is the compound $\gamma \gamma$ system.
Moreover, we require each photon to be as central as the $\gamma\gamma$ system itself.
For the 5-plet, we find that the di-$\gamma$ search can be stronger than the mono-$\gamma$ in presence of large systematic uncertainties, where suppressing the SM background is more important. For the 3-plet, which has a smaller EW charge, the signal yield is too much affected by the requirement of a second emission to be competitive with the mono-V. In both cases, the values of $S/B$ for the excluded di-$\gamma$ signal are much larger than for the mono-$\gamma$ signal, and systematic errors thus have a smaller impact. Details of the results are reported in \tabl~\ref{tab:Majorana} in the Appendix.

Second, we consider the double $W$ emission
\begin{equation}
    \ell^+\ell^- \to \chi^i \chi^{-i\mp 2} + W^\pm W^\pm   \, ,\label{ssdw}
\end{equation}
which holds a potentially very clean signature due to the two same-sign $W$ bosons.
We focus on leptonically decaying $W$ bosons to ensure that their charge can be accurately tracked. 
A potential SM background consists in events with two lost charged particles, with the leading contribution being
\begin{equation}
    \ell^+ \ell^- \to W^- W^- W^+ W^+\,,
\end{equation}
where two $W$ bosons of same sign are lost. This background is however negligible, as pairs of $W$ bosons with opposite charge tend to be radiated from the same external leg and to be collinear: requiring only one of two collinear $W$ bosons to be within detector acceptance reduces the rate to negligible levels. The other possible background is given by events with a misidentified charge,
\begin{subequations}
\begin{align}
    &  \ell^+ \ell^- \to W^- W^+ (\mathrm{mistag}) \; \nu \bar{\nu} \;, \label{ssdwBG} \\
    &  \ell^+ \ell^- \to W^- W^+ (\mathrm{mistag})\; \ell^+ \ell^-\;, \label{ssdwBGee}
\end{align}
\end{subequations}
where in the second case the charged final-state leptons are lost along the beam line.
Requiring $p_{\mathrm{T},W\!W}\gtrsim \sqrt{s}/10$ makes the process in \eqref{ssdwBGee} subdominant with respect to the $\nu \bar{\nu}$ background \eqref{ssdwBG}.
On top of this $p_T$ cut, we do not apply further selection cuts, and simply require the two $W$ bosons to be within the geometrical acceptance of the detector, $|\eta_W | < 2.5$.
As an estimate for the charge misidentification probability we take $\epsilon_{\mathrm{misid}}=10^{-3}$.

Due to the negligible background contamination, the same-sign di-$W$ signal has a much higher signal-to-noise ratio than the mono-V channels and even than the di-photon signal, reaching up to $S/B \sim \mathcal{O}(1)$. This makes this channel very robust against systematic uncertainties, and particularly effective for large $n$-plets $n\geq 5$ at higher energies due to their large EW charge. This signature may be one of the most robust and convincing signal of $n=5$ multiplets at colliders. 
Further sources of background and a proper characterization of the missing (transverse) momentum in this reaction depend on detector performances, as well as on the knowledge of the initial state of the collision to be used in the computation of kinematic variables. We leave a careful evaluation of these aspects to future work.

\medskip

We summarize the results of all the mono-V and di-V signatures discussed above in \fig~\ref{fig:barchart}, where we show the $95\%$ C.L.\ exclusion on $M_\chi$ for real fermion 3-plets and 5-plets, together with the $5\sigma$ discovery potential, at two benchmark muon colliders. We also show the combined reach from all these missing mass channels. The bands with different shadings correspond to different systematic uncertainties. One can see that the inclusive mono-$W$ yields the strongest exclusion for both the 3-plet and the 5-plet. The main effect of di-V searches is to reduce the impact of systematic uncertainties.
A 14~TeV muon collider with the benchmark luminosity of \eqref{eq:lumi} would be able to probe a thermally-produced Majorana 3-plet WIMP, while a center-of-mass energy of slightly above  30~TeV is needed to probe the thermal freeze-out mass with missing energy searches in the case of the 5-plet.

Scalar WIMPs have lower production cross-sections. Missing mass searches do not allow to put stringent constraints on their mass, nor to probe the masses required for thermal freeze-out. We provide more details on the collider signatures, and results for real scalars in \appendixname~\ref{app:scalarWIMPs}.

\subsection{Disappearing tracks}\label{sec:collider_dt}

A second handle to tag the production of EW WIMPs at colliders is the detection of tracks from the charged states in the $n$-plet.
As discussed in \sect~\ref{sec:WIMP}, the decay of $\chi^{\pm}\to \chi^0 \pi^{\pm}$ has a lifetime of roughly $c\tau_{\chi^+}\simeq 48\text{ cm}/(n^2-1)$, which is sufficiently long-lived to give rise to reconstructed tracks of length $\mathcal{O}$(cm) for $n=3,5$ that can be observable at colliders. The resulting tracks from these processes are somewhat too short for regular track reconstruction to work efficiently and they will show up as disappearing tracks (DTs), with missing hits in the outermost layers of the tracker and with little or no activity in the calorimeter and the muon chamber. States with higher electric charge in larger multiplets decay promptly to $\chi^{\pm}$, and eventually contribute to the number of disappearing tracks.

\begin{figure*}
\centering
Disappearing tracks --- Majorana 3-plet \hfil\qquad\qquad\qquad\qquad\quad Disappearing tracks --- Majorana 5-plet \\[-8pt]
\includegraphics[width=0.45\textwidth]{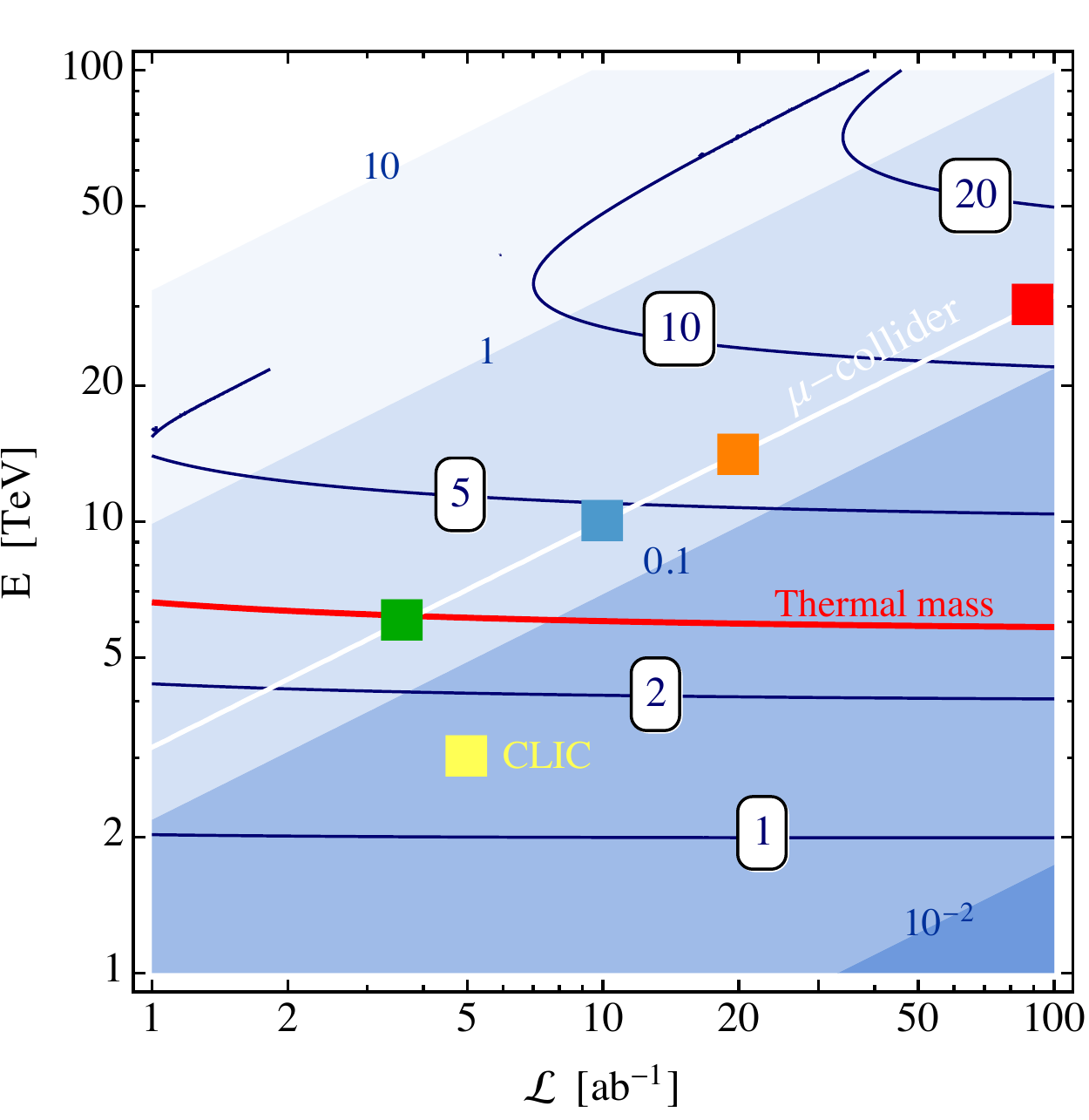}\hfill%
\includegraphics[width=0.45\textwidth]{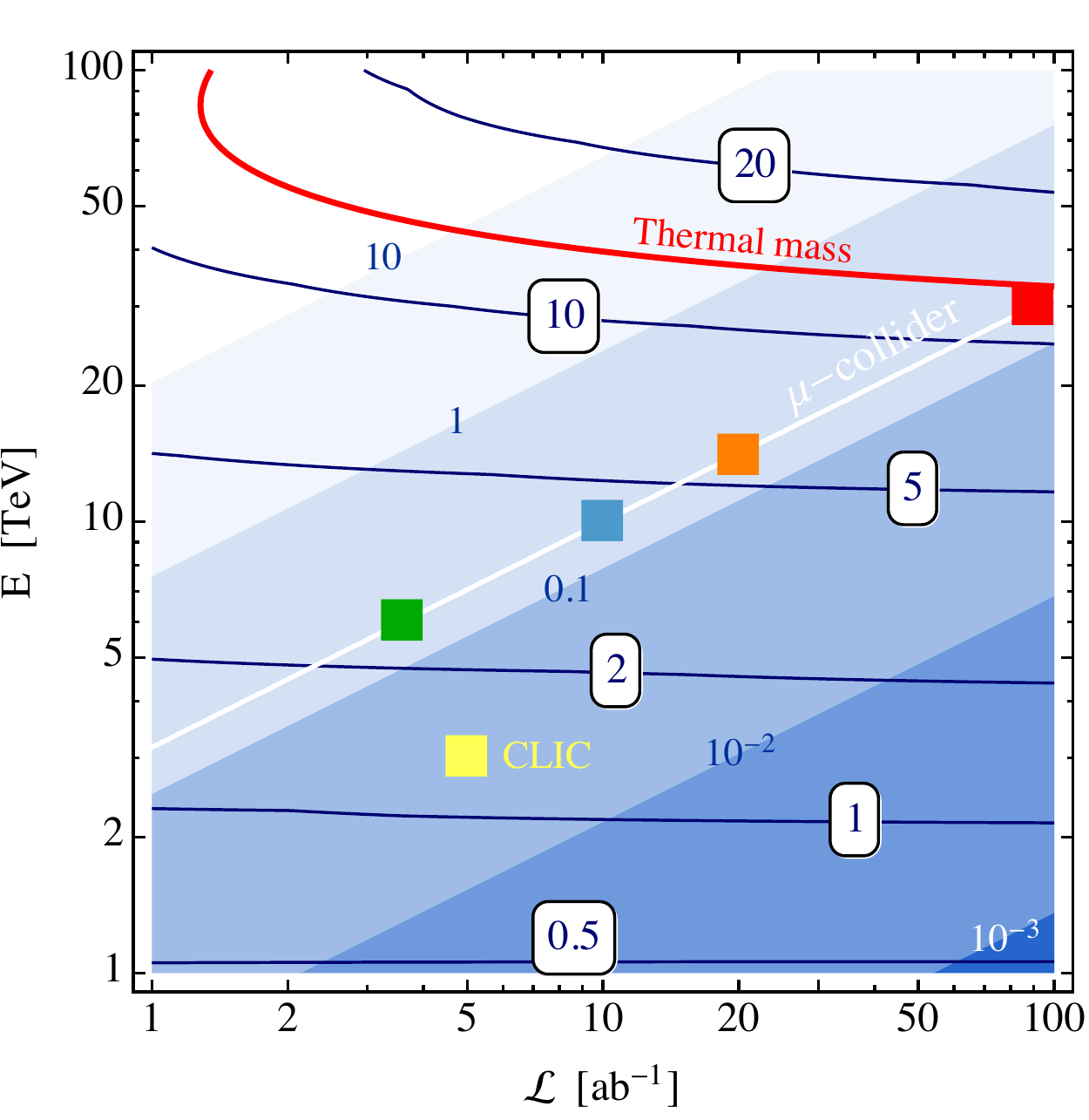}
\caption{Same as \fig~\ref{fig:lumi_vs_energy}, but for disappearing track searches in mono-$\gamma$ events.
{\bf Left:} Majorana 3-plet. {\bf Right:} Majorana 5-plet.\label{fig:lumi_vs_energy_tracks}}
\end{figure*}

A full-detector level study has shown that a high energy lepton collider like CLIC at $\sqrt{s}=3\text{ TeV}$ can reconstruct them sufficiently well to separate them from other sources of look-alike short tracks \cite{UlricheSchoorLCWS2019,LLP20Brondolin}. A recent study \cite{2102.11292v1} has attempted a first evaluation of the performance of this type of search at a multi-TeV muon collider. A main source of worry and a main difference with respect to $e^+e^-$ machines is the abundant number of tracker hits from underlying event activity due to the muon beam decay and to the resulting secondary particles from the interactions with the machine and detector materials. These hits can accidentally  become a potentially severe source of background for searches aimed at highlighting the presence of short tracks of BSM origin. 
We do not enter in the details of these issues here, and simply follow the analysis of~\cite{2102.11292v1}, which is based on a simulation of beam-induced background at 1.5~TeV, and recast their results for the EW 3-plet and the 5-plet. We remind that the background from decaying muons is expected to decrease at higher energies, making our estimate conservative in this sense.

We consider mono-photon events with disappearing tracks, and search for events compatible with a WIMP signal.
Following \cite{2102.11292v1}, we distinguish two event-selection strategies to hunt for disappearing tracks: i) events with at least a disappearing track with $p_{\mathrm{T}}>300 \text{ GeV}$ and a hard photon with $E_\gamma>25 \text{ GeV}$; ii) events with a hard photon, and two disappearing tracks 
originating from the same point along the beam axis.
To estimate the reach we work in the cut-and-count scheme as in \eqref{eq:Significance}, and ignore systematic uncertainties. 
Further details are summarized in \appendixname~\ref{app:DT} for completeness. 

The result of our recast is shown in the last two columns of \fig~\ref{fig:barchart} for Majorana 3-plets and 5-plets at two benchmark colliders, and in \fig~\ref{fig:lumi_vs_energy_tracks} as a function of collider energy and luminosity.
One can see that DTs are especially powerful in the case of the 3-plet, where the reach goes almost up to the kinematical threshold. In particular, an EW 3-plet WIMP of mass as predicted by thermal freeze-out can be discovered already at a $6$ TeV muon collider as suggested in \cite{2102.11292v1,Han:2020uak}.
For higher $n$-plets DT substantially loose exclusion power because the lifetimes of the $\chi^{\pm}\to \chi^0 \pi^{\pm}$  decay become shorter. For the 5-plet the DT reach is comparable to the combined reach of the MIM searches.

As discussed in more detail in the Appendix, DT searches are particularly important to probe scalar WIMPs, since the lower production cross-sections have no significant impact on these almost background-free searches. Disappearing tracks might be the only direct signature of scalar WIMPs at collider experiments.

\section{WIMP direct and indirect detection}\label{sec:directandindirect}

In this Section we briefly summarize the opportunities of the future experimental program in direct and indirect detection in light of the mass predictions derived in \tabl~\ref{table:summary}. 

\subsection{Indirect Detection}

\begin{figure}[htp!]
\includegraphics[width=0.49\textwidth]{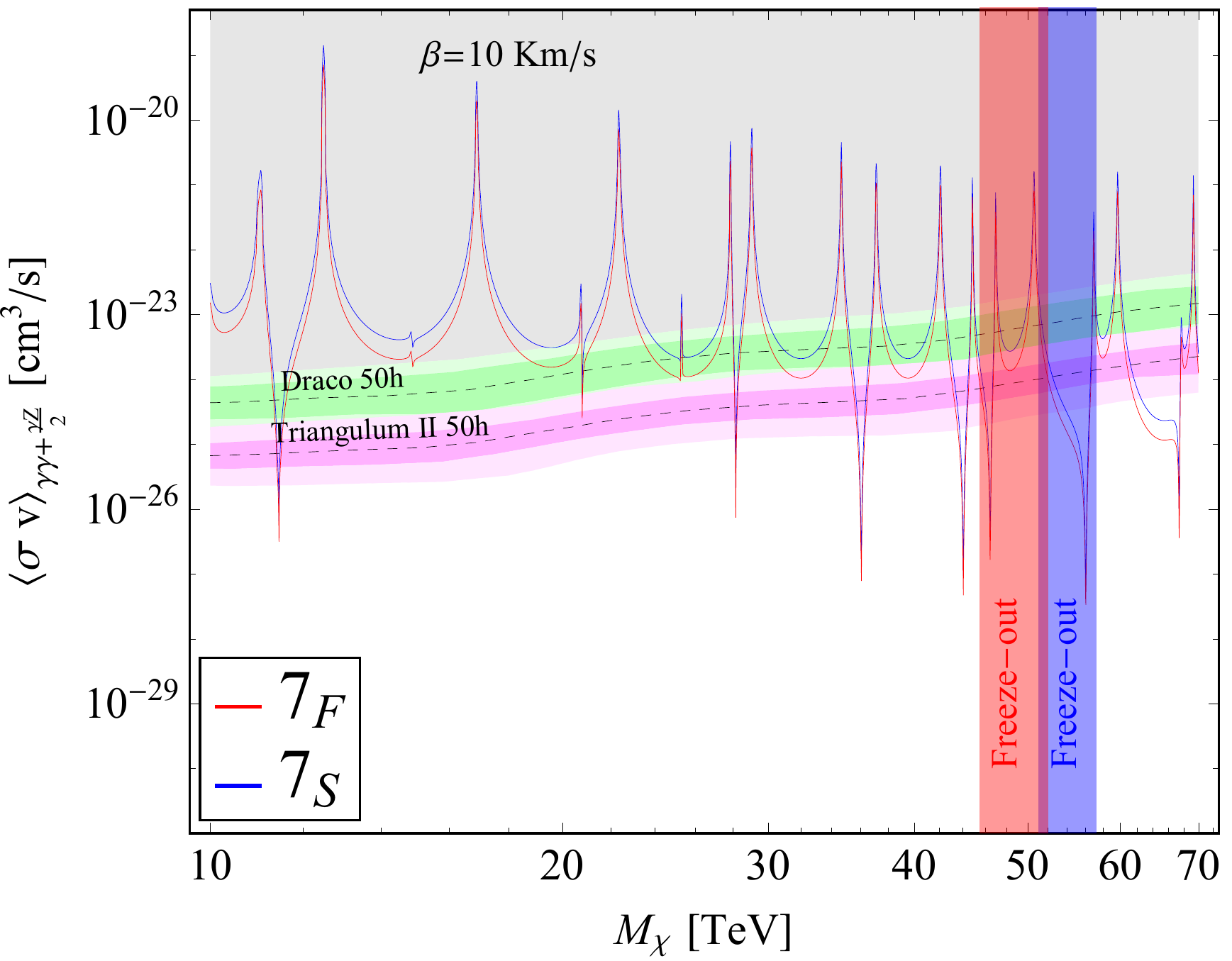}
\caption{Expected CTA sensitivities (dashed black lines) with 68\% and 95\% CL intervals derived as in Ref.~\cite{Lefranc:2016fgn} assuming 50 hours observation time towards Draco (green) and Triangulum II (magenta). We show the SE annihilation cross-section into the channels that contribute to the monocromatic gamma line signal (i.e. $\gamma\gamma$ an $\gamma Z$) for a scalar 7-plet (blue) and a fermionic 7-plet (red). The vertical bands show the predicted thermal masses for the scalar 7-plet (blue) and the fermionic 7-plet (red), where the theory uncertainty is dominated by the neglected NLO contributions  (see \tabl~\ref{fig:summary}).}
\label{fig:id}
\end{figure}

The current and upcoming ground-based Cherenkov telescopes are in a very good position to probe heavy WIMP $n$-plets with $n>5$, which would be inaccessible otherwise. Indeed, these telescopes are designed to detect very high energy gamma-rays (i.e. $E_\gamma \gtrsim 100$ GeV) coming from different astrophysical objects and they are therefore sensitive to the gamma-ray signal from the annihilations of EW $n$-plets. The typical spectrum is characterized at very high energy by gamma-ray lines, peaking at the DM mass $E_\gamma \simeq M_\chi$, from the loop-induced annihilations into $\gamma\gamma$ and $\gamma Z$. The cross-section in this channel is largely boosted by the SE (see e.g.~\cite{Cirelli:2007xd,Cirelli:2015bda,Garcia-Cely:2015dda}) and can raise above the gamma-ray continuum from the showering, hadronization and decays of the electroweak gauge bosons~\cite{Cirelli:2010xx}. 

From the astrophysical point of view, the reach of high energy gamma lines searches depends very much on which portion of the sky the telescopes will be pointed at. In finding the optimal choice, a balance has to be found between the maximization of photon flux at Earth and the control over the systematical uncertainties. Two very well studied astrophysical targets are the Galactic Center (GC)~\cite{Lefranc:2016fgn,Rinchiuso:2018ajn} and the Milky Way's dwarf Spheroidal galaxies (dSphs)~\cite{Lefranc:2016fgn}. In the GC, the uncertainties are dominated by the importance of the baryonic physics in the inner most region of the Milky Way which comes together with the poor knowledge of the DM distribution at the center of the Milky Way~\cite{Iocco:2015xga,2016MNRAS.463..557W,Pato:2015dua,2016MNRAS.463.2623H}. On the contrary, dSphs stands out as very clean environments to search for high energy $\gamma$-lines only residually affected by systematics related to the determination of their astrophysical parameters in the presence of limited stellar tracers~\cite{Lefranc:2016dgx,Ullio:2016kvy}. 

Motivated by the above considerations, we show a very preliminary analysis of ID signals coming from annihilations of the WIMP 7-plet. We focus on the CTA prospects by considering 50h of observations time towards two dSph targets in the northern hemisphere: the classic dSph Draco and the ultra-faint one Triangulum II. Notice that the DM properties of Draco come from hundreds of stellar tracers, while those from Triangulum II are based on just 13 tracers, making the latter more speculative and subject to large systematics in the determination of the geometrical $J$-factor~\cite{Hayashi:2016kcy}. Hence, the reach of Draco should be taken as the baseline reach for CTA.

Our analysis is simplified because the signal shape we consider is essentially a single line at $E_\gamma\simeq M_\chi$. Consistently we take the CTA prospects derived in Ref.~\cite{Lefranc:2016fgn} for a pure line. We ignore the contributions of the continuum spectrum, the extra features of the spectral shape induced by the resummation of EW radiation and the contribution of the BSF to the photon flux. While neglecting BSF is justified if we focus on very high energy photons, a careful computation of the $\gamma+X$ cross-section, where $X$ is any other final state would be needed to precisely assess the experimental sensitivity~\cite{Baumgart:2017nsr}. In the last decade, many different groups have investigated the impact of large Sudakov logarithms and large collinear logarithms on the indirect detection reach, focusing mainly on the case of the fermionic 3-plet~\cite{Hryczuk:2011vi,Ovanesyan:2014fwa,Baumgart:2014saa,Baumgart:2015bpa,Ovanesyan:2016vkk,Ovanesyan:2016vkk}. The inclusion of these effects has been shown to increase the reach of $\sim 20\div30\%$ for the 3-plet~\cite{Lefranc:2016fgn,Abdalla:2018mve,Rinchiuso:2018ajn} and it is expected to be even more important for higher DM masses. 

In \fig~\ref{fig:id} we overlay the SE annihilation cross-section for the 7-plets at $v=10\text{ km}/\text{sec}$ against the CTA experimental reaches. In order to compute the SE in this velocity regime, we took advantage of the parametrization introduced in \cite{Garcia-Cely:2015dda} and used the full expressions for the SE at leading order, including EW breaking effects. The SE saturate already at $v\simeq10^{-3}\div10^{-2}$ far away from the resonances. As we can see, both a 50 hour observation of Triangulum II and of Draco have good chances to detect the high energy $\gamma$ line in the 7-plet annihilation spectrum. 

As we see from \fig~\ref{fig:id}, given the strong mass-dependence of the features of the SE cross-section, a major source of theoretical uncertainty on the reach of indirect detection is still the determination of the 7-plet thermal mass. Therefore, a full computation of the thermal relic mass including NLO effects is required together with a careful computation of the $\gamma+X$ cross-section along the lines of Ref.s~\cite{Hryczuk:2011vi,Ovanesyan:2014fwa,Baumgart:2014saa,Baumgart:2015bpa,Ovanesyan:2016vkk,Ovanesyan:2016vkk} to careful assess the indirect detection reach for the 7-plet.  

Independently on our current inability of making a conclusive statement because of the large theory uncertainties, it is clear that large $n$-plets are a perfect target for future Cherenkov telescopes which deserves further theoretical study. A complementary open phenomenological question is if the low energies gamma lines at $E_\gamma\simeq E_B$ associated to BSF can be actually disentangled from the continuum (see \cite{Mitridate:2017izz,Mahbubani:2019pij} for preliminary work in this direction). An analogous question can be asked for monocromatic neutrinos from BS annihilations.

\subsection{Direct Detection}

\begin{figure*}[htp!]
\includegraphics[width=0.45\textwidth]{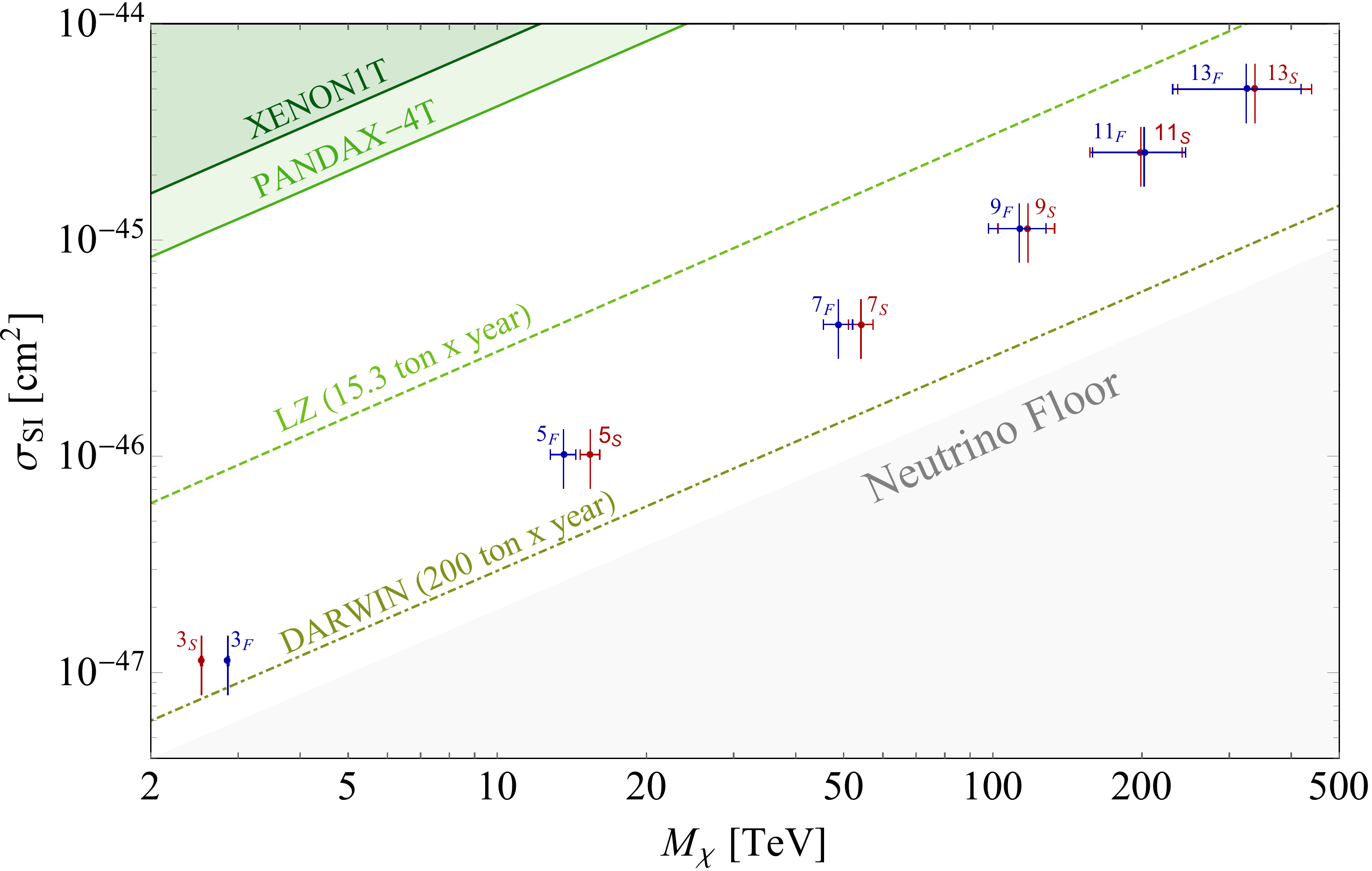}\qquad\qquad
\includegraphics[width=0.45\textwidth]{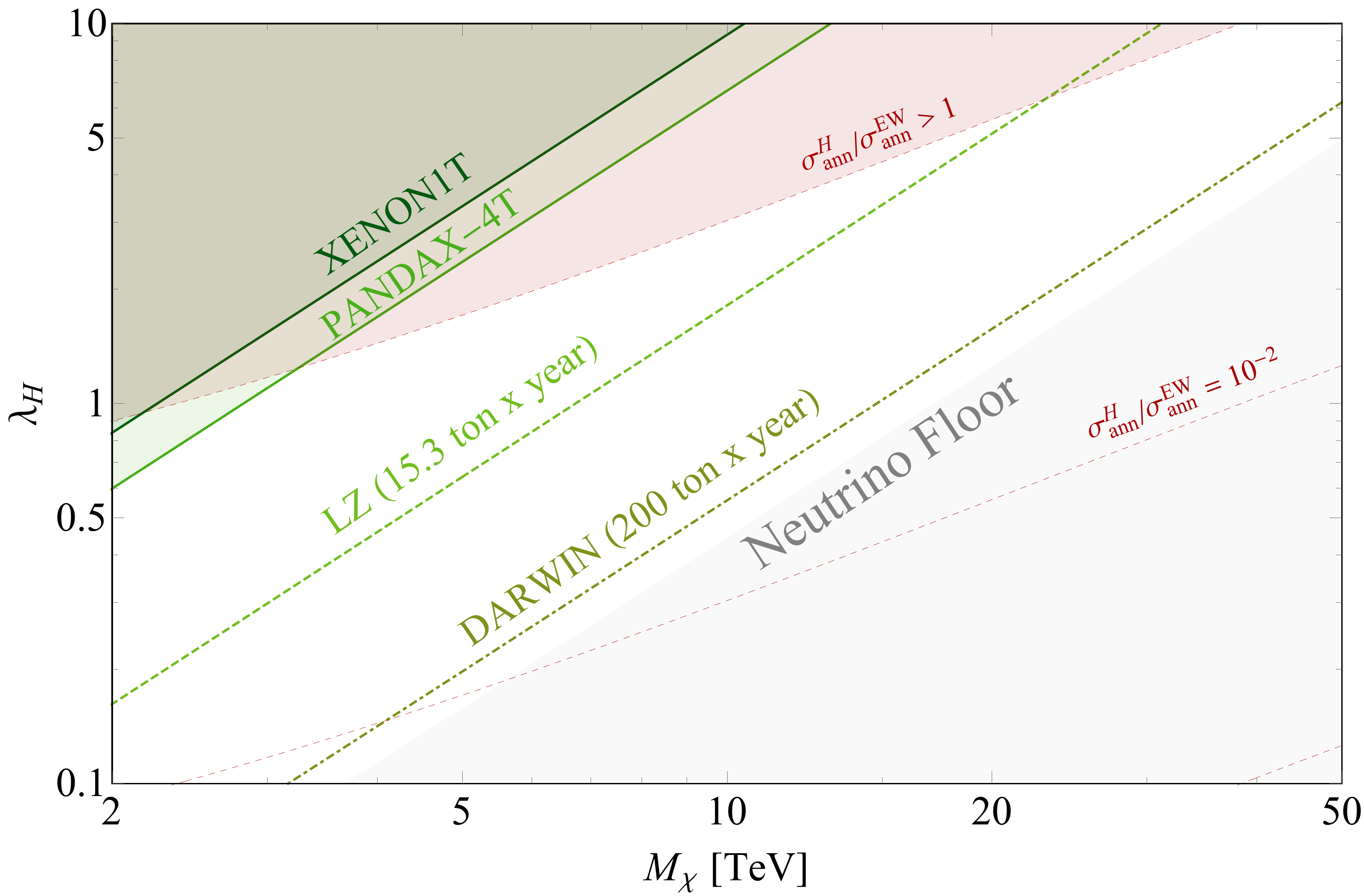}
\caption{ In dark green we show the present contraints from XENON-1T~\cite{Aprile:2018dbl} and  PandaX-4T~\cite{PandaX:2021osp}, the green dashed line shows the reach of LZ~\cite{Mount:2017qzi} and the  brown green dot-dashed line the ultimate reach of DARWIN~\cite{Aalbers:2016jon}. The light gray region show the neutrino floor for 200 ton/year exposure derived in Ref.~\cite{Billard:2013qya}. {\bf Left:} Expected spin independent (SI) direct detection cross-section for Majorana $n$-plets (red) and for real scalar $n$-plets (blue) (assuming the Higgs portal coupling  $\lambda_H=0$). The vertical error bands correspond to LQCD uncertainties on the elastic cross-section in \eqref{elasticSIDD} while the horizontal error band comes from the theory determination of the WIMP freeze out mass. {\bf Right:} Current and future reach on the Higgs portal quartic $\lambda_H$ defined in \eqref{eq:scalarWIMP} for scalar DM. In the shaded dark red region the quartic modifies the freeze-out cross-section by $\mathcal{O}(1)$ or more. The dashed red contours indicate smaller ratios of the Higgs-portal and the EW annihilation cross-sections.}
\label{fig:dd}
\end{figure*}

For $Y=0$ the elastic scattering of DM with the nuclei is induced by EW loop diagrams first computed in \cite{Hisano:2004pv,Hisano:2010fy}. After EW gauge bosons are integrated out, the structure of the UV effective Lagrangian describing the DM interactions reads
\begin{equation*}\label{LagrUVDD}
\mathscr{L}^{\text{SI}}_{\text{eff}}= \bar{\chi} \chi\left( f_q m_q \bar{q} q+  f_G G_{\mu\nu}G^{\mu\nu}\right)+ \frac{g_q}{M_{\chi}} \bar{\chi} i \partial^{\mu} \gamma^{\nu} \chi \mathcal{O}^q_{\mu\nu}\, ,
\end{equation*}
where we focus on the DM spin independent (SI) interactions with quarks and gluons~\cite{DelNobile:2013sia}. The quark twist-2 operator is defined as $\mathcal{O}^q_{\mu\nu} \equiv \frac{i}{2} \bar{q} \left( D_{\mu} \gamma_{\nu} + D_{\nu} \gamma_{\mu} - g_{\mu\nu}\slashed{D}/2   \right)q$. The Wilson coefficients of the operators for general EW $n$-plets with $Y=0$ have been computed in Ref.~\cite{Hisano:2011cs} and at the leading order in $M_\chi/m_{W,h}\gg1$ read
\begin{align}
\label{coeffUV1}
&f_q^{\text{EW}} \simeq \frac{(n^2-1)\pi}{16} \frac{\alpha_2^2}{m_Wm_h^2} \, ,\\
&f_G^{\text{EW}} \simeq - \frac{(n^2-1)}{192}\frac{\alpha_2^2\alpha_s}{m_W}\left( \frac{\sum_{q}  \kappa_q}{m_h^2} + \frac{1}{m_W^2} \right)\, ,\\
&g_q^{\text{EW}} \simeq - \frac{(n^2-1)\pi}{24} \frac{\alpha_2^2}{m_W^3}\, ,\label{coeffUV3}
\end{align}
where $m_h=125\text{ GeV}$ is the SM Higgs mass, $q\in(c,b,t)$ and $\kappa_c=1.32$, $\kappa_b=1.19$, $\kappa_t=1$.  

Following Ref.~\cite{DelNobile:2013sia}, starting from the UV DM interactions we derive the IR interaction of DM with the nucleons. All in all, the SI elastic cross-section per nucleon in the limit $M_\chi\gg m_N$ reads
\begin{equation}
\label{elasticSIDD}
\sigma_{\text{SI}}^{\text{EW}} \simeq \frac{4}{\pi} m_N^4 \vert k_N^{\text{EW}} \vert^2,
\end{equation}
where $m_N$ is the nucleon mass and $k_N^{\text{EW}}$ is defined as 
\begin{equation*}
k_N^{\text{EW}} =  \sum_{q} f_q^{\text{EW}} f_{Tq} + \frac{3}{4} (q(2) + \bar{q}(2)) g_q^{\text{EW}} - \frac{8 \pi}{9 \alpha_s} f_{TG} f_G^{\text{EW}}\, .
\end{equation*}
with the dimensionless nucleon form factors defined as $f_{Tq} = \langle N\vert  m_q \bar{q} q \vert N\rangle / m_N$, $f_{TG} = 1-\sum_{q} f_{Tq}$ with $q\in(u,d,s)$ and $\langle N(p)\vert \mathcal{O}^q_{\mu\nu} \vert N(p)\rangle=\frac{1}{m_N} (p_{\mu}p_{\nu} - \frac{1}{4} m_N^2 g_{\mu\nu})  (q(2) + \bar{q}(2))$, where $q(2)$ and $\bar{q}(2)$ are the second moments of the parton distribution functions for a quark or antiquark in the nucleon taken from \cite{Hisano:2011cs}. Notice that we choose a different set of values for the nucleon form factors with respect to previous studies~\cite{Hisano:2015rsa} which explain the difference in our results. In particular, we take the FLAG average of the lattice computations in the case of $N_f=2+1+1$ dynamical quarks~\cite{Aoki:2019cca,Alexandrou:2014sha, Freeman:2012ry}.  

By propagating LQCD uncertainties on the elastic cross-section (\ref{elasticSIDD}), we obtain the vertical uncertainties on the SI cross-section predictions in \fig~\ref{fig:dd}. We find the partial accidental cancellation between the one loop and the two loop contribution to reduce the elastic cross-section up to 30\%. The horizontal bars represent the uncertainties coming from the computation of the thermal masses through the relic abundance. As shown in the plot, while all the WIMP cross-sections lie above the Xenon neutrino floor as computed in \cite{Billard:2013qya} but only a very large exposure experiment like DARWIN~\cite{Aalbers:2016jon} would be able to probe the heavy thermal WIMPs.

Spin dependent (SD) interactions of DM with the nuclei are also induced by EW loops 
\begin{equation}
\mathscr{L}^{\text{SD}}_{\text{eff}}=d_q (\bar{\chi} \gamma^{\mu} \gamma_5 \chi) (\bar{q} \gamma_{\mu} \gamma_5 q),\quad\! d_q\simeq - \frac{(n^2-1)\alpha_2^2\pi}{24 {m_W M_\chi}},
\end{equation}
where the Wilson coefficient was computed in Ref.~\cite{Hisano:2011cs} and we expanded it at zeroth order in $M_\chi/m_{h}\gg1$. The corresponding SD cross-section is too small to be probed even at a very large exposure experiment like DARWIN. 

Finally, we comment on the new opportunities for direct detection that arise for scalar DM. Here, a non-zero Higgs portal quartic in \eqref{eq:fermionWIMP} leads to a new contribution to the SI DM scattering cross-section with the nuclei, which again in the $M_\chi\gg m_N$ limit reads 
\begin{equation}
\sigma_{\text{SI}}^{\text{H}}=\frac{4}{\pi} m_N^4 \vert k_N^{\text{H}} \vert^2\ ,
\end{equation}
where 
\begin{equation}
k_N^{\text{H}}\simeq \frac{\lambda_H f_N}{4 m_h^2 M_\chi}\ , 
\end{equation}
with $f_N\simeq0.31$ obtained from lattice QCD results (see \cite{Katayose:2021mew} for a more detailed discussion on the scalar triplet). In the right panel of \fig~\ref{fig:dd} we show 
the regions of parameter-space where the Higgs-portal interaction can be tested in direct detection. The requirement of not significantly affecting the freeze-out dynamics bounds the 
%
annihilation cross-section induced by the Higgs portal to be smaller than the EW cross-section,
$\sigma_{\text{ann}}^H/\sigma_{\text{ann}}^{\text{EW}}\lesssim1$, which results in an upper bound on the quartic coupling $\lambda_H$ shown by the red shading in \fig~\ref{fig:dd}. An estimate for this bound can be obtained by comparing the hard annihilation cross-sections, and reads $\lambda_H^2\lesssim(n^2-3)(n^2-1)g_2^4/8$.  Interestingly, XENON1T and PANDAX-4T already exclude a large part of the region where the Higgs portal induces $\mathcal{O}(1)$ modifications of the freeze-out predictions, while LZ will completely exclude this possibility.  

\section{Conclusions}

After many years of hard experimental and theoretical work, the possibility that Dark Matter is part of an EW multiplet is still open and deserves theoretical attention in view of the future plans for experimental searches. In this paper we made a first step in sharpening the theoretical predictions computing all the calculable thermal WIMP masses for real EW representations with vanishing hypercharge. 
We included both Sommerfeld enhancement and bound-state-formation effects at LO in gauge boson exchange and emission. Our results are summarized in \tabl~\ref{table:summary}.

We find that the largest calculable SU(2) $n$-plet at LO is the 13-plet, which is as heavy as 350 TeV. Stronger requirements about the perturbativity of the EW sector up at high scales can further lower the number of viable candidates. We consistently assign a theory error to our predictions by estimating the NLO corrections to the SE. The latter dominate the theory uncertainty for $n\geq7$, while for $n=5$ the error is dominated by the approximate treatment of EW symmetry-breaking effects in the computation of the BSF cross-sections. 

Given the updated mass predictions from thermal freeze-out, we re-examined various phenomenological probes of WIMP DM.

High energy lepton colliders in the 10\,--\,30 TeV range, such as a future muon collider, can directly produce EW multiplets with $n\leq5$. In order to probe a Majorana fermion with $n=3$ ($n=5$) with missing-mass searches,
a collider with at least $\sqrt{s}\sim 12$~TeV ($\sqrt{s}\sim 35$~TeV) and the baseline integrated luminosity of \eqref{eq:lumi} would be required. The highest mass reach is obtained by means of an inclusive mono-$W$ search. 

Interestingly, disappearing tracks originating from the decay of the singly-charged state into the neutral one are robust predictions of real EW multiplets with $Y=0$, and ameliorate the sensitivity for the 3-plet compared to missing-mass searches. For the 5-plet we find the expected sensitivity of disappearing tracks to be very similar to the one of missing-mass searches due to the shorter average lifetime of the tracks.

Scalar WIMPs can not be probed through missing-mass searches, due to their smaller production cross-section. However, disappearing tracks searches are very powerful tests even for scalar multiplets, thanks to their very low background contamination. This signature is therefore a crucial ingredient to fully explore the parameter space of thermally produced WIMP Dark Matter at future colliders.

Heavy EW WIMPs with $n>5$ are too heavy to be produced at colliders. However, they are perfect targets for indirect detection at upcoming ground-based Cherenkov telescopes like CTA. More theoretical work is necessary to make a robust forecast both on the determination of the photon spectrum for large $n$-plets and on improved precision predictions for the freeze-out masses.   

Finally, large-exposure liquid Xenon experiments like DARWIN can in principle probe all the relevant EW WIMPs through their weak interaction with nuclei. Scalar WIMPs can further be tested through their Higgs-portal quartic interaction. Interestingly, $\mathcal{O}(1)$ modification of the thermal freeze-out masses due to the Higgs portal are already partially excluded by the XENON1T and PANDAX-4T results, and will be completely excluded by LZ.  

A natural continuation of the work done here would be to consider complex EW multiplets. For vanishing hypercharge both the cosmology and the phenomenology will be very similar to the ones discussed here. The suppression of the annihilation cross-section, resulting in lower thermal masses, together with the enhancement of the production cross-section at colliders will favour the direct exploration of complex multiplets at colliders. More interestingly, EW multiplets with nonzero hypercharge, such as the Higgsino in supersymmetric models, are also phenomenologically viable if the DM elastic scattering with nucleons is suppressed or kinematically forbidden. 
Classifying the predictions of this class of models would give a complete picture on EW DM multiplets. We hope to come back to this open issues in the near future.


\section*{Acknowledgments}
{\small We are grateful to Brando Bellazzini, Kallia Petraki, Luca di Luzio, Guido Martinelli, Federico Meloni, Michele Redi, Filippo Sala, Alessandro Strumia, Ryosuke Sato, Juri Smirnov, Alfredo Urbano, Andrea Wulzer for interesting discussions. DB, MC, and LV are partially supported by the PRIN 2017L5W2PT, and by the INFN grant FLAVOR.}

\appendix

\section{Bound States Dynamics}\label{app:boundstates}

\begin{figure*}[htp!]
    \centering
\includegraphics[ width= 0.99 \linewidth]{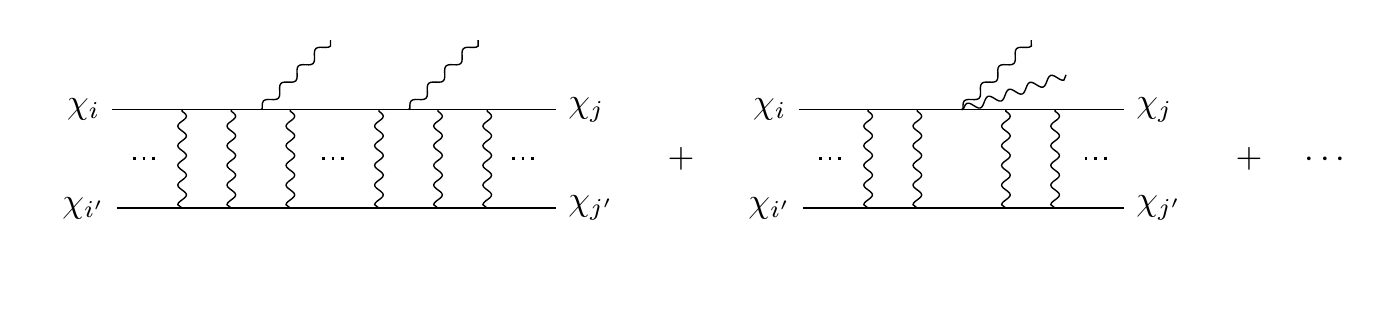}
    \caption{Examples of diagrams controlling the BS effective Hamiltonian at next-to-leading order in gauge boson emission The first diagram corresponds to the second order Born approximation for the dipole operators in \eqref{eq:eff_ham}, where the resummation of the vector boson insertions between the two emission reconstructs the wave function of an intermediate BS or scattering state. The second diagram, instead, is obtained from the $\mathcal{O}(A^2)$ terms in the interaction Hamiltonian, at leading order in the Born approximation.}
    \label{fig:abelian}
\end{figure*}

In this Appendix we discuss in detail the dynamics of Bound State Formation (BSF). First we discuss in \sect~\ref{app:LOandNLO} the general features of BSF at leading order (LO) in gauge boson emission and at next-to-leading order (NLO) in gauge boson emission.  Then we detail in \sect~\ref{eq:7plet}  the 7-plet BS dynamics, focusing on the differences with the 5-plet case.

\subsection{Bound State formation at LO and at NLO}\label{app:LOandNLO}

At leading order, bound states form through the emission of a single vector boson $V^a$: $\chi_i + \chi_j \rightarrow \text{BS}_{i'j'} +V^a$. The non-relativistic limit of the amplitude can be recast in the form of an effective interaction Hamiltonian, such that the full amplitude can be obtained from its matrix element with the wave function of the initial and final two-particle states (reconstructed from the resummation of the ladder diagrams). The leading order contribution to this effective hamiltonian comes in the form of electric dipole interaction terms \cite{Mitridate:2017izz,Harz:2018csl}:

\begin{widetext}
\begin{equation}
\label{eq:eff_ham}
\mathscr{H}_I^{\text{LO}} = -\frac{g_2}{M_\chi}\left(\vec{A}^a(\vec{x}_1)\cdot \vec{p}_1 T^a_{i'i}\delta_{j'j}+\vec{A}^a(\vec{x}_2)\cdot \vec{p}_2 \overline{T}^a_{j'j}\delta_{i'i}\right)+g_2\alpha_2\left(\vec{A}^a(0)\cdot\hat{r}e^{-M_a r}\right)T^b_{i'i}\overline{T}_{j'j}^cf^{abc}\ ,
\end{equation}
\end{widetext}
where the first to terms are a simple generalization of the standard QED dipole interaction while the last one is a purely non-abelian term which arises from vector boson emission from a vector line.

The computation of the transition amplitudes from \eqref{eq:eff_ham} simplifies if we assume the SU(2)$_L$-invariant limit. This approximation applies when the DM (BS) de Broglie wavelength is much smaller than the range of the Yukawa interaction $1/m_W$ and therefore for $z\leq(M_\chi/m_W)^2$.  In this regime the Yukawa potential is well approximated by the Coulomb one which turns out to be a good approximation to describe WIMP freeze-out. The BS dynamics can then be understood by using isospin selection rules while the main consequence of having finite vector masses is to provide an energy threshold to the emission of a single massive boson in the formation or the decay of a BS.

Since $\alpha_{\eff}\sim n^2$, increasing the dimensionality of the DM multiplet enhances next to leading order (NLO) processes in gauge boson emission such as $\chi_i + \chi_j \rightarrow \text{BS}_{i'j'} +V^a+V^b$. These could be in principle relevant for both the computation of the thermal mass and the saturation of the perturbative unitarity bound.

The main NLO contributions to BSF come from diagrams like the ones in \fig~\ref{fig:abelian} and are essentially of two types: i) the first diagram is essentially the second order Born approximation of the LO Hamiltonian, with the intermediate state being a free or a BS; ii) the second diagram, where the two emitted vectors come from the same vertex, is generated by the effective Hamiltonian at order $\mathcal{O}(A^2)$. The latter contains terms of the form
\begin{equation}
\mathscr{H}_I^{\text{NLO}} \supset \frac{g_2^2}{2M_\chi}T^aT^b\!\left[\vec{A}^a\cdot\vec{A}^b + \frac{(\vec{p}\cdot \vec{A}^a)(\vec{p}\cdot\vec{A}^b)}{M^2_\chi}\right],
\end{equation}
where we focus here on the abelian part of the hamiltonian, postponing a full study for a future work.  Given the above Hamiltonian and the LO one in \eqref{eq:eff_ham} we can estimate the corresponding contribution to the double emission BSF cross-section as: 
\begin{align}
&\sigma_{\text{BSF}}^{\text{LO}}\vrel\simeq  \frac{2\pi\alpha_\eff}{M^3_\chi}\Delta E\,\\
&\sigma_{\text{BSF}}^{\text{NLO}}\vrel\simeq \frac{g_\chi^2}{8M_\chi^2\vrel}\left(\frac{\Delta E}{M_\chi}\right)^3\ ,\label{eq:NLObs}
\end{align}
where $g_\chi=1$ for Majorana fermions ($g_\chi=2$) for real scalars. In the LO estimate, a factor $\frac{2}{\alpha_\eff M_\chi}\frac{2\pi \alpha_\eff}{\vrel}$ comes from the overlap integral while a factor $\frac{\Delta E}{8\pi}$ from the two-body phase space. Similarly, in the NLO estimate a factor $\frac{1}{2}\frac{\Delta E^3}{256\pi^3}$ comes from the 3-body phase space, taking into account the two identical final vectors, and $\left(\frac{2}{\alpha_{\eff} M_\chi}\right)^3\frac{2\pi\alpha_{\eff}}{\vrel}$ from the overlap integrals between the wave functions. From the above formula we derive the scaling of the NLO corrections in \eqref{eq:1loopBSF}.

We now discuss the contributions from second order Born expansion whose general expression is given by
\begin{equation}
\label{eq:sig2V}
(\sigma \vrel)_{2V} =\frac{2^6\alpha_2^2}{3^3\pi M_\chi^4} \int\!\!\mathrm{d}\omega \omega(E_n-\omega)\left|\mathcal{C}_{\text{BS}}+\mathcal{C}_{\text{free}}\right|^2\ ,
\end{equation}
where we defined 
\begin{widetext}
\begin{subequations}
\label{eq:cint}
\begin{align}
&\mathcal{C}_{\text{BS}}=\sum_m\left(\frac{1}{E_n-E_m-\omega+i\Gamma_{\text{dec},m}}+\frac{1}{\omega-E_m+i\Gamma_{\text{dec},m}}\right)\mathcal{I}_{\vec{q}m}\mathcal{I}_{mn}\ ,\\
&\mathcal{C}_{\text{free}}=\int\frac{\mathrm{d}^3k}{(2\pi)^3}\left(\frac{1}{E_n-\omega+\frac{k^2}{M_\chi}+i\epsilon}+\frac{1}{\omega-\frac{q^2}{M_\chi}+\frac{k^2}{M_\chi}+i\epsilon}\right)\mathcal{I}_{\vec{q}\vec{k}}\mathcal{I}_{\vec{k}f}\ ,
\end{align}
\end{subequations}
\end{widetext}
with $\mathcal{I}_{if}$ being the overlap integrals between the states $i$ and $f$, the index $m$ running over all intermediate BS and the $k$-integral running over all the intermediate scattering states. 

Starting from $\mathcal{C}_{\text{BS}}$, the intermediate BS are rather narrow resonances because
\begin{equation}
\Gamma_{\text{dec}}\sim \alpha_{\eff}^3 E_B \ll E_B\ ,
\end{equation}
where $E_B$ is a typical binding energy. This quick estimate, supported by the full numerical computation, suggests that $\mathcal{C}_{\text{BS}}$ contribution is fully captured in the Narrow Width Approximation (NWA) for the intermediate BS. Therefore, neglecting the interference terms, one gets
\begin{equation}
(\sigma \vrel)_{2V}=\sum_m (\sigma \vrel)_{1V,m}\mathrm{BR}_{m\rightarrow n}\ ,
\end{equation}
which is exactly the single emission result.

To estimate the contribution from $\mathcal{C}_{\text{free}}$ we need to estimate $\mathcal{I}_{\vec{q}\vec{k}}$ which encodes the contribution from intermediate continuum states. For simplicity, we stick to the abelian contribution which reads 
\begin{equation}
\mathcal{I}_{\vec{q}\vec{k}}=\int r^2\mathrm{d}rR_{\vec{k},1}\partial_r R_{\vec{q},0}\ .
\end{equation}
The integral above can be split into small and large $r$ regions, roughly separated by the Bohr radius $a_0=\frac{1}{\alpha_\eff M_\chi}$
\begin{equation}
\begin{split}
\mathcal{I}_{\vec{q}\vec{k}}= &\int_0^{a_0} r^2\mathrm{d}rR_{\vec{k},1}\partial_r R_{\vec{q},0}+\int_{a_0}^{\infty} r^2\mathrm{d}rR_{\vec{k},1}\partial_r R_{\vec{q},0}\\
\sim & \frac{1}{\alpha_\eff M_\chi \sqrt{k q}}+\frac{q}{(M_\chi \alpha_\eff)^2}\delta(q-k)\ ,
\end{split}
\end{equation}
which plugged into \eqref{eq:cint} gives an estimate to $\mathcal{C}_{\text{free}}$. All in all, plugging these estimates in \eqref{eq:sig2V} and replacing $q=M_\chi \vrel$ we get that the contribution from NLO exchange of continuum states behaves similarly to the ones estimated in \eqref{eq:NLObs}  up to subleading terms in the $\vrel<\alpha_\eff$ regime.

In conclusion, NLO corrections to BSF are suppressed by $\sim\alpha_\eff^3/64\pi$ with respect to the LO ones. As a consequence, the leading NLO contributions to the total annihilation cross-section are the ones correcting the LO SE. The latter are log-enhanced as detailed in \eqref{eq:1loop} and first computed in \cite{Beneke:2020vff} for the fermionic 3-plet. It would be interesting to extend these computation to higher EW $n$-plets. 

\begin{figure*}
\includegraphics[width=0.44\textwidth]{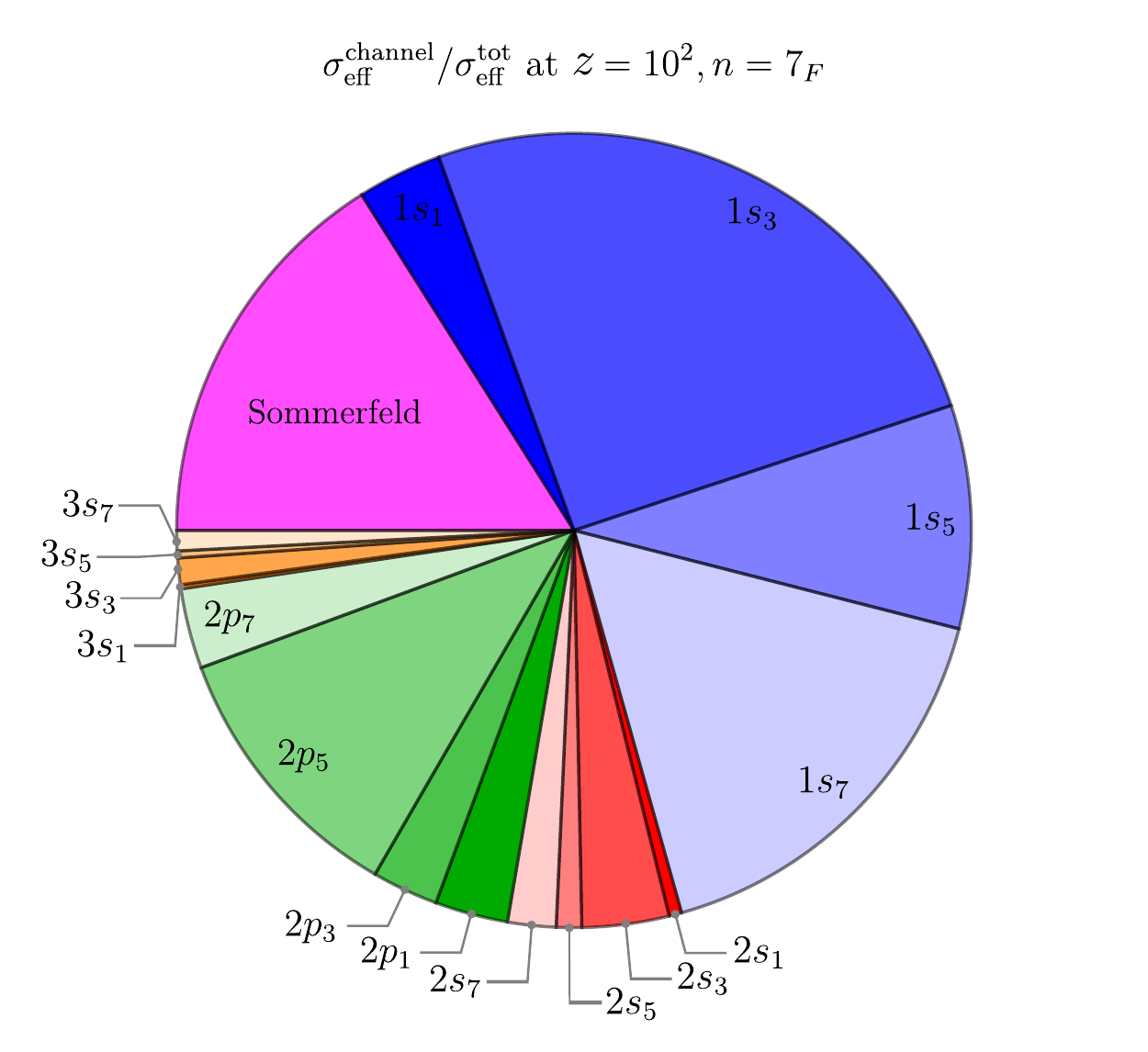}\qquad\qquad
\includegraphics[width=0.44\textwidth]{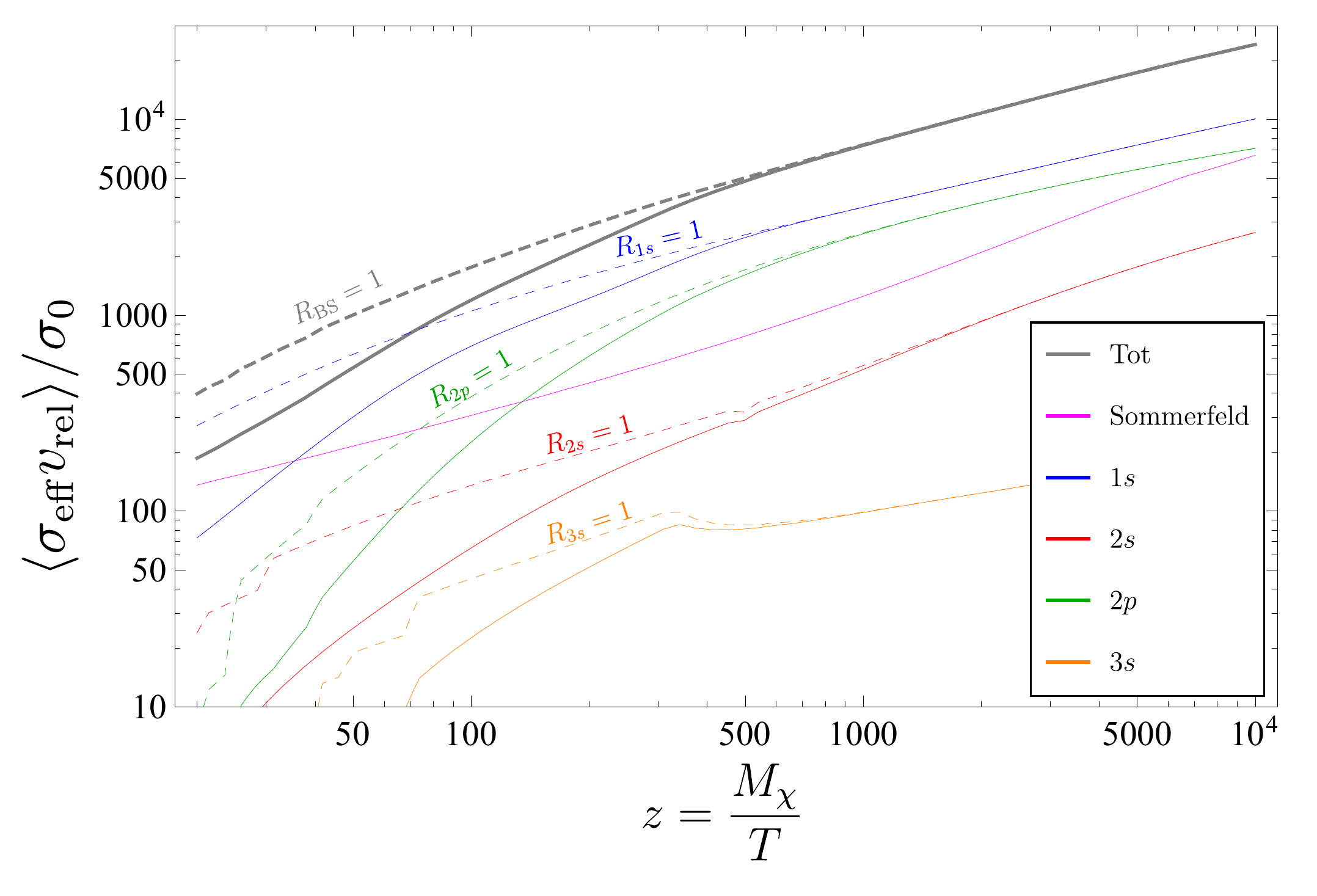}
\caption{{\bf Left:} Piechart showing the contributions to the 7-plet effective annihilation cross-section of each single BS channel, together with the SE, at fixed $T=10^{-2} M_\chi$ (i.e. $z=10^2$). {\bf Right:} Temperature dependence of the different contributions to the 7-plet effective cross-sections. Each BS channels is summed over the different isospins. } 
\label{fig:pie}
\end{figure*}
\begin{figure}
\includegraphics[width=0.44\textwidth]{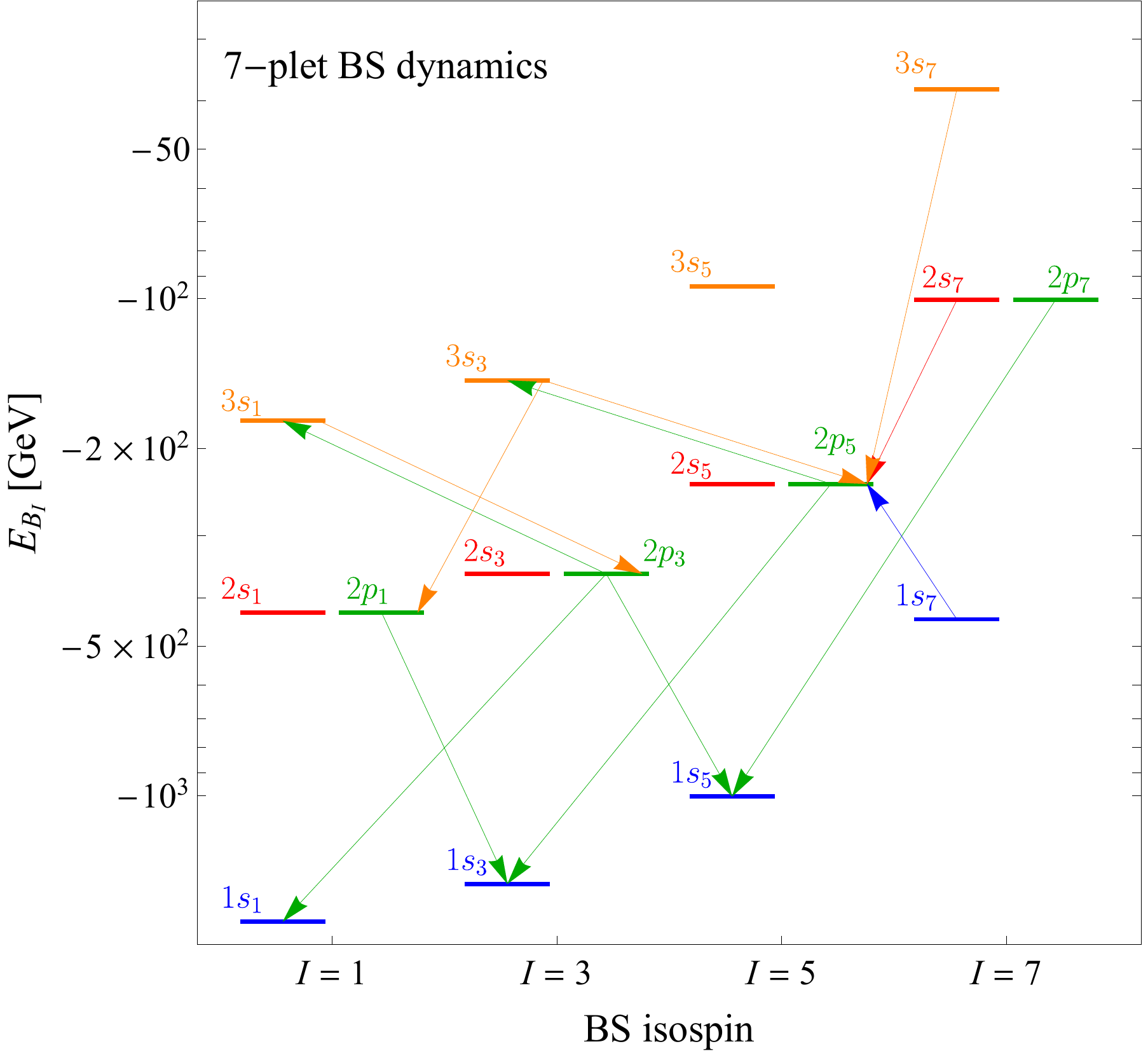}
\caption{BS energy levels for the 7-plet: blue $n=1$, red $n=2$ with $L=0$, green $n=2$ with $L=1$ and orange $n=3$. The arrows indicate the decay rate of each state. BS with no lines attached have an annihilation rate at least one order of magnitude larger than the decay rate.\label{fig:BSdynamics}}
\end{figure}

\subsection{The 7-plet Bound States in detail}\label{eq:7plet}

The 7-plet has a richer bound states dynamics with respect to the 5-plet, essentially because of the additional layers of isospin and energy levels. As we will discuss here, keeping track of this dynamics is crucial to compute correctly the relic abundance. 

The left panel of \fig~\ref{fig:pie} shows the relative importance of the different BS and the SE to the effective cross-section at fixed $T=10^{-3} M_\chi$. As we can see, BSF accounts for most of the total cross-section. Compared to the 5-plet case, the new attractive isospin channels with $I=7$ give a sizeable contribution to the 7-plet cross-section as well as the $2p$ states which were instead irrelevant for the 5-plet.
In the right panel of \fig~\ref{fig:pie} we show how the details of the bound state dynamics are especially important at temperatures around the freeze-out (i.e. $z=10^2$) where the effects of BS breaking due to interactions with the plasma are non negligible. This can be seen by comparing the behavior of the full computation of the effective cross-section (solid lines) against the BSF cross-section with zero ionization rate (i.e. $R_{BS}=1$ in the notation of \eqref{eq:rbs}). In particular taking $R_{BS}=1$ yields an overestimate of the final thermal mass of about 6 TeV. Interestingly, we see that for $z=10^3$ all the BSF rates approach the $R_{BS}=1$ limit, signalling that the ionization rate is already heavily Boltzmann suppressed. 

We now illustrate the details of the BS dynamics for the 7-plet. The general computation outlined around \eqref{eq:boltzmann} is in general cumbersome, but it simplifies singling out the specific features of each BS. These are summarized in \fig~\ref{fig:BSdynamics}. We now discuss them in turn, going from the largest to the smallest binding energy.

\begin{figure*}[ht!]
\begin{centering}
\includegraphics[width=0.46\textwidth]{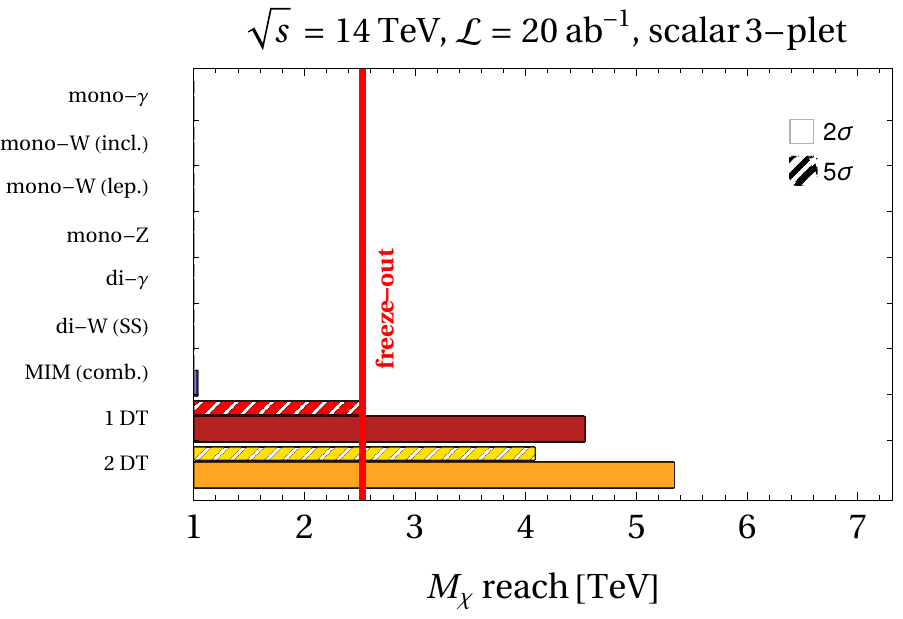}\qquad\qquad
\includegraphics[width=0.46\textwidth]{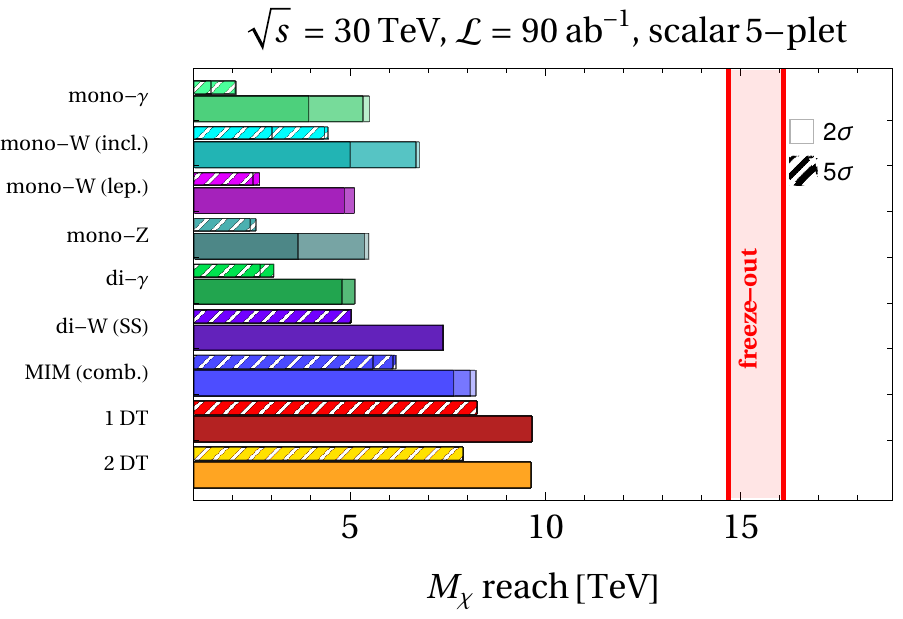}
\end{centering}
\caption{Different bars show the reach at $2\sigma$ (full wide) and at $5\sigma$ (hatched thin) on the WIMP mass at a muon collider with baseline luminosity given by \eqref{eq:lumi} for the different search channels discussed in \sect~\ref{sec:missingp}:
mono-gamma, inclusive mono-W, charged mono-W, mono-Z, di-gamma, same-sign di-W , the combination of all these MIM channels (blue). We also show the reach of disappearing tracks as discussed in \sect~\ref{sec:collider_dt}:  at least 1 disappearing track (red), or  exactly 2 tracks (orange).  All the results are obtained assuming systematic uncertainties to be: 0 (light), $1\permil$ (medium), or 1\% (dark). The vertical red lines show the freeze-out prediction band. {\bf Left:} Scalar 3-plet for $\sqrt{s} = 14\,{\rm TeV}$  {\bf Right:} Scalar 5-plet for $\sqrt{s} = 30\,{\rm TeV}$.\label{fig:barchartscalar}}
\end{figure*}

\begin{enumerate}
\item $1s_I$ and $2s_I$ states with isospin $I\leq 5$ annihilate fast into pairs of SM vectors and fermions, with rates $\Gamma_{\text{ann}}\simeq \frac{\alpha_{\eff}^5}{n_B^2} M_\chi$. Since their decay rate can be neglected, the effective cross-section can easily be obtained from \eqref{eq:singleBS}. 
\item The $1s_7$ BS cannot decay directly into SM pairs because of its large isospin so that its annihilation rate arises at NLO in gauge boson emission. Similarly, the decay to lower $1s$ states can only go through NLO processes or velocity-suppressed magnetic transitions. As a consequence, this BS can only be excited to $2p_5$ at LO, and its effective cross-section can be written in terms of the one of the $2p_5$:
\begin{equation}
\label{eq:1s7}
R_{1s_7}(z)=\frac{\langle\Gamma_{1s_7\rightarrow 2p_5}\rangle}{\langle\Gamma_{1s_7\rightarrow 2p_5}\rangle+\langle\Gamma_{1s_7,\text{break}}\rangle}R_{2p_5}(z)\ , 
\end{equation}
where the excitation rate can be written in terms of the decay rate $\Gamma_{I\rightarrow J}\simeq g_I/g_J\Gamma_{J\rightarrow I}e^{-\frac{\Delta E}{T}}$  times the probability of finding a vector in the plasma with energy of order $\Delta E$. Because of the small energy required from the plasma compared to ionization, excitations still occur long after the ionizations have gone out of equilibrium. 
\item The $2s_7$ has a suppressed annihilation rate to SM like the $1s_7$, but it quickly decays to the $2p_5$ at LO in vector boson emission so that we have
\begin{equation}
\label{eq:2s7}
R_{2s_7}(z)=\frac{\langle\Gamma_{2s_7\rightarrow 2p_5}\rangle}{\langle\Gamma_{2s_7\rightarrow 2p_5}\rangle+\langle\Gamma_{2s_7,\text{break}}\rangle}R_{2p_5}(z)\ . 
\end{equation}
\item The annihilation rates into SM state of the $2p_I$ BS are suppressed by $\alpha_{\text{eff}}^{2}$ compared to the ones of the $2s_I$ BS. Their dynamics is then dominated by the decay (excitation) rates into lower (higher) $s-$orbital BS which scale as $\Gamma_{\text{dec}}\sim \alpha_{\eff}^5M_\chi$.  A simple example of this dynamics is provided by the two-state system $2p_1-1s_3$ where $2p_1$  dominantly decays to $1s_3$, which promptly annihilates to SM. The effective cross-section of $2p_1$ reads
\begin{equation}
\label{eq:doubleBS}
R_{2p_1}(z)=\frac{\langle\Gamma_{2p_1\rightarrow 1s_3}\rangle}{\langle\Gamma_{2p_1\rightarrow 1s_3}\rangle+\langle\Gamma_{2p_1,\text{break}}\rangle}R_{1s3}(z)\ ,
\end{equation}
as we would intuitively expected. The other $2p$ states have more intricated chains, which involve also excitations $3s$ states.
\item We also include $3 s_I$ BS which annihilate directly to SM for $I\leq 5$ and decay into $p-$orbitals states. 
\end{enumerate}
Finally, we checked that $p$ states with $n>2$, $s$ states with $n>3$, and BS with $I=9$ have a negligible impact on the cosmological evolution.  

\section{More on WIMPs at future lepton colliders}\label{app:collider}

\subsection{The scalar WIMPs}\label{app:scalarWIMPs}

\begin{figure*}
\begin{centering}
\includegraphics[width=0.46\textwidth]{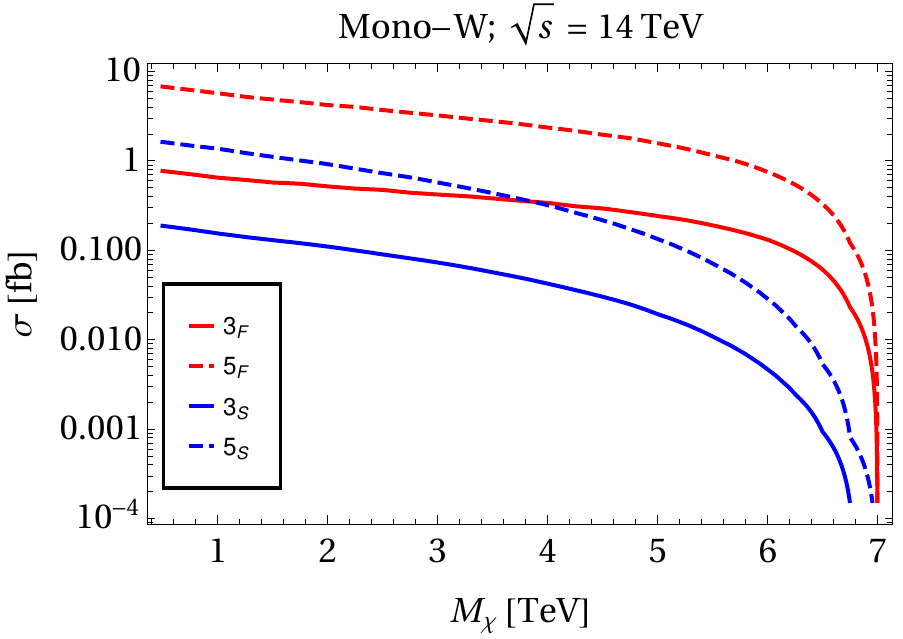}
\qquad\qquad
\includegraphics[width=0.46\textwidth]{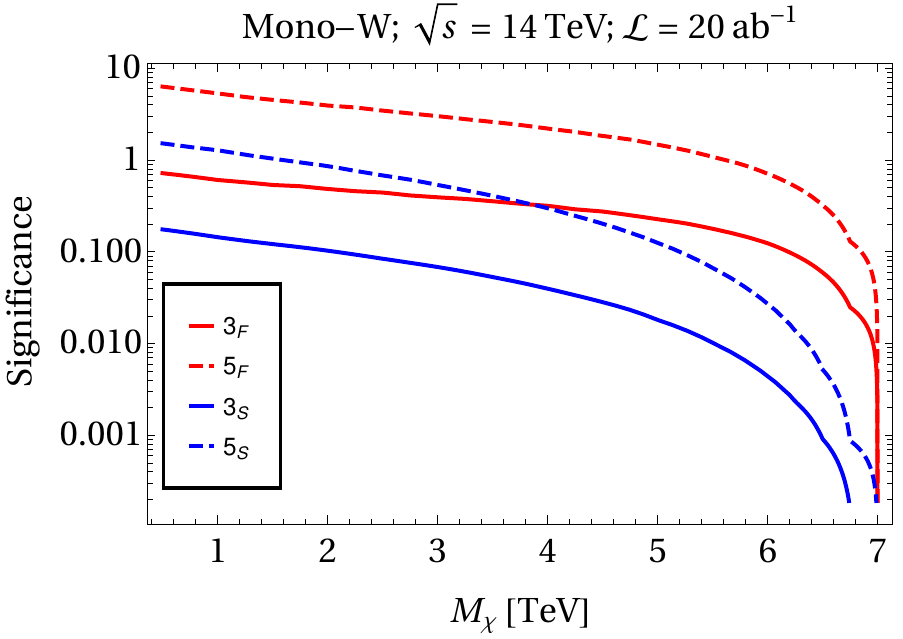}
\end{centering}
\caption{{\bf Left:} Drell-Yan Mono-W cross-section for $\sqrt{s} = 14\,{\rm TeV}$.  {\bf Right:} Significance of the mono-$W$ search
for $\sqrt{s} = 14\,{\rm TeV}$. In both plots, the only cuts applied are $|\eta_W|<2.5$ (geometric acceptance) and $\mathrm{MIM}>2M_\chi$. \label{fig:monow_reach}}
\end{figure*}

\begin{figure*}[ht]
\centering
Mono-$W$ reach --- Scalar 3-plet \hfil\qquad\qquad\qquad\qquad\qquad\,\, Mono-$W$ reach --- Scalar 5-plet \\[-8pt]
\includegraphics[width=0.45\textwidth]{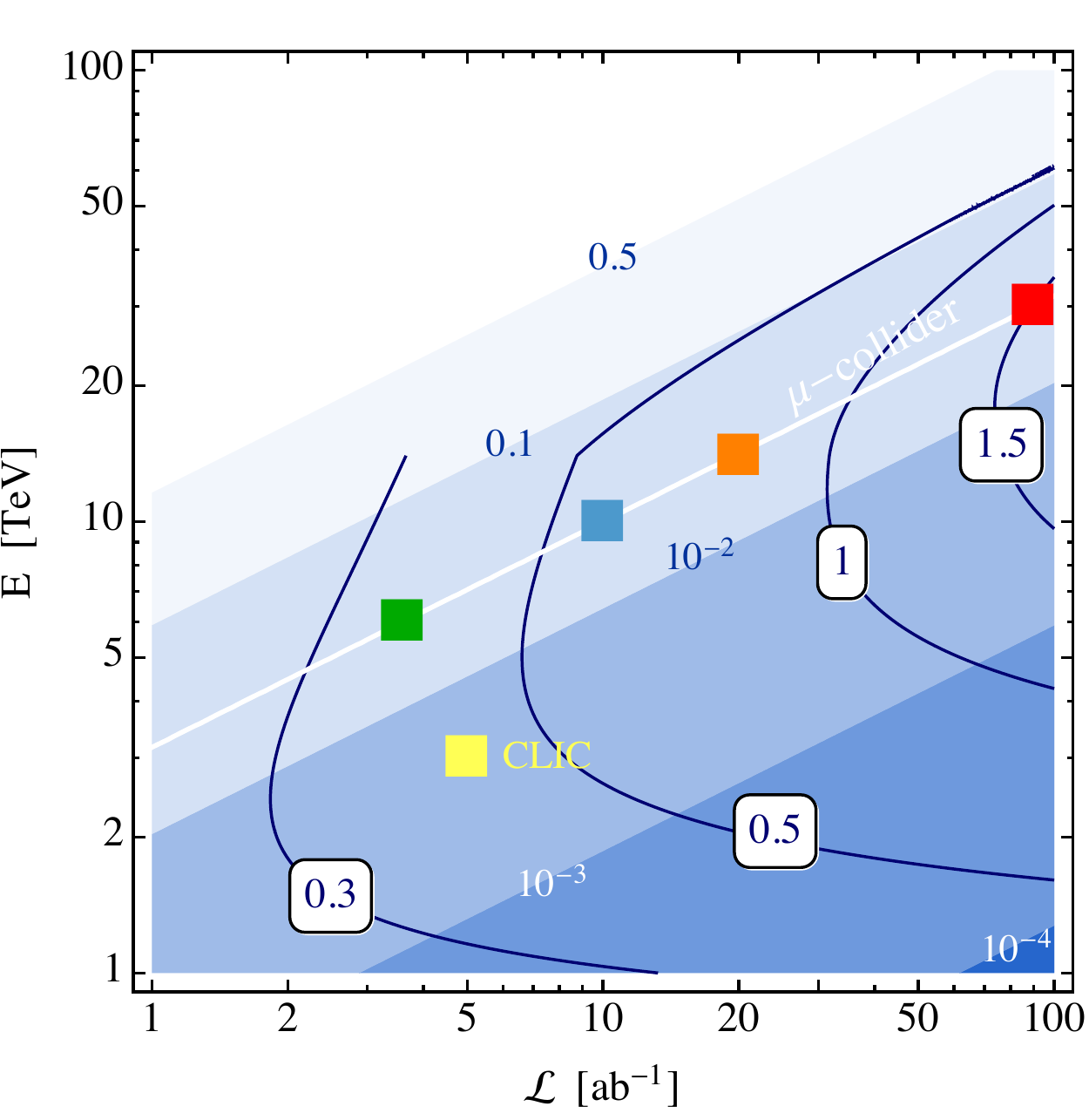}\hfill%
\includegraphics[width=0.45\textwidth]{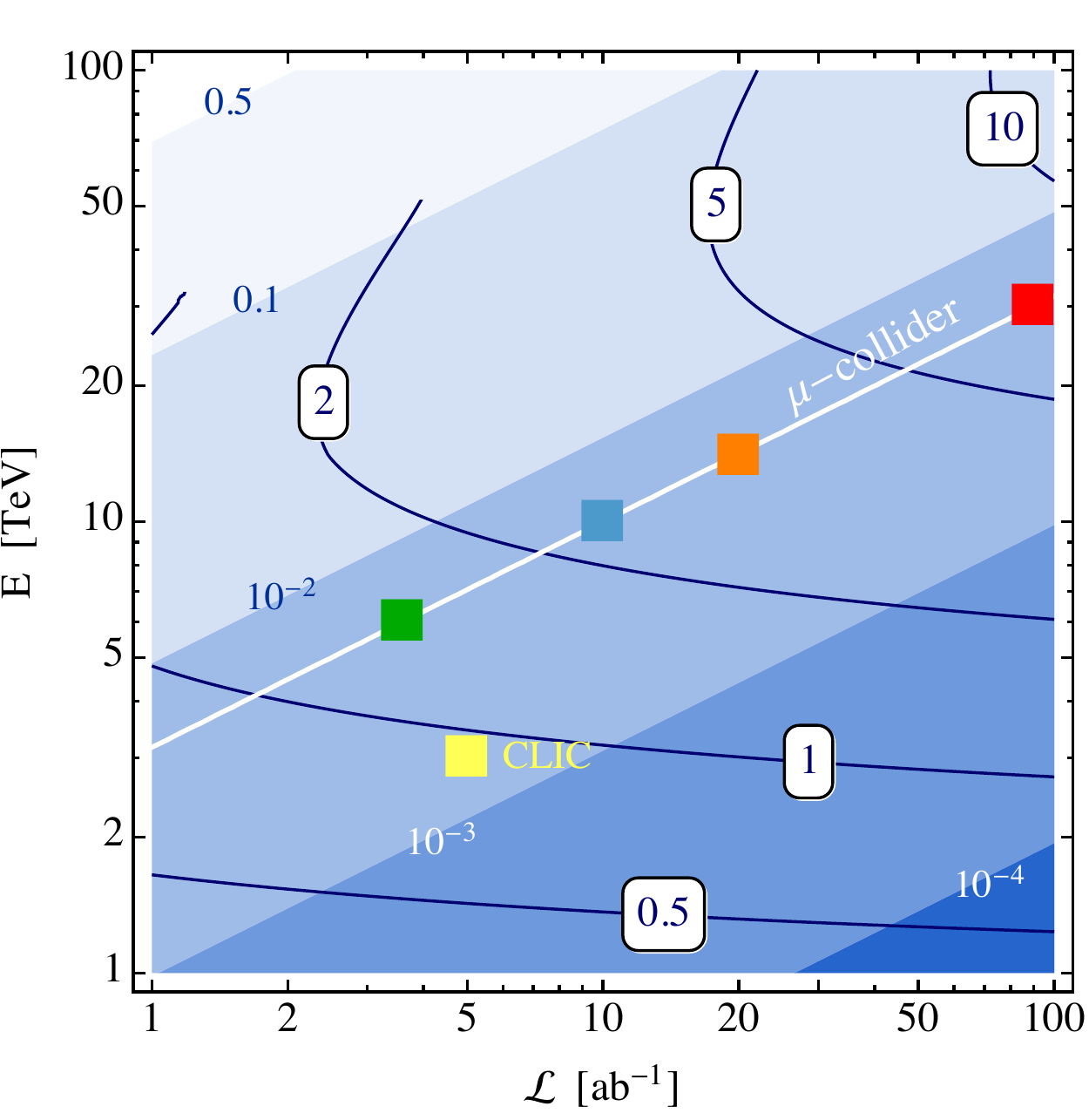}
\caption{Same as \fig~\ref{fig:lumi_vs_energy}, but for real scalar WIMPs.
{\bf Left:} Scalar 3-plet. {\bf Right:} Scalar 5-plet.\label{fig:lumi_vs_energy_scalar} }
\end{figure*}

\begin{figure*}
\begin{centering}
\includegraphics[width=0.46\textwidth]{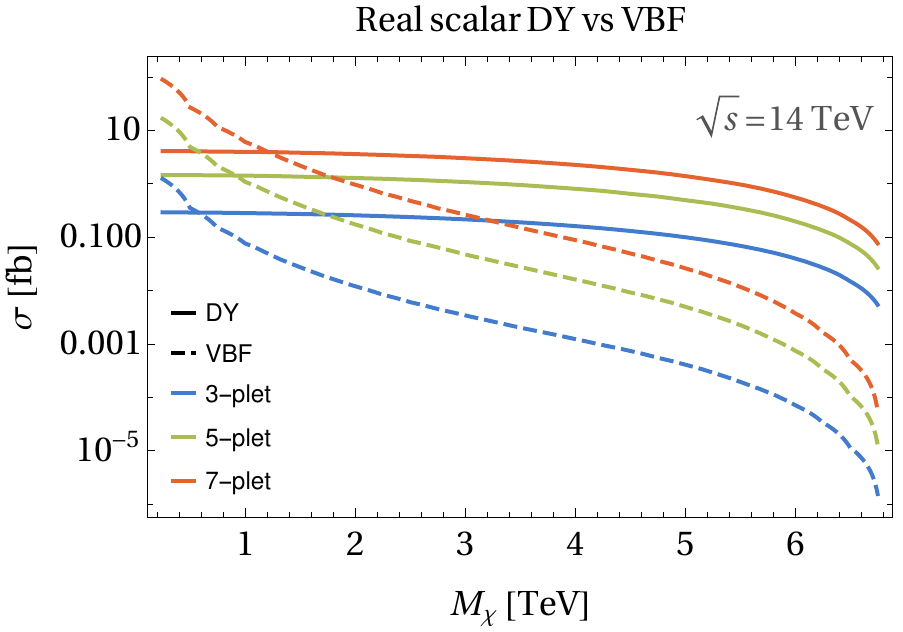}
\qquad\qquad
\includegraphics[width=0.46\textwidth]{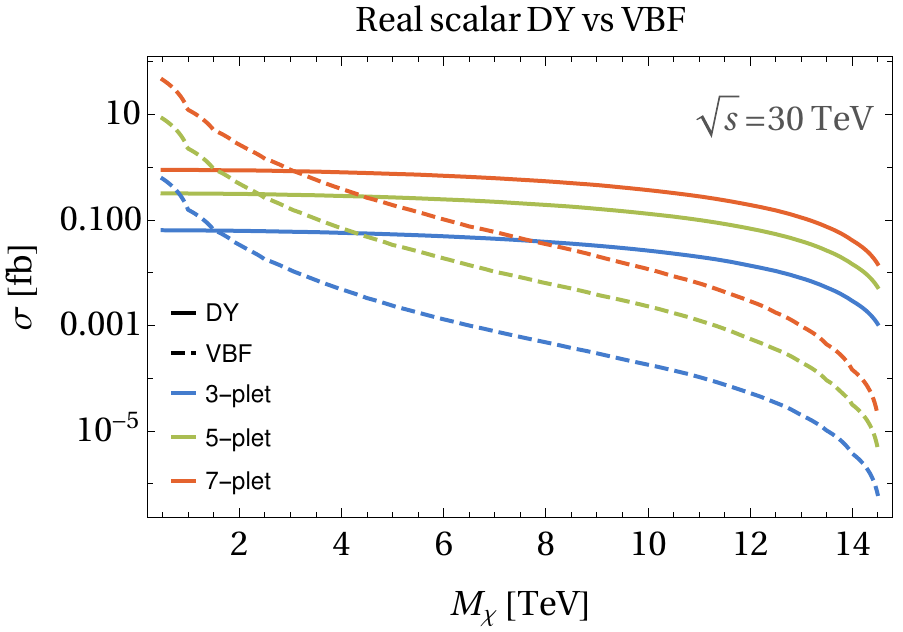}
\end{centering}
\caption{Drell-Yan and $W$-fusion $\chi \chi$ production as a function of $M_\chi$. {\bf Left:} Scalar $3$-plet cross-section for $\sqrt{s} = 14\,{\rm TeV}$.  {\bf Right:} Scalar $5$-plet cross-section for $\sqrt{s} = 30\,{\rm TeV}$.\label{fig:dyvsvbf_scalar}}
\end{figure*}

Probing scalar WIMPs with typical missing mass searches is quite hard.  This is due to multiple reasons: i) the scalar production cross-sections are roughly one order of magnitude smaller than for fermions with same $n$, as shown on the left of \fig~\ref{fig:monow_reach}. A factor of $4$ suppression comes from the lower number of degrees of freedom for  scalar final states, while the remaining suppression comes from a velocity suppressed production cross-section compared to the fermionic case. Since the reach is a very slow function of the mass of the WIMP $M_\chi$, as shown in the right panel of \fig~\ref{fig:monow_reach}, a reduction of the signal cross-section implies a drastic change in the reach. ii) The scalar WIMPs have typically larger freeze-out masses compared to fermionic WIMPs with same EW charge $n$.

All in all, scalar WIMPs give dimmer signals at colliders and are generically heavier than fermionic WIMP. It is thus not surprising that the results expected from collider searches of scalar WIMPs, shown in \fig~\ref{fig:barchartscalar}, are far less exciting than those for fermions in \fig~\ref{fig:barchart}. 
 The overall picture in the landscape of possible beam energy and luminosity options for a future very high energy lepton collider is displayed in \fig~\ref{fig:lumi_vs_energy_scalar}. At variance with the fermionic case presented in \fig~\ref{fig:lumi_vs_energy}, the potential to probe scalar WIMPs with mono-X signals is very limited. Details on the optimized analyses we carried out are given in \tabl~\ref{tab:Scalar}. 
 
 We stress that our results are based purely on Drell-Yan production of $\chi$, which accounts perfectly for the total production rate of WIMPs of mass comparable with $\sqrt{s}$. For significantly lighter WIMPs it is possible to add further production modes and discovery channels, such as production by vector boson fusion and mono-muon channels studied for lighter fermionic WIMPs~\cite{Han:2020uak}, which may result in a bound for light enough scalar WIMPs.
In \fig~\ref{fig:dyvsvbf_scalar} we plotted the cross-sections for scalar $\chi \chi$ production in $W$-fusion (as a representative for VBF modes) and Drell-Yan as a function of $M_\chi$. It can be seen that the VBF cross-section decreases quickly, while DY remains almost constant except near the kinematic threshold. In particular, for the real scalar $5$-plet at $\sqrt{s}=30$ TeV our DY $2\sigma$ reaches can be trusted, as the VBF contribution is smaller than 10\% of the DY one. For the scalar triplet at $\sqrt{s}=14$ TeV, the inclusion of VBF modes is not expected to improve the reach for masses $\gtrsim 1$ TeV. 

\begin{figure*}
\centering
Disappearing tracks --- Scalar 3-plet \hfil\qquad\qquad\qquad\qquad\quad\,\, Disappearing tracks --- Scalar 5-plet \\[-8pt]
\includegraphics[width=0.45\textwidth]{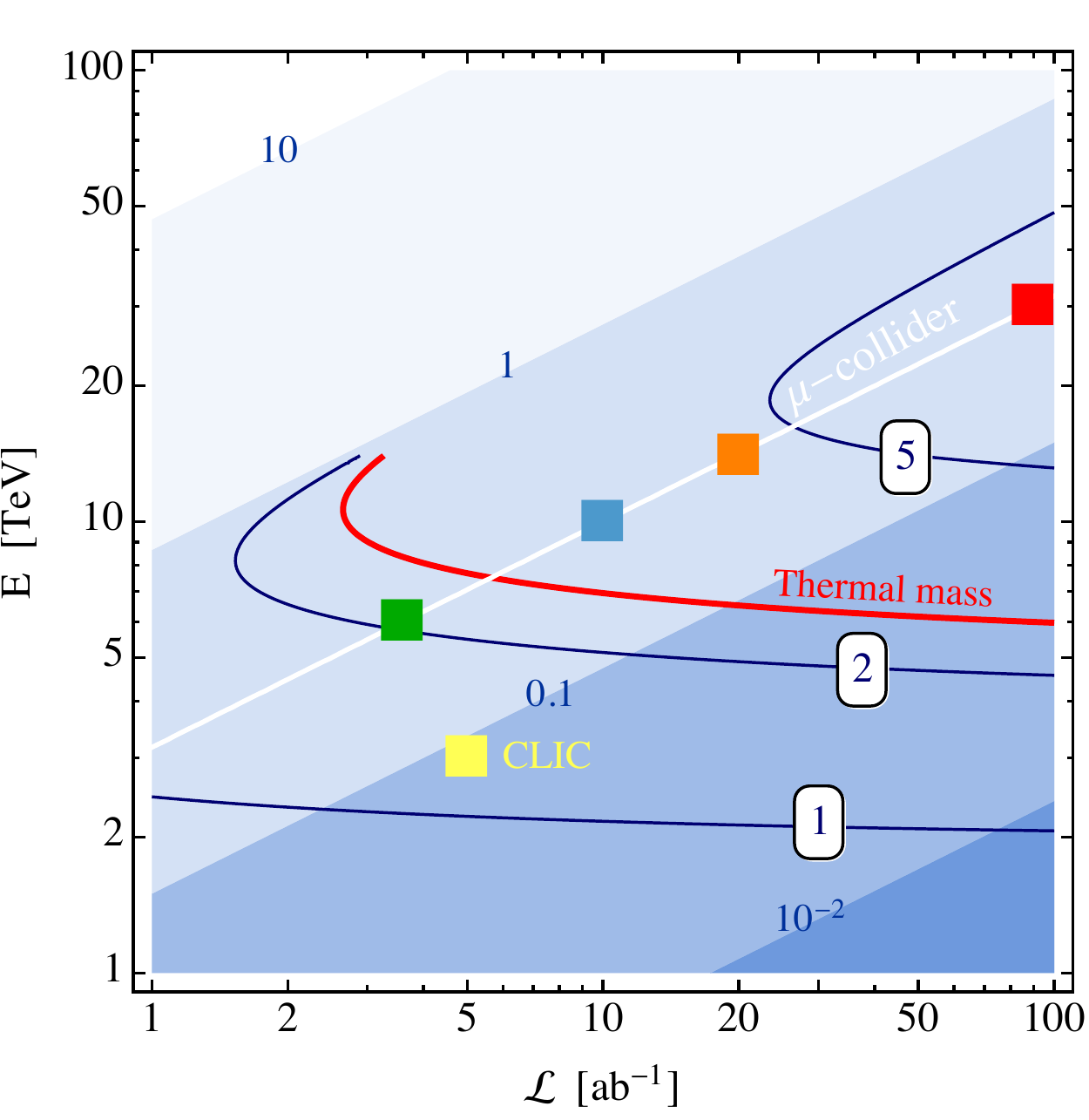}\hfill%
\includegraphics[width=0.45\textwidth]{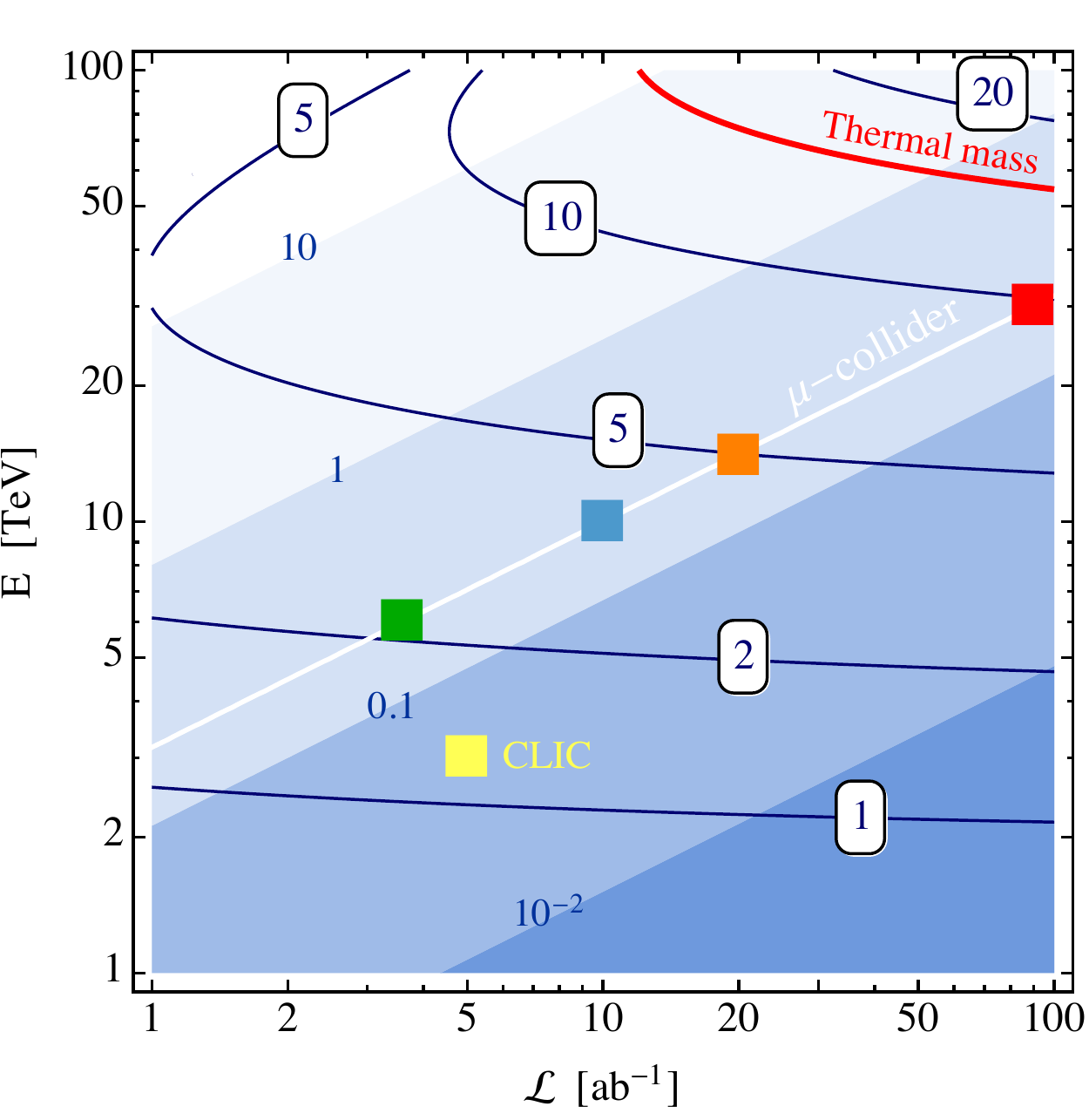}
\caption{Same as \fig~\ref{fig:lumi_vs_energy_tracks}, but for real scalar WIMPs.
{\bf Left:} Scalar 3-plet. {\bf Right:} Scalar 5-plet.\label{fig:lumi_vs_energy_scalartracks}  }
\end{figure*}

It is remarkable that for real scalars the mass splitting between charged and neutral states in the $n$-plet is dominated by EW interactions. Indeed, no splitting term with the Higgs can be written at the quartic level, due to the antisymmetry of the SU(2) contraction. By hypercharge conservation, and assuming the scalar does not get any extra VEV, the leading terms contributing to the mass splitting are dimension $6$ in the SM. Therefore the stub-track prediction is robust and does not depend on peculiar UV completions of the model.
Results for searches of scalar WIMPs from stub-track analyses are reported in \fig~\ref{fig:lumi_vs_energy_scalartracks}. Details on the recasting of results contained in  Ref.~\cite{2102.11292v1} to obtain our results are given in the following \appendixname~\ref{app:DT}.

\subsection{Details of the missing momentum analyses}\label{app:tables}

In \tabl~\ref{tab:Majorana} and \tabl~\ref{tab:Scalar} we provide the results of the optimized cuts for all the considered mono-V and di-V channels, for the case of a Majorana $n$-plet, or a real scalar, respectively. The optimization was carried out in an equally spaced $25 \times 12$ grid in the rectangle 
$\left[0, \sqrt{s}/2 \right] \times \left[0, 2.4 \right]$ in the $p_{T,X}^{\rm{cut}}-\eta_X^{\rm{cut}}$ plane.

We also report the expected number of signal events, the signal-to-noise ratio, and the value of the mass that can be excluded at 95\% C.L. We provide results for muon colliders with $\sqrt{s} = 3, 14, 30$~TeV with integrated luminosity as in \eqref{eq:lumi}, and for systematic uncertainties $\epsilon_{\rm sys} = 0, 1\permil, 1\%$.

Among all the channels considered, the only background that needs some careful treatment is the mono-$W$ one. We split this background in two contributions. For pseudo-rapidities of the final state lost muon $\eta_\mu>\eta_{\mathrm{match}}$ (computed with respect to the direction of the initial state muon with the same charge),
we compute the cross-section of the process $\gamma \mu^\mp \to W^\mp \nu$, using the improved Weizs\"acker-Williams approximation~\cite{Frixione:1993yw}. For $2.5 < \eta_\mu <\eta_{\mathrm{match}}$, we compute the full hard process $\mu^- \mu^+ \to W^\mp\nu \ell^\pm$. The values used for $\eta_{\mathrm{match}}$ are $5.4,7.0,7.5$ for $\sqrt{s}=3,14,30$ TeV, respectively. These values are such that the two background contributions are the same in the pseudorapidity region $\left( \eta_{\mathrm{match}},\eta_{\mathrm{match}}+0.2 \right)$ for the lost muon.

\begin{table*}
\footnotesize
\begin{centering}%
\renewcommand{\arraystretch}{1.2}
\begin{tabular}{|c|c|c||c|c|c|c|c||c|c|c|c|c|}
\cline{4-13}
\multicolumn{3}{c|}{} & \multicolumn{5}{c||}{Majorana 3-plet} & \multicolumn{5}{c|}{Majorana 5-plet}\\
\cline{2-13}
\multicolumn{1}{c|}{} & $\sqrt{s}$ & $\epsilon_{\rm sys}$ & $\eta_{X}^{\rm cut}$& $p_{T,X}^{\rm cut}$ {[}TeV{]} & $S_{95\%}$ & $S_{95\%}/B$& $M_{95\%}$ {[}TeV{]} & $\eta_{X}^{\rm cut}$& $p_{T,X}^{\rm cut}$ {[}TeV{]} & $S_{95\%}$ & $S_{95\%}/B$& $M_{95\%}$ {[}TeV{]}\\
\cline{2-13}
\multicolumn{13}{c}{}\\[-9pt]
\hline
\multirow{9}{*}{\rotatebox{90}{Mono-$\gamma$}} & \multirow{3}{*}{3 TeV} & 0 & 2.4 & 0.18  & 1007 & 0.004 & 0.72 & 2.4 & 0.0 & 3038 & 0.001 & 1.4 \\
 & & 1\permil & 2.2 & 0.24 & 746 & 0.006 & 0.67 & 1.2 & 0.0 & 3683 & 0.003 & 1.3 \\
 & & 1\% & 1.2 & 0.78 & 107 & 0.05 & 0.58 & 0.6 & 0.3 & 639 & 0.02 & 1.1 \\
\cline{2-13}
& \multirow{3}{*}{14 TeV} & 0 & 1.6 & 2.5 & 360 & 0.01 & 2.2 & 2.2 & 0.28 & 3693 & 0.001 & 5.5 \\
& & 1\permil & 1.6 & 2.8 & 323 & 0.01 & 2.2 & 1.2 & 0.84 & 1300 & 0.004 & 5.2 \\
& & 1\% & 1.0 & 4.5 & 108 & 0.05 & 1.9 & 0.8 & 2.8 & 331 & 0.03 & 4.4 \\
\cline{2-13}
& \multirow{3}{*}{30 TeV} & 0 & 1.2 & 7.8 & 174 & 0.02 & 4.4 & 1.6 & 1.8 & 1795 & 0.002 & 11 \\
& & 1\permil & 1.2 & 7.8 & 175 & 0.02 & 4.4 & 1.0 & 2.4 & 1312 & 0.004 & 11 \\
& & 1\% & 1.2 & 8.4 & 190 & 0.03 & 4.0 & 0.8 & 6.0 & 455 & 0.03 & 8.8 \\
\hline 
\hline
\multirow{9}{*}{\rotatebox{90}{Mono-$W$ (inclusive)}} & \multirow{3}{*}{3 TeV} & 0 & 1.6 & 0.36 & 842 & 0.005 & 0.79 & 2.2 & 0.06 & 5625 & 0.0007 & 1.2 \\
 & & 1\permil & 1.4 & 0.48 & 534 & 0.008 & 0.78 & 1.0 & 0.24 & 1649 & 0.004 & 1.2 \\
 & & 1\% & 1.0 & 0.84 & 172 & 0.04 & 0.64 & 0.6 & 0.54 & 515 & 0.02 & 1.0 \\
\cline{2-13}
& \multirow{3}{*}{14 TeV} & 0 & 1.6 & 2.0 & 819 & 0.005 & 3.4 & 1.8 & 0.56 & 5325 & 0.0008 & 5.5 \\
& & 1\permil & 1.6 & 2.2 & 665 & 0.007 & 3.3 & 1.0 & 1.4 & 1342 & 0.004 & 5.2 \\
& & 1\% & 0.8 & 4.2 & 155 & 0.04 & 2.8 & 1.2 & 2.5 & 635 & 0.03 & 4.4 \\
\cline{2-13}
& \multirow{3}{*}{30 TeV} & 0 & 1.4 & 5.4 & 696 & 0.006 & 6.7 & 1.8 & 1.8 & 3946 & 0.001 & 12 \\
& & 1\permil & 1.4 & 5.4 & 606 & 0.007 & 6.7 & 1.4 & 2.4 & 2771 & 0.003 & 11 \\
& & 1\% & 1.0 & 9.0 & 211 & 0.03 & 5.2 & 0.8 & 5.4 & 813 & 0.02 & 9.3 \\
\hline  
\hline
\multirow{9}{*}{\rotatebox{90}{Mono-$W$ (leptonic)}} & \multirow{3}{*}{3 TeV} & 0 & 1.4 & 0.6 & 88 & 0.05 & 0.64 & 2.4 & 0.12 & 1175 & 0.003 & 1.1 \\
 & & 1\permil & 1.4 & 0.6 & 88 & 0.05 & 0.64 & 1.6 & 0.24 & 506 & 0.009 & 1.1 \\
 & & 1\% & 1.4 & 0.6 & 97 & 0.05 & 0.6 & 1.4 & 0.42 & 261 & 0.03 & 1.0 \\
\cline{2-13}
& \multirow{3}{*}{14 TeV} & 0 & 1.4 & 3.1 & 92 & 0.05 & 2.6 & 1.6 & 1.1 & 610 & 0.007 & 5.0 \\
& & 1\permil & 1.4 & 3.1 & 92 & 0.05 & 2.6 & 1.6 & 1.1 & 642 & 0.007 & 4.9 \\
& & 1\% & 1.2 & 3.4 & 77 & 0.06 & 2.5 & 1.4 & 2.0 & 308 & 0.03 & 4.5 \\
\cline{2-13}
& \multirow{3}{*}{30 TeV} & 0 & 1.2 & 7.8 & 72 & 0.06 & 5.1 & 1.6 & 2.4 & 642 & 0.006 & 10 \\
& & 1\permil & 1.2 & 7.8 & 72 & 0.06 & 5.1 & 1.4 & 3.0 & 442 & 0.01 & 10 \\
& & 1\% & 1.2 & 7.8 & 65 & 0.07 & 5.0 & 1.2 & 5.4 & 177 & 0.04 & 9.4 \\
\hline  
\hline
\multirow{9}{*}{\rotatebox{90}{Mono-$Z$}} & \multirow{3}{*}{3 TeV} & 0 & 1.4 & 0.72 & 330 & 0.02 & 0.37 & 1.4 & 0.0 & 1798 & 0.002 & 1.2 \\
 & & 1\permil & 1.4 & 0.72 & 277 & 0.02 & 0.36 & 1.0 & 0.0 & 1946 & 0.003 & 1.2 \\
 & & 1\% & 1.2 & 0.9 & 127 & 0.04 & 0.29 & 0.6 & 0.48 & 563 & 0.02 & 0.9 \\
\cline{2-13}
& \multirow{3}{*}{14 TeV} & 0 & 1.2 & 3.6 & 263 & 0.02 & 1.1 & 1.2 & 0.28 & 4458 & 0.001 & 5.0 \\
& & 1\permil & 1.4 & 3.4 & 273 & 0.02 & 1.1 & 0.6 & 1.4 & 827 & 0.006 & 4.8 \\
& & 1\% & 0.8 & 5.3 & 82 & 0.06 & 0.9 & 0.4 & 3.1 & 260 & 0.03 & 3.7 \\
\cline{2-13}
& \multirow{3}{*}{30 TeV} & 0 & 1.8 & 5.4 & 470 & 0.01 & 2.1 & 1.0 & 1.8 & 2515 & 0.002 & 10 \\
& & 1\permil & 1.6 & 6.0 & 443 & 0.01 & 1.9 & 0.8 & 3.0 & 1159 & 0.005 & 9.8 \\
& & 1\% & 0.8 & 11 & 80 & 0.06 & 1.5 & 0.2 & 6.0 & 267 & 0.03 & 7.5 \\
\hline 
\hline
\multirow{9}{*}{\rotatebox{90}{Di-$\gamma$}} & \multirow{3}{*}{3 TeV} & 0 & 2.4 & 0.42 & 106 & 0.04 & 0.31 & 2.4 & 0.0 & 509 & 0.008 & 1.2 \\
 & & 1\permil & 2.4 & 0.42 & 106 & 0.04 & 0.31 & 1.8 & 0.0 & 404 & 0.01 & 1.2 \\
 & & 1\% & 2.4 & 0.48 & 84 & 0.07 & 0.29 & 1.0 & 0.12 & 160 & 0.04 & 1.1 \\
\cline{2-13}
& \multirow{3}{*}{14 TeV} & 0 & 2.2 & 2.8 & 71 & 0.07 & 1.3 & 1.4 & 0.56 & 331 & 0.01 & 4.8 \\
& & 1\permil & 2.2 & 2.8 & 71 & 0.07 & 1.3 & 1.4 & 0.56 & 332 & 0.01 & 4.8 \\
& & 1\% & 2.0 & 3.6 & 58 & 0.08 & 1.2 & 1.0 & 1.4 & 125 & 0.04 & 4.5 \\
\cline{2-13}
& \multirow{3}{*}{30 TeV} & 0 & 2.4 & 6.6 & 103 & 0.04 & 2.6 & 1.6 & 1.2 & 580 & 0.007 & 9.9 \\
& & 1\permil & 2.4 & 6.6 & 103 & 0.04 & 2.5 & 1.6 & 1.2 & 574 & 0.008 & 9.9 \\
& & 1\% & 2.4 & 9.0 & 47 & 0.1 & 2.4 & 1.0 & 2.4 & 274 & 0.03 & 9.0 \\
\hline  
\hline
\multirow{9}{*}{\rotatebox{90}{Di-$W$ (same-sign)}} & \multirow{3}{*}{3 TeV} & 0 & 2.5 & 0.3 & 6 & 2.6 & 0.32 & 2.5 & 0.3 & 5 & 3.9 & 1.0 \\
 & & 1\permil & 2.5 & 0.3 & 6 & 2.6 & 0.32 & 2.5 & 0.3 & 5 & 3.9 & 1.0 \\
 & & 1\% & 2.5 & 0.3 & 6 & 2.6 & 0.32 & 2.5 & 0.3 & 5 & 3.9 & 1.0 \\
\cline{2-13}
& \multirow{3}{*}{14 TeV} & 0 & 2.5 & 1.5 & 10 & 0.66 & 1.7 & 2.5 & 1.5 & 9 & 0.89 & 4.8 \\
& & 1\permil & 2.5 & 1.5 & 10 & 0.66 & 1.7 & 2.5 & 1.5 & 9 & 0.89 & 4.8 \\
& & 1\% & 2.5 & 1.5 & 10 & 0.66 & 1.7 & 2.5 & 1.5 & 9 & 0.89 & 4.8 \\
\cline{2-13}
& \multirow{3}{*}{30 TeV} & 0 & 2.5 & 3 & 14 & 0.4 & 3.7 & 2.5 & 3 & 12 & 0.52 & 10 \\
& & 1\permil & 2.5 & 3 & 14 & 0.4 & 3.7 & 2.5 & 3 & 12 & 0.52 & 10 \\
& & 1\% & 2.5 & 3 & 14 & 0.4 & 3.7 & 2.5 & 3 & 12 & 0.52 & 10 \\
\hline
\end{tabular}
\end{centering}
\caption{95\% C.L.\ reach on the mass of a Majorana 3-plet and 5-plet from the various mono-X channels. The excluded number of signal events $S_{95\%}$ and the relative precision $S_{95\%}/B$ are also given, together with the values of the optimal event selection cuts on $\eta_X$ and $p_{T,X}$, where $X$ is either the single vector boson or the compound diboson system for Di-$W$ and Di-$\gamma$. The numbers are shown for different collider energies $E_{\rm cm}$ and systematic uncertainties $\epsilon_{\rm sys}$. \label{tab:Majorana}}
\end{table*}

\begin{table*}
\footnotesize
\begin{centering}%
\renewcommand{\arraystretch}{1.2}
\begin{tabular}{|c|c|c||c|c|c|c|c||c|c|c|c|c|}
\cline{4-13}
\multicolumn{3}{c|}{} & \multicolumn{5}{c||}{Scalar 3-plet} & \multicolumn{5}{c|}{Scalar 5-plet}\\
\cline{2-13}
\multicolumn{1}{c|}{} & $\sqrt{s}$ & $\epsilon_{\rm sys}$ & $\eta_{X}^{\rm cut}$& $p_{T,X}^{\rm cut}$ {[}TeV{]} & $S_{95\%}$ & $S_{95\%}/B$& $M_{95\%}$ {[}TeV{]} & $\eta_{X}^{\rm cut}$& $p_{T,X}^{\rm cut}$ {[}TeV{]} & $S_{95\%}$ & $S_{95\%}/B$& $M_{95\%}$ {[}TeV{]}\\
\cline{2-13}
\multicolumn{13}{c}{}\\[-9pt]
\hline
\multirow{9}{*}{\rotatebox{90}{Mono-$\gamma$}} & \multirow{3}{*}{3 TeV} & 0 & 1.2 & 0.9 & -- & -- & --  & 0. & 1.6 & 2749 & 0.002 & 0.79 \\
& & 1\permil & 1.2 & 0.9 & -- & -- & -- & 1.4 & 0.18 & 916 & 0.005 & 0.72 \\
& & 1\% & 1.2 & 0.9 & -- & -- & -- & 0.8 & 0.54 & 252 & 0.03 & 0.53 \\
\cline{2-13}
& \multirow{3}{*}{14 TeV} & 0 & 1. & 5.0 & -- & -- & -- & 1.2 & 1.4 & 809 & 0.005 & 2.6 \\
& & 1\permil & 1. & 5.0 & -- & -- & -- & 1.2 & 1.7 & 619 & 0.007 & 2.5 \\
& & 1\% & 1. & 5.0 & -- & -- & -- & 0.8 & 3.6 & 201 & 0.03 & 2.0 \\
\cline{2-13}
& \multirow{3}{*}{30 TeV} & 0 & 1. & 9.6 & -- & -- & -- & 1.2 & 4.8 & 447 & 0.009 & 5.5 \\
& & 1\permil & 1. & 9.6 & -- & -- & -- & 1.2 & 4.8 & 459 & 0.009 & 5.3 \\
& & 1\% & 0.8 & 11 & -- & -- & -- & 0.6 & 7.8 & 186 & 0.04 & 3.9 \\
\hline
\hline
\multirow{9}{*}{\rotatebox{90}{Mono-$W$ (inclusive)}} & \multirow{3}{*}{3 TeV} & 0 & 1.4 & 0.72 & 213 & 0.02 & 0.23 & 1.4 & 0.36 & 881 & 0.005 & 0.76 \\
& & 1\permil & 1.4 & 0.78 & 213 & 0.02 & 0.22 & 1.2 & 0.48 & 523 & 0.008 & 0.74 \\
& & 1\% & 1. & 0.96 & 118 & 0.04 & 0.2 & 0.8 & 0.78 & 197 & 0.03 & 0.64 \\
\cline{2-13}
& \multirow{3}{*}{14 TeV} & 0 & 1.2 & 4.2 & 181 & 0.02 & 0.82 & 1.6 & 1.7 & 1016 & 0.004 & 3.3 \\
& & 1\permil & 1.2 & 4.2 & 160 & 0.02 & 0.82 & 1.2 & 2.2 & 642 & 0.007 & 3.2 \\
& & 1\% & 0.8 & 5.0 & 80 & 0.06 & 0.72 & 0.6 & 3.6 & 256 & 0.03 & 2.6 \\
\cline{2-13}
& \multirow{3}{*}{30 TeV} & 0 & 1.2 & 9. & 160 & 0.03 & 1.5 & 1.4 & 3.6 & 988 & 0.004 & 6.8 \\
& & 1\permil & 1.2 & 9. & 160 & 0.03 & 1.5 & 1.2 & 5.4 & 605 & 0.007 & 6.7 \\
& & 1\% & 1. & 10 & 103 & 0.05 & 1.3 & 0.6 & 10 & 103 & 0.05 & 5.0 \\
\hline
\hline
\multirow{9}{*}{\rotatebox{90}{Mono-$W$ (leptonic)}} & \multirow{3}{*}{3 TeV} & 0 & 1.2 & 0.84 & -- & -- & -- & 1.6 & 0.48 & 149 & 0.03 & 0.6 \\
& & 1\permil & 1.2 & 0.84 & -- & -- & -- & 1.6 & 0.48 & 150 & 0.03 & 0.6 \\
& & 1\% & 1.2 & 0.84 & -- & -- & -- & 1.2 & 0.6 & 96 & 0.05 & 0.58 \\
\cline{2-13}
& \multirow{3}{*}{14 TeV} & 0 & 1.2 & 4.2 & -- & -- & -- & 1.6 & 2.2 & 178 & 0.02 & 2.5 \\
& & 1\permil & 1.2 & 4.2 & -- & -- & -- & 1.6 & 2.2 & 178 & 0.02 & 2.5 \\
& & 1\% & 1.2 & 4.2 & -- & -- & -- & 1.2 & 3.4 & 82 & 0.06 & 2.3 \\
\cline{2-13}
& \multirow{3}{*}{30 TeV} & 0 & 1. & 10.2 & 30 & 0.2 & 0.94 & 1.4 & 6.0 & 139 & 0.03 & 5.1 \\
& & 1\permil & 1. & 10.2 & 30 & 0.2 & 0.94 & 1.4 & 6.0 & 131 & 0.03 & 5.1 \\
& & 1\% & 1. & 10.2 & 31 & 0.2 & 0.93 & 1.4 & 6.6 & 107 & 0.05 & 4.9 \\
\hline
\hline
\multirow{9}{*}{\rotatebox{90}{Mono-$Z$}} & \multirow{3}{*}{3 TeV} & 0 & 1.4 & 0.72 & -- & -- & -- & 1.2 & 0.0 & 1737 & 0.002 & 0.76 \\
& & 1\permil & 1.4 & 0.72 & -- & -- & -- & 0.8 & 0.18 & 1049 & 0.005 & 0.71 \\
& & 1\% & 1. & 1.0 & -- & -- & -- & 0.8 & 0.72 & 245 & 0.03 & 0.57 \\
\cline{2-13}
& \multirow{3}{*}{14 TeV} & 0 & 1.4 & 3.4 & -- & -- & -- & 1. & 1.4 & 996 & 0.004 & 2.9 \\
& & 1\permil & 1.4 & 3.4 & -- & -- & -- & 0.6 & 1.4 & 815 & 0.006 & 2.7 \\
& & 1\% & 0.8 & 5.3 & -- & -- & -- & 0.6 & 3.9 & 209 & 0.03 & 2.1 \\
\cline{2-13}
& \multirow{3}{*}{30 TeV} & 0 & 1.4 & 7.2 & -- & -- & -- & 1.2 & 3.0 & 1207 & 0.003 & 5.5 \\
& & 1\permil & 1.4 & 7.8 & -- & -- & -- & 1. & 4.2 & 669 & 0.007 & 5.4 \\
& & 1\% & 0.8 & 11 & -- & -- & -- & 0.6 & 7.2 & 340 & 0.03 & 3.7 \\
\hline
\hline
\multirow{9}{*}{\rotatebox{90}{Di-$\gamma$}} & \multirow{3}{*}{3 TeV} & 0 & 1.8 & 0.78 & -- & -- & -- & 1.4 & 0.0 & 318 & 0.01 & 0.63 \\
& & 1\permil & 1.8 & 0.78 & -- & -- & -- & 1.2 & 0.0 & 285 & 0.02 & 0.63 \\
& & 1\% & 1.8 & 0.78 & -- & -- & -- & 1. & 0.18 & 116 & 0.05 & 0.58 \\
\cline{2-13}
& \multirow{3}{*}{14 TeV} & 0 & 2.2 & 3.6 & -- & -- & -- & 1.0 & 1.4 & 117 & 0.04 & 2.6 \\
& & 1\permil & 2.2 & 3.6 & -- & -- & -- & 1.0 & 1.4 & 117 & 0.04 & 2.6 \\
& & 1\% & 2.2 & 3.9 & -- & -- & -- & 1.0 & 1.4 & 135 & 0.04 & 2.5 \\
\cline{2-13}
& \multirow{3}{*}{30 TeV} & 0 & 2.4 & 9.0 & -- & -- & -- & 1.4 & 3.0 & 224 & 0.02 & 5.1 \\
& & 1\permil & 2.4 & 9.0 & -- & -- & -- & 1.4 & 3.0 & 225 & 0.02 & 5.1 \\
& & 1\% & 2.4 & 9.0 & -- & -- & -- & 1. & 4.2 & 116 & 0.05 & 4.8 \\
\hline
\hline
\multirow{9}{*}{\rotatebox{90}{Di-$W$ (same-sign)}} & \multirow{3}{*}{3 TeV} & 0 & 2.5 & 0.3 & -- & -- & -- & 2.5 & 0.3 & 6 & 2.9 & 0.7 \\
& & 1\permil & 2.5 & 0.3 & -- & -- & -- & 2.5 & 0.3 & 6 & 2.9 & 0.7 \\
& & 1\% & 2.5 & 0.3 & -- & -- & -- & 2.5 & 0.3 & 6 & 2.9 & 0.7 \\
\cline{2-13}
& \multirow{3}{*}{14 TeV} & 0 & 2.5 & 1.5 & -- & -- & -- & 2.5 & 1.5 & 10 & 0.71 & 3.4 \\
& & 1\permil & 2.5 & 1.5 & -- & -- & -- & 2.5 & 1.5 & 10 & 0.71 & 3.4 \\
& & 1\% & 2.5 & 1.5 & -- & -- & -- & 2.5 & 1.5 & 10 & 0.71 & 3.4 \\
\cline{2-13}
& \multirow{3}{*}{30 TeV} &  0 & 2.5 & 3 & -- & -- & -- & 2.5 & 3 & 14 & 0.42 & 7.4 \\
& & 1\permil & 2.5 & 3 & -- & -- & -- & 2.5 & 3 & 14 & 0.42 & 7.4 \\
& & 1\% & 2.5 & 3 & -- & -- & -- & 2.5 & 3 & 14 & 0.42 & 7.4 \\
\hline
\end{tabular}
\par\end{centering}
\caption{\label{tab:Scalar} Same as \tabl~\ref{tab:Majorana} but for scalar 3-plet and 5-plet. A `--' indicates that no 95\% C.L.\ exclusion is possible.}
\end{table*}

\subsection{Recasting the disappearing tracks}\label{app:DT}

We   recast the two search strategies discussed in Ref.~\cite{2102.11292v1} that exploit the presence of a single short reconstructed disappearing track or a two-track analysis that require at least one of them to be a short disappearing track, in addition to a trigger photon.  The requirements  are summarized in \tabl~\ref{tab:dtcuts} from Ref.~\cite{2102.11292v1}.

\begin{table}[b!]
\renewcommand{\arraystretch}{1.2}
\begin{center}
\begin{tabular}{|c|c|c|}
\hline
 & Single track (1T) & Double track (2T)\\ 
 \hline
 $E_\gamma$ & $>25$ GeV & $>25$ GeV\\
 \hline
 $p_{\rm T}$ leading track & $>300$ GeV & $>20$ GeV \\
 \hline
 $p_{\rm T}$ subleding track & $/$ & $>10$ GeV \\
 \hline
 $\theta$ leading track & $\frac{2\pi}{9}<\theta<\frac{7\pi}{9}$ & $\frac{2\pi}{9}<\theta<\frac{7\pi}{9}$\\
 \hline
 $\Delta z$ tracks & $/$ & $<0.1$ mm \\
 \hline
\end{tabular}
\caption{Event selections in the two signal regions considered in the original work~\cite{2102.11292v1}.\label{tab:dtcuts}}
\end{center}
\end{table}

{\bf Single-track search.}\;
For the single-track analysis we take the background cross-section quoted in \cite{2102.11292v1}. This rate is mainly determined by  the combinatorial of track reconstruction induced by beam-induced backgrounds.\footnote{As acknowledged in~\cite{2102.11292v1}, this estimate of the background is quite conservative because it is based on detailed beam dynamics simulation for $\sqrt{s}=1.5$~TeV. Due to the relativistic dilution of muon decays, we expect smaller background cross-section at higher $\sqrt{s}$.}
To determine the rate of the single-track events, we compute the mono-photon cross-section doubly differential in the polar angles of the charged particles $\chi_1,\chi_2$. This $d\sigma/d\theta_{1}d\theta_{2}$ is obtained at LO in perturbation theory with \mgfive\ and is further reweighted to take into account angular and distance sensitivity to stub-tracks reported in Ref.~\cite{2102.11292v1}. Let $P(\theta_1)$ be the probability that the particle $\chi_1$ is reconstructed as a track:
\begin{equation}\label{eq:pint}
	P(\theta,r_{\rm min},r_{\rm max})\!=\!\!\int_{r_{\rm min}}^{r_{\rm max}} \!\frac{ \mathrm{d}r \, \epsilon_{\mathrm{rec}}(r,\theta)}{c\tau \beta\gamma \sin \theta} e^{-r/(c\tau\beta\gamma \sin \theta)},\!
\end{equation} 
where $r$ is the transverse radius and $\epsilon_{\mathrm{rec}}(r,\theta)$ is the probability to reconstruct as a track a particle travelling at an angle $\theta$ that decayed at a transverse radius $r$ given in \fig~11 of Ref.~\cite{2102.11292v1}. 
For single tracks $ \epsilon_{\mathrm{rec}}(r,\theta)$ is $0$ outside the interval $r \in [50\,\mathrm{mm}, 127\,\mathrm{mm}]$, and outside $\pi/6<\theta<5\pi/6$. The radial condition reflects the fact that tracks can only be reconstructed if the particles make at least 4 hits in the vertex detector, which for the considered geometry means that the particle must travel at least a minimum distance of 50 mm in the detector, while the upper limit stems from the disappearing condition of the track. The latter condition will be relaxed in the 2-tracks search.
With the knowledge of $\epsilon_{\mathrm{rec}}$ the integral in \eqref{eq:pint} can be performed numerically.
As per \tabl~\ref{tab:dtcuts}, the hard cross-section $\sigma_{S,\gamma}$ is subject to trigger requirements: the leading observed track is required to have 
\begin{equation}\label{ptleading}
p_{\rm T}>300\textrm{ GeV}
\end{equation}
to help discriminate it against fake tracks, and it must lie within the cone
\begin{equation}
\frac{2\pi}{9}<\theta < \frac{7\pi}{9}.\label{thetaleading}
\end{equation}
In our recast, due to lack of a detailed tracking and detector simulation, these cuts are implemented at parton level on the DM particles momenta, which leads us to overestimates the number of events that pass the selection. To account for this effect we assume that only a fraction $\epsilon_{\mathrm{tran}}$ of the events with parton $p_{\rm T} >300$ GeV gives a track whose $p_{\rm T}$ fulfils the same conditions. The transfer factor $\epsilon_{\mathrm{tran}} \approx 0.5$ is estimated from the $p_{\rm T}$ distribution of $\chi$ obtained at generator level, and track $p_{\rm T}$ distribution given in Ref.~\cite{2102.11292v1}. We assume that tracks with $p_{\rm T}>300$~GeV can only come from $\chi$ with $p_{\rm T}> 300$~GeV.
To properly avoid over-counting events with two reconstructed tracks, we divide the final state phase space into two non-overlapping regions that require different reconstruction constraints:
\begin{enumerate}[leftmargin=20pt]
\item Both $\chi$ fulfil the conditions to be considered as leading track (Eq.s~(\ref{ptleading}) and (\ref{thetaleading})). In this case both tracks are subject to the detection and reconstruction efficiencies $\epsilon_{\mathrm{tran}}$ and  $\epsilon_{\mathrm{rec}}\left(\theta,r\right)$. These events may give rise to zero, one, or two reconstructed stub-tracks. We count events with at least one stub-track.
\item Exactly one $\chi$ fulfils the conditions to be considered as leading track. Only events in which this track is reconstructed according to detection and reconstruction efficiencies $\epsilon_{\mathrm{tran}}$ and $\epsilon_{\mathrm{rec}}\left(\theta,r\right)$ are counted.   The fate of the sub-leading $\chi$ (if any) is irrelevant.
\end{enumerate}
The largest contribution to the single-track cross-section comes from events in region i), where both DM particles satisfy the $p_{\rm T}$ and $\theta$ requirements to be considered as a leading track. The preference for this configuration reflects the approximate 2-body kinematics of the mono-$\gamma$ events with small $p_{\rm T}$. 
In order to understand the nature of signal we can split it into two further sub-categories with: a) exactly one reconstructed track which fulfils the conditions \eqref{ptleading} and \eqref{thetaleading}; b) exactly 2 reconstructed stub-tracks, of which at least one fulfils the same conditions. The respective rates are given by:
\begin{equation*}
 \frac{\mathrm{d^{2}}\sigma_{S,\gamma}^{\mathrm{1T}}}{\mathrm{d}\cos\theta_{1}\mathrm{d}\cos\theta_{2}} \cdot \begin{cases}
\epsilon_{\mathrm{tran}}2P(\theta_{1})(1-P(\theta_{2})) & \text{1 track},\\
\big(1-(1-\epsilon_{\mathrm{tran}})^{2}\big)P(\theta_{1})P(\theta_{2}) & \text{2 tracks},
\end{cases}
\end{equation*}
where the hard cross-section $\sigma_{S,\gamma}^{\mathrm{1T}}$ 
is restricted to the phase-space region where both $\chi$ particles fulfil the requirements of Eq.s~(\ref{ptleading}) and (\ref{thetaleading}).
The boost factor $\beta \gamma$ and the angular distribution are both taken from a MC sample with cuts only on the photon at generator level. 

The resulting number of events is used to compute the reach on the DM mass reported in \fig~\ref{fig:barchart}, according to \eqref{eq:Significance} with $\epsilon_{\mathrm{sys}}=0$.

Interestingly, the results obtained from the MC sample can also be understood semi-analytically thanks to the simple kinematics of the mono-photon process.
Given that the photon tends to be soft, the kinematics of the three body process is not too different from direct production of a pair of oppositely charged DM particles without the photon. Therefore a very good analytic approximation of the above results can be obtained, with the $\chi$ boost factor and flight directions approximated by the ones for pair-produced DM particles with energy $\sqrt{s}/2$,
%
\begin{equation}
\beta \gamma \approx \sqrt{\frac{s}{4M_\chi^2}-1} \;,\qquad\quad \theta_1 = \pi + \theta_2.
\end{equation}
The angular distribution can also be computed analytically in the 2-body limit,
$$\frac{1}{\sigma_{S,\gamma}}\frac{\mathrm{d} \sigma_{S,\gamma}}{\mathrm{d}\cos \theta}\propto\begin{cases} 1 + 4\frac{M_\chi^2}{s} + \left(1 -4\frac{M_\chi^2}{s}\right) \cos^2\theta\,, & \mathrm{fermion},\\
\sin^2 \theta\,, & \mathrm{scalar}.\end{cases}$$
Results obtained using the MC 3-body angular distributions are in good agreement with the ones obtained with this analytic two-body approximation.

\medskip

{\bf Double-track search.}\;
The signal of the double tracks is computed by requiring both DM particles to be reconstructed as tracks. The rate in this case is
\begin{equation}\label{eq:doubletrackrate}
\frac{\mathrm{d^2} \sigma_{\mathrm{S, \gamma}}^{\mathrm{2T}}}{\mathrm{d}\cos \theta _1\mathrm{d} \cos \theta_2} P(\theta_1) P(\theta_2)\;.
\end{equation}
We additionally require the two tracks to originate from points that are close to each other along the direction of the beam axis, $\Delta z < 0.1$~mm (see \tabl~\ref{tab:dtcuts}). This effectively reduces the background to negligible levels.
In this limit, we use 4 signal events as a conservative estimate of the 95\% C.L. exclusion for a Poissonian counting.

The angular cuts on the tracks are the same as in the single track case, while the $p_{\rm T}$ cuts are much milder: $p_{\rm T}>10,20$ GeV for the sub-leading and leading tracks, respectively. In this case the mismatch between the $p_{\rm T}$ of the reconstructed track and the $p_{\rm T}$ of the charged $\chi$ obtained at generator level is negligible.
The additional cuts do not affect significantly the signal events.
Note that, following Ref.~\cite{2102.11292v1}, the disappearing condition is required on at least one track, i.e.\ this analysis includes in the signal all events in which the second track extends   up to a transverse radius of $r=1153$ mm. Following Ref.~\cite{2102.11292v1}, we assumed for such long tracks a reconstruction efficiency equal to the tracks decaying between $101 \; \mathrm{mm}<r<127 \;\mathrm{mm}$.
Also for double tracks, the result obtained using the MC sample $\beta \gamma$ and $\theta$ distributions are in agreement with the ones computed analytically in the 2-body limit.

We remark that for SU(2) triplets the double track analysis has a higher exclusion power than the single track analysis, whereas for $n\ge 5$ it has a lower reach. This is due to the shorter life-time $\tau_{\chi} \propto 1/n^{2}$ of larger multiplets, that suppresses the exponential decay factor of \eqref{eq:pint} twice in the double-track rate.

\bibliography{wimp,wimpcol}

\end{document}